\documentclass[usenatbib,useAMS]{mn2e}
\usepackage{epsfig}
\usepackage{amsmath}
\usepackage{amssymb}
\usepackage{ifthen}
\usepackage{subfigure}
\usepackage{rotating}
\usepackage{url}
\usepackage{varioref}
\usepackage{verbatim}

\newcommand{\ctoe}{cell centre to interface}
\newcommand{\atoe}{average to interface}
\newcommand{\atoc}{average to centre}
\newcommand{\ctoa}{centre to average}

\newcommand{\tf}{t_\text{F}}
\newcommand{\err}{\mathcal{E}}
\newcommand{\Lonenorm}{\text{$L_1$}}
\newcommand{\Exp}[2]{\mathrm{e}\text{#1}#2}
\newcommand{\SWENO}[1]{WENO-IFV}
\newcommand{\SWENOuf}[1]{WHAM}
\newcommand{\SWENOnos}{WHAM-IS}
\newcommand{\plotnotation}{The numerical solution is shown with connected dots, and the analytical solution is shown with the solid line}

\newcommand{\RedeclareMathOperator}[2]{\renewcommand{#1}{}\let#1\relax\DeclareMathOperator{#1}{#2}}
\DeclareMathOperator{\MINMOD}{MINMOD}

\defcitealias{lis03}{LW03}
\newcommand{\lwt}{\citetalias{lis03}}
\newcommand{\lwp}{\citepalias{lis03}}
\defcitealias{zha06}{ZM06}
\newcommand{\zhat}{\citetalias{zha06}}
\newcommand{\zhap}{\citepalias{zha06}}

\newcommand{\delone}[1]{\ensuremath{\Delta_{\!R}#1}}
\newcommand{\delonea}[1]{\ensuremath{\Delta_{\!A}#1}}
\RedeclareMathOperator{\det}{Det}
\newcommand{\polyconst}{\Gamma}
\newcommand{\dt}{\Delta t}
\newcommand{\vp}{v_0'}
\newcommand{\den}{\ensuremath{1+\vp t}}
\newcommand{\p}{\partial}
\newcommand{\ppx}{\frac{\p}{\p x}}
\newcommand{\lhs}{l.h.s.\hbox{}}
\newcommand{\rhs}{r.h.s.\hbox{}}
\newcommand{\duF}{{^*}\!\!F}
\newcommand{\MA}{\textsc{ma}}
\newcommand{\EM}{\textsc{em}}
\newcommand{\myfrac}[2]{{^{#1}\!/\hspace{-1pt}_{#2}}}
\newcommand{\simplefrac}[2]{#1/#2}
\newcommand{\gdet}{\sqrt{-g}}
\newcommand{\tensor}[3]{{{#1}^{#2}}_{\!#3}}
\newcommand{\Dx}{{\Delta x^1}}
\newcommand{\Dy}{{\Delta x^2}}

\newcommand{\Dt}{{\Delta t}}
\newcommand{\norm}[1]{\ensuremath{\lVert#1\rVert}}
\newcommand{\avg}[1]{\ensuremath{\langle#1\rangle}}
\newcommand{\abs}[1]{\ensuremath{\lvert#1\rvert}}

\newcommand{\Abs}[1]{\ensuremath{\left\lvert#1\right\rvert}}

\newcommand{\etc}{etc.}
\newcommand{\ie}{i.e.\hbox{}}
\newcommand{\eg}{e.g.\hbox{}}

\newcommand{\wrt}{w.r.t.\mbox{}}
\newcommand{\cf}{c.f.\hbox{}}
\newcommand{\vs}{vs.\hbox{}}
\newcommand{\Order}{\mathcal{O}}
\newcommand{\smoothest}{smoothest}

\newcommand\apj{\rmfamily{ApJ}}%
\newcommand\apjl{\rmfamily{ApJ}}%
\newcommand\apjs{\rmfamily{ApJS}}%
\newcommand\apss{\rmfamily{Ap\&SS}}%
\newcommand\aap{\rmfamily{A\&A}}%
\newcommand\mnras{\rmfamily{MNRAS}}%
\newcommand\prd{\rmfamily{Phys.~Rev.~D}}%
\title[WHAM: A WENO-based general relativistic scheme]{WHAM: A WENO-based general relativistic numerical scheme I: Hydrodynamics}

\author[A. Tchekhovskoy, J.~C. McKinney, \& R. Narayan]
   {Alexander Tchekhovskoy,$^1$\thanks{\hbox{E-mail: atchekho@cfa.harvard.edu~(AT);} \hbox{jmckinney@cfa.harvard.edu~(JCM);} \hbox{rnarayan@cfa.harvard.edu~(RN)}}
    Jonathan C. McKinney,$^2$\footnotemark[1]
    Ramesh Narayan$^2$\footnotemark[1]\\
  $^1$Harvard-Smithsonian Center for Astrophysics, 60 Garden Street, MS 10,
    Cambridge, MA 02138, USA \\
  $^2$Institute for  Theory and Computation, Harvard-Smithsonian Center for Astrophysics,
  60 Garden Street, MS 51, Cambridge, MA 02138, USA
  }

\date{Accepted 2007 April 18. Received 2007 April 3; in original form 2006 December 29}

\pagerange{\pageref{firstpage}--\pageref{lastpage}} \pubyear{2006}

\begin{document}
\label{firstpage}

\maketitle

\begin{abstract}

Active galactic nuclei, x-ray binaries, pulsars, and gamma-ray bursts are all
believed to be powered by compact objects surrounded by relativistic plasma
flows driving phenomena such as accretion, winds, and jets. These flows are
often accurately modelled by the relativistic magnetohydrodynamics (MHD)
approximation. Time-dependent numerical MHD simulations have
proven to be especially insightful, but one regime that remains difficult
to simulate is when the energy scales (kinetic, thermal, magnetic) within the plasma
become disparate. We develop a numerical scheme that significantly
improves the accuracy and robustness of the solution in this regime. We use a
modified form of the WENO method to construct a finite-volume general relativistic
hydrodynamics code called WHAM that converges at fifth order. We avoid (1)
field-by-field decomposition by adaptively reducing down to $2$-point stencils
near discontinuities for a more accurate treatment of shocks, and (2) excessive
reduction to low order stencils, as in the standard WENO formalism, by maintaining
high order accuracy in smooth monotonic flows. Our scheme
performs the proper surface integral of the fluxes, converts cell averaged
conserved quantities to point conserved quantities before performing the
reconstruction step, and correctly averages all source terms. We demonstrate
that the scheme is robust in strong shocks, very accurate in smooth flows, and
maintains accuracy even when the energy scales in the flow are highly disparate.

\end{abstract}

\begin{keywords}
 accretion, accretion discs -- black hole physics --  
 galaxies: jets -- hydrodynamics -- magentohydrodynamics (MHD) -- methods: numerical
\end{keywords}

\section{Introduction}\label{sec_intro}
Astrophysical systems containing compact objects are the brightest and
most efficient engines in the Universe. Enormous
amounts of energy in the form of radiation and outflows are released by accreting black hole systems
at the heart of active galactic nuclei~(AGN,~\citealt{kro99}),
gamma-ray bursts~(GRBs, \citealt{woo93}), and black hole X-ray
binaries~(XRBs, \citealt{lew95}).  Pulsars and their associated
plerions are powered by the relativistic spin-down of highly
magnetized neutron stars \citep{gol69}.
These systems involve strong magnetic fields, relativistic flows, and
strong gravity.  A general framework that has proven to
be accurate in this regime is the ideal magnetohydrodynamic
approximation~(MHD)~\citep{phi83}.

Significant uncertainty has remained in how black hole accretion systems
are powered and produce relativistic jets.  Several seminal works have
outlined plausible key mechanisms. \citet{bal91} showed that the
magnetorotational instability~(MRI) drives the turbulent transport of
angular momentum within an accretion disk.  The MRI generates a dynamo
that amplifies any weak magnetic field in the accreting
plasma. \citet{bp82} found self-similar MHD solutions that
describe winds launched magnetocentrifugally from thin magnetized
accretion disks. \citet{bz77} described a model
that shows how black hole spin energy could be used to power a relativistic jet.
Despite these important advances, it remains unclear how the various mechanisms
coexist and how effective they are in real astrophysical systems.

Time-dependent numerical simulations have demonstrated that an accreting rotating
black hole produces
a disc wind and a pair of relativistic jets as a \emph{natural} outcome of
quasi-steady magnetized disc accretion. \citet{dev03} identified the general morphology of black
hole accretion flows to consist of a disk, a corona, a coronal outflow, and an
evacuated funnel where the magnetized jet develops.
\citet{mck04} showed that the magnetized funnel jet is well-described by
the~\citeauthor{bz77} solution.  Following up on the pioneering simulations by \citet{koi02} and
\citet{koi03}, \citet{kom05} showed that even without a disk the natural outcome
is that black hole spin energy is extracted via the~\citeauthor{bz77} effect.
\citet{hirose04} and \citet{mck05} showed that
the field geometry associated with the Blandford-Znajek effect is the
most stable large-scale field geometry in the presence of a turbulence disk around a black hole.
\citet{km07} showed that the Blandford-Znajek effect operates even for $a=1$ despite
concerns that at high black hole spin the field might get expelled from the horizon.
\citet{mck06jf} demonstrated that the disc models studied by \citet{dev03} and
\citet{mck04} produce a self-consistent magnetized jet with a Lorentz factor of $\gamma \sim 10$ and
an opening angle of $\theta\sim 5^\circ$, demonstrating that the
ideal MHD approximation is sufficient to explain many relativistic
astrophysical jets.  Finally, \citet{mn06a,mck06ff}
showed that the Blandford-Payne magnetically-driven wind does not
properly describe the wind from turbulent disks near black holes and
that the low density magnetized jet from the black hole is collimated
by the disk corona and wind rather than being self-collimated.

Many of the results described above were obtained from simulations that used
numerical methods based upon so-called `conservative schemes,' which have been
proven to be accurate and robust for modeling relativistic flows
\citep{alo99,alo00,font03,mm03,leis04,alo05}. The main advantage of these schemes is that they conserve
the integrals of motion up to machine precision, and this explicit conservation enables them to
accurately resolve discontinuities in the flow. However, systems with kinetic,
thermal, or magnetic energy scales differing by orders of magnitude pose
serious difficulties for conservative numerical schemes. This happens in, \eg,
highly supersonic jets and winds where the internal energy is much smaller than
the kinetic energy. Since conservative schemes evolve the values of momenta and
total energy (the sum of the kinetic, internal, and magnetic energies), even a
relatively small error in the evolution of these quantities can destroy the
accuracy of such flows.  In the hydrodynamic case, this is referred to as the high Mach number
problem~\citep{ryu93,fen04}.  A similar problem occurs in the highly magnetized
MHD regime, where the equations become stiff when the effective magnetic
``Mach'' number $M_\text{mag}\equiv\sqrt\sigma \gg 1$, where $\sigma$ is the comoving magnetic
energy density per unit rest mass energy density. In this paper we focus on
solving the hydrodynamic high Mach number problem, although the ultimate goal
is to eventually apply our method to the MHD equations to study black hole and
neutron star systems that may have $M_\text{mag} \gtrsim 10^3$.

There are alternatives to evolving the total energy equation and some of them
provide a reasonably accurate treatment of high Mach number flows.
One can use a ZEUS-type method~\citep{zeus1} that solves the internal energy
evolution equation with an artificial viscosity~\citep[\eg, ][]{haw00, igu03,
dev03a}. However, the schemes by~\citet{haw00, dev03a} lose energy generated in
reconnecting current sheets and the scheme by~\citet{igu03}, which uses an
artificial resistivity, still does not strictly conserve energy. \citet{buc06}
employ a method that solves the entropy evolution equation and does not allow
any shocks in the solution.

Another approach is the `dual-energy formalism' that involves switching from
evolving the total energy to evolving the internal energy (or entropy)
equations of motion for a Mach number larger than a threshold value $M_\text{th}$.
For instance, $M_\text{th}\sim10$ for \citet{ryu93},
$M_\text{th}\sim5$ for \citet{bry95}, and $M_\text{th}\sim50$ for \citet{fen04}. For example,
\citet{ryu93} switch from evolving the energy to evolving the entropy equation
when (1)~the Mach number is greater than $\sim 10$ and (2)~either the shock
strength is $\lesssim 1/3$ or the flow is diverging. This criterion can fail to
capture the flow properly in the presence of weak shocks (which could dominate
the energy dissipation). Also, once the entropy equation is used to determine
the pressure, the pressure could be secularly underestimated if a shock is
building up.

Finally, there are Lagrangian-type schemes such as by~\citet{tra04}, who use a
moving grid technique for treating high Mach number flows (see
section~\ref{sec_test_1d_sod} for further discussion). However, their scheme
requires a sophisticated treatment of the advection terms in the equation of
motion.  There also exist specialized schemes that are independent
of the Mach number \citep{mar00}.

We describe a general relativistic hydrodynamic
scheme that solves the total energy equation yet
provides an accurate and robust solution in the difficult regime
of high Mach number flows. This is achieved by
(1) being finite-volume (conserves integrals of motion to machine accuracy),
(2) being fifth order accurate,
(3) maintaining fifth order accuracy in monotonic regions
    (which avoids the excessive reduction that occurs in the
    standard WENO formalism),
(4) using adaptive stencil reduction near discontinuities, and
(5) obtaining fluxes from reconstruction of primitive, rather than \eg\ conserved, quantities. Our scheme performs the proper
surface integral of the fluxes and converts cell averaged conserved quantities
to point conserved quantities before the reconstruction step. It also averages
the source terms that appear on the right hand side of the conservation
equations. These averaging and de-averaging procedures enable our
scheme to have an error term that is fifth order in grid cell spacing for
an arbitrary metric and coordinate system. Our scheme
uses a modified form of the WENO method~\citep[see, \eg, ][]{shu97} for spatial
discretisation and the method of lines approach~\citep[see, \eg,][]{lev91} with
Runge-Kutta stepping~\citep{numrec, shu97} for time discretisation. We
eliminate the expensive eigenvector decomposition step used by other high order
methods by adaptively reducing down to $2$-point stencils in the vicinity of
discontinuities. This reduction technique gives a similar accuracy to
eigenvector decomposition without the additional computational expense. On the
other hand by maintaining high order accuracy in smooth monotonic flows,
we are able to avoid excessive reduction to low order stencils that occurs in
the standard WENO formalism.

Our numerical scheme is called WHAM, which stands for WENO high accuracy
magnetohydrodynamics. In section~\ref{sec_equations} we describe the equations
that determine the relevant physics (ideal MHD, single fluid approximation) in
the form used by the scheme. This is followed by a detailed description of the
finite-volume numerical scheme in section~\ref{sec_numerical} and of the fifth
order WENO-type reconstruction procedure in section~\ref{sec_reconstruction}.
Section~\ref{sec_numtests} presents results of an extensive set of tests for
code verification. Section~\ref{sec_limitations} describes the limitations of
the numerical scheme, section~\ref{sec_applications} outlines some possible
applications, and section~\ref{sec_conclusions} summarizes our results and
concludes. In the Appendix we describe the method for reducing the stencil size
near discontinuities, preserving the fifth order of reconstruction in
monotonic regions, and other implementation details.

\section{Governing GRMHD equations} \label{sec_equations}

While this paper focuses on a new purely hydrodynamical numerical
scheme, many of the tools we develop can be applied to a full MHD
scheme.  Therefore we keep our discussion as general as possible.
Here we outline the general relativistic magnetohydrodynamic (GRMHD)
equations of motion in the form of conservation laws.
Throughout,
we use the Einstein summation convention, where Greek letters run
from $0$ to $3$ and Latin letters from $1$ to $3$.  The units are such
that $G = c = 1$.

The continuity equation that describes baryon number conservation is
given by
\begin{subequations} \label{eqs_grmotion}
\begin{equation} \label{eq_continuity}
(\rho u^{\mu})_{;\mu} = 0,
\end{equation}
where $\rho$ is mass density in the fluid frame, $u^\mu$ is the
$4$-velocity of the fluid and the `$;\mu$' subscript denotes the
covariant derivative with respect to $x^\mu$.  The
energy-momentum conservation equation is given by
\begin{equation}  \label{eq_enmom}
\tensor{T}{\mu}{\nu;\mu} = 0,
\end{equation}
where $T^{\mu\nu}$ are the components of the energy-momentum tensor.
Finally, the source-free part of Maxwell's equations describes the
evolution of the fields,
\begin{equation}  \label{eq_maxwell}
\tensor{\duF}{\mu\nu}{;\nu} = 0,
\end{equation}
where $\duF^{\mu\nu} = \myfrac{1}{2} \, \epsilon^{\mu\nu\kappa\lambda}
F_{\kappa\lambda}$ are the components of the dual of the
electromagnetic tensor \citep{mtw}.
\end{subequations}

The energy-momentum tensor can be written as a sum of matter and
electromagnetic contributions, $T^{\mu\nu} = T^{\mu\nu}_\MA + T^{\mu\nu}_\EM$,
with
\begin{align}
T^{\mu\nu}_\MA &= (\rho+u_g+p_g) u^\mu u^\nu + p_g \, g^{\mu\nu}, \\
    T^{\mu\nu}_\EM &= F^{\mu\kappa} \tensor{F}{\nu}{\kappa} -
    \myfrac{1}{4} \, g^{\mu\nu} F^{\kappa\lambda} F_{\kappa\lambda},
\end{align}
where $u_g$ and $p_g$ are the internal energy and pressure in the
fluid frame, and $F^{\mu\nu}$ is the electromagnetic field
tensor. Introducing the comoving magnetic field $4$-vector, $b^\nu =
u_\mu \duF^{\mu\nu}$, in ideal MHD it can be shown that
\begin{equation}
\tensor{T}{\mu}{\nu} =
  (\rho+u_g+p_g+b^2)u^\mu u_\nu
  + (p_g+\myfrac{1}{2} \, b^2) \, {\delta^\mu}_{\!\nu}
  - b^\mu b_\nu
\end{equation}
\citep{gam03}.

For numerical purposes we write down equations \eqref{eq_continuity} -- \eqref{eq_maxwell} in
a coordinate basis. The continuity equation \eqref{eq_continuity} takes
the form \citep{lan2},
\begin{subequations}  \label{eqs_grmotioncoordbasis}
\begin{equation} \label{eq_coord_continuity}
\partial_t( \gdet \rho u^t) + \partial_i( \gdet \rho u^i) = 0,
\end{equation}
where $g = \det g_{\mu\nu}$ is the metric determinant and the `$t$' index denotes the
time  component.  Similarly, the energy-momentum equation \eqref{eq_enmom}
takes the form
\begin{equation}  \label{eq_coord.enmom}
  \partial_t (\gdet\, \tensor{T}{t}{\nu})
  + \partial_i(\gdet\, \tensor{T}{i}{\nu})
  = \gdet\, \tensor{T}{\kappa}{\lambda}  \tensor{\Gamma}{\lambda}{\nu\kappa},
\end{equation}
where $\tensor{\Gamma}{\lambda}{\nu\kappa}$ are the connection coefficients.
Finally, with the introduction of $3$-vectors of magnetic and electric fields, $B^i
\equiv \duF^{it}$ and $E^i \equiv F^{it} / \gdet$,
equation \eqref{eq_maxwell} is equivalent to the induction equation,
\begin{align}
\partial_t( \gdet \, B^i ) &= 
  -  \partial_j[\gdet(\epsilon^{ijk} E_k)]  \notag\\
  &= - \partial_j[\gdet(B^i v^j - B^j v^i)]
  , \label{eq_coord_maxwell}
\end{align}
where $v^i = u^i / u^t$ is the $3$-velocity, plus the no-monopoles constraint,
\begin{equation} \label{eq_nomonopoles}
\partial_i(\gdet \, B^i) = 0.
\end{equation}
In deriving the last equality in equation~\eqref{eq_coord_maxwell} we have used the
ideal MHD approximation $E_i = - \epsilon_{ijk} v^j B^k \equiv - (\mathbf{v}
\times \mathbf{B})_i$.

\end{subequations}

Equations \eqref{eq_coord_continuity}--\eqref{eq_nomonopoles} determine the
evolution of a plasma system for given initial and boundary conditions. We derive
numerical discretisations of these equations in the next section.

\section{Numerical scheme} \label{sec_numerical}

We develop a numerical scheme based upon a solution to a general set
of conservation laws.

\subsection{Vector form of conservation laws}
The equations of motion \eqref{eq_coord_continuity}--\eqref{eq_coord_maxwell} can be
written in the form of a vector conservation law:
\begin{equation}
\label{eq_conserve}
  \p_t \mathbf{U}(\mathbf{P})
+ \p_i \mathbf{F}^i(\mathbf{P})
= \mathbf{S}(\mathbf{P}), \\
\end{equation}
where the vector of conserved quantities is
\begin{align}
\label{eq_integrals}
\mathbf{U}(\mathbf{P}) &\equiv \gdet (
        \rho u^t,
        \tensor{T}{t}{t}+\rho u^t,
        \tensor{T}{t}{j},
        B^k), \\
\intertext{the vector of fluxes in the $i$th direction is}
\label{eq_fluxes}
\mathbf{F}^i(\mathbf{P}) &\equiv \gdet (
  \rho u^i,
  \tensor{T}{i}{t}+\rho u^i,
  \tensor{T}{i}{j},
  B^i {v}^k-{B}^k v^i\makebox[0pt]{$\phantom{_j}$}), \\
\intertext{and the vector of source terms is}
\label{eq_sources}
\mathbf{S}(\mathbf{P}) &\equiv
   \gdet(
       0, \,
       \tensor{T}{\nu}{\lambda} \tensor{\Gamma}{\lambda}{\mu\nu}, \,
       0, \, 0, \, 0), \\
\intertext{where, according to the convention, $i,j,k$ run from $1$ to $3$ and
$\lambda,\mu,\nu$ run from $0$ to $3$
so the vectors $\mathbf{U}$, $\mathbf{F}^i$, and $\mathbf{S}$
have $8$ components each.
We have analytically removed the mass energy density from the conserved quantity
that corresponds to the total energy, the second
component of $\mathbf{U}$, so that the method remains accurate in the
limit of nonrelativistic flows (see Appendix~\ref{sec_cancellaton}).
As a set of primitive quantities we choose}
\label{eq_primitives}
\mathbf{P} &\equiv (\rho, \, u_g, \, \tilde{u}^j, \, B^k),
\end{align}
where the spatial components of the relative $4$-velocity \hbox{$\tilde{u}^j \equiv u^j- \gamma
 \eta^j$}, in which \hbox{$\gamma = -u^\mu \eta_\mu$} is the Lorentz
 factor of the flow as measured in the normal observer frame, and
 $\eta_\mu = (-\alpha,0, 0,0)$ is the $4$-velocity of a zero angular
 momentum observer in the coordinate basis for axisymmetric
 space-times; here $\alpha = (-g^{tt})^{-1/2}$~\citep{nob06}.  One can
 show that $\tilde u^t = 0$ and $\gamma = \sqrt{1+\tilde u^i \tilde
 u_i}$.  Using the relative $4$-velocity (as compared to the usual
 $4$-velocity) has $2$ advantages: ($1$) its interpolation always
 leads to a physical state and ($2$) even in rotating (\eg, Kerr)
 space-times its spatial components determine the unique value of $u^t
 = \gamma/\alpha$.

Note that system~\eqref{eq_conserve} does not contain the no-monopoles
constraint~\eqref{eq_nomonopoles}.  This constraint is considered in
future work, see section~\ref{sec_limitations}.

\subsection{Motivation for a consistent finite volume scheme}
\label{sec_hubble1d}

Eulerian conservative numerical schemes keep track of the values of conserved quantities
for each grid cell. In this section we show the importance of accounting for
the difference between the cell averaged and cell centred values of these
quantities for maintaining high accuracy in studies of highly supersonic
hydrodynamic flows. In such flows small relative errors in the total energy can
lead to large relative errors in the internal energy. This is known as the high
Mach number problem~\citep{ryu93,tra04,fen04}. We show that even though the errors
are second order in grid cell spacing, their magnitude is proportional to the
square of the Mach number and so these errors can destroy the accuracy of
numerical models.

Consider a numerical method based upon piecewise linear interpolations of
primitive quantities, and consider a flow in which
$\rho \propto \text{const}$, $v\propto x$, and
the internal energy is negligible relative to the kinetic energy, $u_g\ll \rho v^2$.
Is a linear interpolation of the conserved energy then sufficient? That is,
can we assume that there is a negligible difference between $E(x) = \rho v^2/2 + u_g \propto x^2$
at the centre $x = x_i$ of a grid cell, and its average value $\avg{E}$ over that grid cell?
The answer is we cannot since for $E(x)$ defined above,
\begin{equation}
\avg{E}-E = \myfrac{1}{24} \, E''(x_i) \, \Delta x^2 = \myfrac{1}{24} \,\rho v^2\bigr|_{x = \Delta x} \gg u_g,
\end{equation}
where $\Delta x$ is the grid cell spacing. Neglecting the difference between $E$ and $\avg{E}$
will manifest as a significant error in the smaller quantity~$u_g$.

In particular, consider a simple nonrelativistic one-dimensional, high Mach number
hydrodynamic flow (no gravity or any other external forces). Suppose
initially the flow has a uniform density and pressure, and its
velocity is a linear function of position. We call this a Hubble-type flow.
The primitive flow
variables (density, velocity, and specific internal energy) at time
$t$ are given by:
\begin{align}
\rho(x,t) &= \frac{\rho_0}{\den},\notag\\
v(x,t) &= \frac{\vp x}{\den}, \label{eq_hubble}\\
u_g(x,t) &= \frac{u_0}{(\den)^\polyconst}, \notag
\end{align}
where $\vp\equiv dv/dx(x,t=0)$. We assume the equation of state $p_g = (\polyconst - 1) u_g$,
where $\polyconst$ is the adiabatic
index of the gas, and we choose the flow parameters $\rho_0$, $u_0$, and $\vp$ such that
the flow is highly supersonic in the grid cell centres, $u_g\ll \rho v^2$.

As we show in Appendix~\ref{sec_hubble_error_analysis},
if one discretises the above equations in space and time to any order but
neglects the difference between $E$ and $\avg E$,
then after one time step one makes a relative error in internal energy,
\begin{equation} \label{eq_hubble_frac_error}
\frac{\err(u_g)}{u_0} \sim - M_\text{min}^2 \delta t,
\end{equation}
where $M_\text{min}^2 \sim \rho_0 \vp^2 \, \Delta x^2/u_0$ is the square
of the minimal value of the Mach number on the grid, $\Delta x$ is the grid cell size, and
$\delta t = \vp\dt$ is the dimensionless time step for the problem for
a computational box with outer edge $x=1$.

At every time step the internal energy uniformly decreases by more than it
should and after several time steps the error will be quite large, and $u_g$ may
even become negative. Even though this error in the internal energy is second
order in space, the coefficient of
this error is proportional to the Mach number of the flow squared. Therefore,
for a large Mach number, any scheme that does not distinguish
between cell averaged and point conserved quantities has to use a
resolution proportional to the Mach number of the flow in order to correctly
capture the evolution of the internal energy over the relevant timescale of
$\delta t \sim 1$. For instance, for $M_\text{min} = 100$ such a scheme
requires roughly a $100$x resolution increase.
See section~\ref{sec_test_1d_hubble} for a numerical verification of
these statements.

\subsection{Numerical grid}
\label{sec_grid}
\begin{figure}
\centering\epsfig{figure=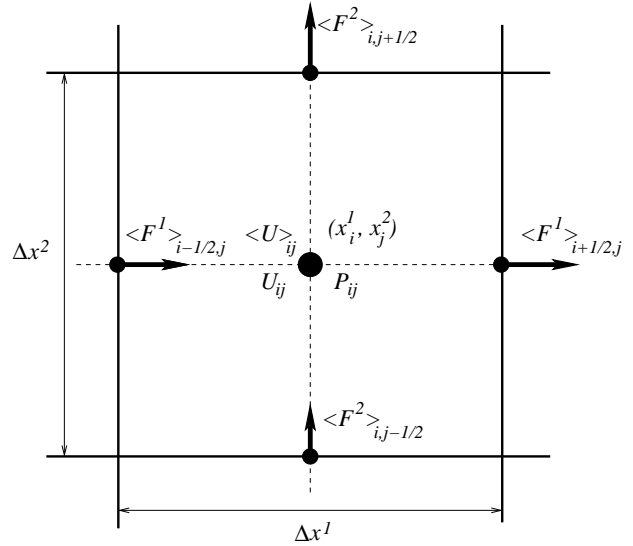,width=\columnwidth}
\caption{Location of quantities within the grid cell $\Delta_{ij}$. Thick solid lines
show the grid of cell interfaces and dashed lines  show the grid of cell centres.
The cell centre is
indicated by the large dot where the primitive $P_{ij}$, point conserved $U_{ij}$,
and average conserved $\avg{U}_{ij}$ quantities are located.  The fluxes are located at the centres
of grid cell interfaces shown by smaller dots. See sections~\ref{sec_grid} -- \ref{sec_fvdiscr}.}
\label{fig_gridcell}
\end{figure}

The numerical scheme is built upon a uniform grid in a coordinate basis, where
for simplicity we consider here the two-dimensional case (the three-dimensional
case is a trivial extension -- also see section~\ref{sec_limitations}). The
grid consists of cells that we define as $\Delta_{ij} = (x^1_{i-1/2},
x^1_{i+1/2})\times (x^2_{j-1/2},x^2_{j+1/2})$, where $(x^1_i,x^2_j,t_n) = (i\Dx,
 j\Dy, n\Dt)$, $\Dx$ and $\Dy$ are the cell spacings (see
figure~\ref{fig_gridcell}), and $\Dt$ is the time step.  For a discussion
of numerical grid generation, see~\citet{tho85}.

An arbitrary smooth transformation can be used to map this computational grid
into physical space as a fixed adaptive mesh that focuses on the regions of
interest \citep[for more detail, see][]{gam03}. The transformation of any
tensorial quantity requires derivatives of the coordinate map that are either
computed analytically or using a simple numerical scheme that results in an
accuracy of roughly the machine accuracy to the $2/3$ power \citep{numrec}. A
similar method is used to compute the connection coefficients. Therefore the
metric identities are of similar accuracy. Both the discretisation of equations
of motion and the reconstruction of functions are independent of the physical
coordinates, and this makes the algorithm simple and general.

\subsection{Finite volume discretisation of the equations}
\label{sec_fvdiscr}

In this section we derive a finite volume discretisation of the
conservation equations \eqref{eq_conserve} for a two-dimensional problem.
This discretisation guarantees that the
integrals of motion are conserved up to machine precision error.
Integrate any component of the vector equation~\eqref{eq_conserve} over a
grid cell $V = \Delta_{ij}$ to obtain
\begin{multline}
 \iint\limits_V \frac{\p U}{\p t} \, dx^1 \, dx^2
+  \iint\limits_V  \frac{\p F^1}{\p x^1} \, dx^1 \, dx^2 \\
+  \iint\limits_V  \frac{\p F^2}{\p x^2} \, dx^1 \, dx^2
=  \iint\limits_V S \, dx^1 \, dx^2.
\end{multline}
Taking the integrals and dividing by grid cell volume $\Dx\Dy$, we obtain
a system of semi-discrete equations~\citep[method of lines, see \eg][]{lev91}:
\begin{multline}  \label{eq_conservediscr_spat}
  \frac{\p \langle U\rangle_{ij}}{\p t}
  + \frac{\langle F^1\rangle_{i+1/2,j} - \langle F^1\rangle_{i-1/2,j}}{\Dx} \\
  + \frac{\langle F^2\rangle_{i,j+1/2} - \langle F^2\rangle_{i,j-1/2}}{\Dy}
  = \langle S\rangle_{ij},
\end{multline}
where the angle brackets denote spatial averaging over the
volume/surface of the grid cell for quantities located inside/at
the surface of the grid cell (Figure~\ref{fig_gridcell}). So far this equation is exact.

We use Runge-Kutta $4$th order accurate time
stepping~\citep[\mbox{RK-$4$}, see][]{numrec} for discretising
the system of ordinary differential equations~\eqref{eq_conservediscr_spat} in time.
Each Runge-Kutta trial step is a first order numerical discretisation of
equation~\eqref{eq_conservediscr_spat} given by:
\begin{multline}  \label{eq_conservediscr_full}
 \frac{\langle U\rangle_{ij}^{n+1} -  \langle U\rangle_{ij}^n}{\Dt}
  + \frac{\langle F^1\rangle_{i+1/2,j} - \langle F^1\rangle_{i-1/2,j}}{\Dx} \\
  + \frac{\langle F^2\rangle_{i,j+1/2} - \langle F^2\rangle_{i,j-1/2}}{\Dy}
  = \langle S\rangle_{ij}.
\end{multline}
The \mbox{RK-$4$} algorithm combines a series of four first-order trial time steps,
which we refer to as substeps,
with different $\Delta t$'s to obtain a $4$th order accurate analog
of~\eqref{eq_conservediscr_full}.
We note that formally we need to use \mbox{RK-$5$} time stepping
in order for our scheme to converge at fifth order.
However, for most problems \mbox{RK-$4$} produces time stepping errors that are much smaller
than the truncation errors due to spatial reconstruction, so \mbox{RK-$4$}
is practically sufficient for attaining fifth order convergence
(\cf\ section~\ref{sec_test_2d_advection} and \citealt{zha06}, hereafter~\zhat; however,
see section~\ref{sec_test_1d_hubble}). We find that  \mbox{RK-$4$}
always gives more accurate results than the \mbox{RK-$3$} scheme from \citet{shu97}, for
both smooth and discontinuous flows.  We solve each Runge-Kutta
substep given by equation~\eqref{eq_conservediscr_full} using the
Godunov technique.

\subsection{High-order Godunov schemes}
\label{sec_highordergodunov}

Godunov schemes perform the evolution in time of a set of grid cell averages of
conserved quantities, $\{\langle{\mathbf{U}}\rangle^{n}_{ij}\}$, by
considering binary interactions between adjacent cells. These interactions are
usually approximated to be one-dimensional. In the simplest case, the first
order Godunov scheme, the distribution of conserved quantity inside each grid
cell is assumed to be a constant. The one-dimensional interaction between two
such distributions is the classical \emph{Riemann problem}: the decay of a
discontinuity between two \emph{constant} distributions~\citep{tor97}. For a given
discontinuity, an exact or approximate \emph{Riemann solution} is obtained for
the flux used to update the conserved quantity and obtain
$\{\langle{\mathbf{U}}\rangle^{n+1}_{ij}\}$ at $t = t_{n+1}$. These
first order schemes are robust but inaccurate for smooth flows or flows with
discontinuities not modelled by the Riemann solver.

Higher-order Godunov schemes attempt to obtain higher accuracy by using higher
order methods to compute both the interface states and the temporal updates.
High-order methods are desirable since typically their computational cost is
much less compared to increasing the numerical resolution for first order
schemes. Higher-order Godunov numerical schemes generate a discontinuity at
each interface as a result of a high-order reconstruction within adjacent grid cells
(see section~\ref{sec_reconstruction}). The left/right interface states are fed
to the Riemann solver that is used to obtain the flux at the interface. However,
such Riemann solutions actually assume the distributions are {\it constant}
within each cell and that the characteristics of the Riemann fan are
linear~\citep[see, \eg, ][]{col84, mig05,zha06}. For example, one of the most
advanced reconstruction schemes is the PPM scheme~\citep{col84} where the
left/right states (fed to a standard Reimann solver) are obtained by an average
over a `domain-of-dependence.' This averaging partially accounts for the
left/right distributions being parabolic, but it still assumes the
characteristics are linear.

Since high-order schemes actually generate {\it non-constant} distributions
in each cell, the Riemann fan is non-linear and so standard Riemann solvers
give an incorrect flux at the interface. This inconsistency leads to incorrect
eigenwave amplitudes and potentially to spurious oscillations. In order to compute
a Riemann solution consistently with the high-order scheme, one should
instead solve the \emph{Generalized Riemann Problem} \citep[GRP, Section 13.4.1
from][]{tor97} that involves non-constant background distributions and so
curved characteristics for the Riemann fan (see also \citealt{tor06}). However, using standard Riemann solvers may
be sufficiently accurate for most problems. For simplicity, like most other
schemes, WHAM uses such a standard Riemann solver.

\subsection{Algorithm outline}
\label{sec_algorithmoutline}

We now describe our implementation of a consistent finite volume scheme, where
by consistent we mean the numerical method takes into account the differences
between point and average conserved quantities, fluxes, and source terms.
Performing the conversion between cell averaged and cell centred values of
conserved quantities is crucial for any test problem with disparate energy
scales, \eg~high Mach number flows. In sections~\ref{sec_hubble1d},
\ref{sec_test_1d_hubble}, \ref{sec_test_1d_sod}, \ref{sec_test_1d_caustics},
and~\ref{sec_test_2d_noh} we give examples of highly supersonic flows that
illustrate the importance of accounting for this difference.

The general procedure of our method is outlined below and can be used with any
definition of the \ctoe\ reconstruction, the averaging and de-averaging of
conserved quantities, the averaging of fluxes, and the averaging of source
terms. Our particular reconstruction methods are described in
section~\ref{sec_reconstruction}.

\begin{figure}
\centering\epsfig{figure=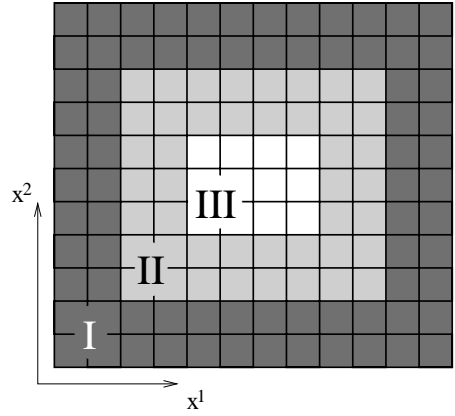,width=0.7\columnwidth}
\caption{Computational grid for an illustrative $4\times3$ resolution.
Grid cells are shown as squares with cell centres being the centres of
these squares.  Upon initialization we set the values of primitive
quantities in the white grid squares (domain III, active grid cells)
according to the initial conditions and in the dark and light shaded
squares (domains I and II, the boundary cells) according to the boundary
conditions; additionally, in domains II and III we compute the average
values of conserved quantities either analytically or numerically.
For domains II and III, at every time step we interpolate the primitive
quantities from cell centres to cell interfaces to obtain the
inter-cell fluxes, average them over the cell faces, and update the
average values of conserved quantities in these domains. Finally, in
domain III, we obtain the values of primitive quantities by converting
the average values of conserved quantities to point values.  In
domains I \& II the primitive quantities are set using the boundary
conditions.  The size of the white active grid cells region can be
arbitrarily large (only limited by memory and speed considerations),
whereas the width of the boundary cell layer surrounding it
(shaded regions) only depends on the order of the scheme used.  }
\label{fig_grid}
\end{figure}

We subdivide the grid into three domains, denoted by Roman numerals in
figure~\ref{fig_grid}. In the standard boundary domain I the primitive
quantities are specified by boundary conditions and in the standard
computational domain~III the conserved quantities are evolved. We also
introduce an additional intermediate domain II where both the conserved
quantities are evolved \emph{and} the values of primitive quantities are
specified according to the boundary conditions. This technique allows our
boundary routines to work only with the primitive quantities in both the
original boundary domain~I and the intermediate boundary domain~II. Without
this additional domain our scheme would require applying the boundary
conditions on the conserved quantities as well. Thus, the introduction of
domain~II simplifies our boundary condition routines, including those that
involve parallel computation. As in the HARM code~\citep{gam03}, our scheme is parallelized using domain
decomposition within the Message Passing Interface (MPI). For
reasonably chosen resolutions per CPU (e.g. $32^2$ to $64^2$) we achieve no less than $70$\% parallel
efficiency for $256$ CPUs for $2$D problems on a cluster with dual $2.0$ GHz
Opteron processors connected by Gigabit ethernet.

The initial conditions are set in domain~III (active grid cells), where we
define primitive quantities at cell centres,~$\{\mathbf{P}^0_{ij}\}$. The
boundary conditions for primitive quantities are set in domains~I~\&~II (boundary grid cells). In
domains I~--~III, we then convert the point primitive quantities to point
conserved quantities~$\{\mathbf{U}^0_{ij}\}$. In domains~II~\&~III, the
conserved quantities are either numerically or analytically averaged over grid
cells to obtain~$\{\langle \mathbf{U}\rangle^0_{ij}\}$.

The cell-averaged conserved quantities are evolved in time in domains II \& III
by a sequence of Runge-Kutta substeps given by \eqref{eq_conservediscr_full}.
For each Runge-Kutta substep the cell-centered primitive quantities
$\{\mathbf{P}^{n+1}_{ij}\}$ and cell-averaged conserved quantities $\{\langle
{\mathbf{U}}\rangle^{n+1}_{ij}\}$ at the new time $t_{n+1}$ are found from
previously computed cell-centered primitive quantities $\{\mathbf{P}^n_{ij}\}$
and cell-averaged conserved quantities $\{\langle
{\mathbf{U}}\rangle^{n}_{ij}\}$ (defined to be at time $t_n$) by the below procedure:

\renewcommand{\theenumi}{Step \roman{enumi}}
\renewcommand{\labelenumi}{Step \roman{enumi}.}
\begin{enumerate}
\item \label{enum_c2e}
Using the \ctoe\ reconstruction on cell centred values of primitive
quantities $\{\mathbf{P}^n_{ij}\}$,
obtain their values on both sides of every cell interface in domains~II~\&~III,
\eg~$\mathbf{P}^n_{i+1/2-0,j}$ and
$\mathbf{P}^n_{i+1/2+0,j}$ for the interface ($i+\myfrac{1}{2}$, $j$);
\item \label{enum_riemann}
Using the approximate HLL Riemann
solver~\citep{gam03}, obtain the flux at the centre of
each interface in domains~II~\&~III,
\eg, $\mathbf{F}^1_{i+1/2,j}(\mathbf{P}^n_{i+1/2-0,j},\mathbf{P}^n_{i+1/2+0,j})$;
\item \label{enum_c2a}
Spatially average the flux through every interface in domains~II~\& III,
using the \ctoa\ reconstruction, to obtain,
\eg, $\langle\mathbf{F}^1\rangle_{i+1/2,j}$;
\item
Find the point values of source terms in domains~I~-- III, $S_{ij}$,
according to equation~\eqref{eq_sources};
\item \label{enum_c2asource}
Average the point values of source terms over grid cells, using the \ctoa\
reconstruction, to obtain the cell-averaged values of source terms
in domains~II~\& III, $\avg{S}_{ij}$;
\item  \label{enum_evolve}
Compute the cell-averaged conserved quantities at the new time in domains
II \& III, $\{\avg{\mathbf{U}}^{n+1}_{ij}\}$, by
equation~\eqref{eq_conservediscr_full};
\item  \label{enum_a2c}
De-average the conserved quantities, using the \atoc\ reconstruction, to obtain the point
values~$\{\mathbf{U}^{n+1}_{ij}\}$, located at grid cell centres, in domain~III;
\item \label{enum_invert}
Obtain the corresponding set of primitive
quantities~$\{\mathbf{P}^{n+1}_{ij}\}$ using a primitive variable
solver~\citet{mig07};
\item \label{enum_boundprim}
Apply the boundary conditions on the primitive quantities to get their values
at grid cell centres in domains~I~\& II;
\end{enumerate}
\renewcommand{\labelenumi}{(\roman{enumi})}
\renewcommand{\theenumi}{(\roman{enumi})}

We refer to the above algorithm as the \mbox{\SWENOuf{n}}~scheme. For the purpose
of demonstrating the importance of distinguishing between average and point values,
we also consider an inconsistent scheme \mbox{\SWENO{n}} that is the same as
WHAM except we disable the averaging and de-averaging procedures (in
the initial conditions and in \ref{enum_c2a}, \ref{enum_c2asource}, and
\ref{enum_a2c}). We also consider yet another scheme, denoted as the
\mbox{\SWENOnos}~scheme, for which we disable only the source integration by
skipping~\ref{enum_c2asource}. This enables us to determine the
effect of not averaging the source terms.\footnote{`IFV'
stands for Inconsistent Finite Volume, and `IS' stands for Inconsistent
Source.}

A feature of our finite volume approach is that conserved quantities are
conserved to machine precision for all time. An alternative is to use the
finite difference approach that evolves the cell centred point values of
conserved quantities, so that the conserved quantities are then conserved to
truncation error~\citep{shu97}. While the finite difference approach can be made to account
for the difference between averaged and point values of conserved quantities,
we find that it is less robust than the finite volume method we have described and less desirable
since it does not conserve the proper quantities.

\section{Reconstruction} \label{sec_reconstruction}

In this section we explain the method WHAM uses to perform the \ctoe\ reconstruction
in \ref{enum_c2e}, the averaging in \ref{enum_c2a} and \ref{enum_c2asource}, and
the de-averaging in \ref{enum_a2c}.

\subsection{Quantities to reconstruct}

High-order schemes are built upon reconstructing some set of quantities from
the cell centre to the cell interface. Commonly either the primitive or
conserved quantities are chosen to be interpolated to obtain the interface
state. We perform the \ctoe\ reconstruction on density and internal energy, and
for the reconstruction of velocity we use a procedure that is more robust in
relativistic shocks~\citepalias[see, \eg, ][]{zha06}. We separately reconstruct
the components of the relative $3$-velocity, $\tilde v^i \equiv \tilde
u^i/\gamma$, and $\gamma$. We then multiply the interpolated value of $\tilde
v^i$ by that of $\gamma$ to get the reconstructed value of the relative
$4$-velocity.

We perform the averaging and de-averaging reconstructions
component-by-component on the conserved quantities and fluxes. To make the
evolution of our scheme failsafe, especially since an arbitrary interpolation
of the conserved quantities and fluxes can lead to an unphysical state, we
implement some features that are more likely to give a physical state (see
Appendix~\ref{sec_failsafeint}).

\subsection{Multi-dimensional reconstruction}

To perform reconstructions in more than one dimension, we use the dimension by
dimension approach. It uses one-dimensional reconstructions, described in the
rest of this section, as building blocks for a multidimensional reconstruction.
This method was found to be faster than a genuine
multidimensional reconstruction and have a similar accuracy~\citep{shi02}. The generalization of the
\ctoe~reconstruction to multi dimensions is straightforward: we perform
one-dimensional \ctoe~reconstructions independently for each dimension.
In several dimensions the \atoc~and \ctoa~reconstructions are performed by
applying one-dimensional reconstructions sequentially dimension by
dimension~\citep{shu97}. We symmetrize this procedure by averaging over all
possible permutations of the sequences for which the one-dimensional
reconstruction can be applied~\citep[see also][]{alo99}. This allows us to
maintain the symmetry of the reconstruction and is our default choice for this
paper. In particular, we maintain \emph{exact} symmetry for two-dimensional
problems.\footnote{The internal registers are often higher precision than the
precision of variables (\eg, 80-bit for Intel 32-bit machines \vs\ 64-bit for
double precision variables) so that one must use a compiler option (\eg, -pc64
for Intel compilers) to disable the use of such internal precision. These compiler
options are only used on those parts of the code that involve operations
with multiple points at the same time.  In the case
of Intel compilers, one must use the -mp compiler option as well in order to
restrict the optimizations to maintain the order of operations.}
For multidimensional problems that do not require the preservation of symmetry,
computational efficiency can be improved by using Strang-type splitting.
Without loss of accuracy, we could do this for the two-dimensional problems described in
sections~\ref{sec_test_2d_jet}, \ref{sec_test_bondi}, and
\ref{sec_test_2d_torus}.  However, we use the same settings throughout the paper
for the sake of showing the ability to use WHAM with one single set of
parameters for a wide range of problems.

\subsection{One-dimensional reconstruction}
\label{sec_1drecon}

Consider a one-dimensional grid along the $x$ axis consisting of grid cells
$\Delta_i \equiv (x_{i-1/2},x_{i+1/2})$ with cell centres located at points
$x_i = x_0 + h \, i$, where $h$ is the grid cell size. Godunov-type schemes
require the conversion of quantities from one type of discretised
representation on the grid to another: \eg, from cell-centred values of density
$\rho_i$ to cell-interface values $\rho_{i+1/2}$.

Since in this example the particular dependence of $\rho(x)$ is unknown
(only known through its discretisation at cell centres $\rho_i$), in order to
get the discretisation of density at cell interfaces, we first reconstruct a
smooth density profile inside each of the grid cells. For this, we use the
values of $\rho_i$ in several grid cells around the grid cell in which the
reconstruction is being performed; this set of grid cells is called the
\emph{stencil}. Then, we combine the reconstructed profiles inside each of the
grid cells to obtain the global reconstruction $\tilde\rho(x)$. Now we can
easily obtain the cell-interface (or any other) representation of density by
evaluating $\tilde\rho(x)$ at the locations of the interfaces (or any other
locations). By construction, the global reconstruction is smooth inside each
grid cell and is in general discontinuous at cell interfaces. For this reason,
we adopt a convention by which $\tilde\rho_{i+1/2} \equiv
\tilde\rho(x_{i+1/2-0})$ and $\tilde\rho_{i-1/2} \equiv \tilde\rho(x_{i-1/2+0})
$, \ie\ the one-half in the indices is replaced with a value infinitesimally
smaller than one-half.

Within piecewise smooth functions the reconstruction is
not unique, and the goal is to find such a reconstruction out of many possible
ones that would give high order of convergence in smooth regions and lead to
minimal spurious oscillations near discontinuities.

Along with the
\ctoe\ reconstruction considered above, this paper makes use of other
reconstruction types that can be formulated similarly.
Section~\ref{sec_typesofreconstruction} below gives more detail.

In the regions where the flow is smooth the reconstruction should be of high
order to ensure fast convergence to the actual solution when the grid is
refined. However, \textit{next to} a kink or a shock, the scheme should avoid
using stencils that contain this non-smooth feature and should instead use
stencils that reside entirely within the smooth regions of the flow: failure to
do so would lead to spurious oscillations.

\subsection{Overview of existing algorithms}

In this section we review some existing methods for reconstruction. The goal of
such methods is to provide higher-order accuracy than a piecewise constant
reconstruction, while avoiding spurious oscillations in the presence of
discontinuities.

Common existing approaches are the following: MINMOD, Monotonized Central~(MC),
van Leer, Piecewise Parabolic Method~(PPM), Essentially
Non-Oscillatory~(ENO) and Weighted Essentially Non-Oscil\-la\-to\-ry~(WENO)
schemes. Throughout the paper, the order of the schemes/reconstructions refers
to the order of their truncation error.

In general, MINMOD, MC, and van Leer schemes are second order in smooth regions
of the flow. Parts of the reconstruction algorithm used in Piecewise Parabolic
Method (PPM) are $4$th order. PPM is a computationally fast reconstruction
method and is able to resolve contact discontinuities well~\citep{col84,gar05,
mig05}. The standard PPM scheme is dimensionally split, resulting in second
order accuracy even for smooth monotonic flows. Also, when used for the
reconstruction of primitive quantities it is in general second order in smooth
monotonic regions because it does not account for the difference between
point and average values of conserved quantities~\citep[\zhat]{blo93}.

The schemes in the above paragraph provide a total variation diminishing (TVD)
reconstruction, \ie\ they do not increase a measure of the number and magnitude
of function extrema with time~\citep{har83}. This produces non-oscillatory and
robust solutions. However, being TVD also means these reconstructions satisfy
the maximum principle that the reconstructed value at an
interface lies in between the values at the two neighboring cell centres.
Therefore, near extrema, where the derivative changes sign, the so-called
\textit{clipping} phenomenon may occur: extrema are flattened thereby increasing
the truncation error of the scheme near smooth extrema and lowering the order of
the scheme there to first order~\citep{har87,jia98}.

Essentially Non-Oscillatory schemes provide a uniformly high order accurate reconstruction,
do not suffer from the clipping phenomenon, provide a total
variation bounded reconstruction (TVB), and give robust solutions for flows with
discontinuities~\citep{har87}. They choose the \smoothest\ stencil
out of a set of possible stencils of a fixed length.
The weakness of this approach is that one needs to choose \textit{one} stencil even if there
are several equivalently smooth ones, e.g., when the function is a constant. Therefore,
unless certain measures are taken \citep[see, e.g.,][]{shu97},
numerical noise, which is always present, may start influencing the stencil
choice procedure. This is referred to as \textit{free stencil adaptation}
and may lead to the numerical noise being amplified. Further, these schemes
require a computationally intensive
eigenvector decomposition to minimize the Gibbs phenomenon near shocks.

Convex Essentially Non-Oscillatory schemes~\citep[CENO,][]{liu98}
are similar to ENO schemes in that they choose one out of the possible stencils
for performing the reconstruction, but include a mechanism for reducing the
stencil size in discontinuities. The latter feature allows these algorithms
to avoid the use of expensive eigenvector decomposition,
making it easier to apply them to general
relativistic problems, as was done by~\citet{and06} and \citet{miz06}, who
achieve third order convergence in smooth flows. However, the CENO
method
can give non-smooth results since the stencil choice may change discretely between
adjacent grid cells.

Weighted Essentially Non-Oscillatory sche\-mes (WENO) are designed to provide
smooth numerical flux and do not suffer from the free stencil adaptation
problem~\citep{shu97}. Instead of using only the \smoothest~stencil, they take
a linear combination with coefficients, called \textit{weights}, of all
possible stencils of a given size. The weights in the linear combination add up
to unity and are distributed in such a way that stencils that contain
discontinuities get extremely small weight. Further, the weights are designed
in such a way that when the function is smooth in all stencils, the weights
become close to the optimal ones so that the resulting linear combination of
the stencils gives a higher order approximation~-- the same order as the one
that the larger, combined, stencil would give. WENO-based schemes are
advantageous because they are able to both capture shocks and accurately
resolve complex smooth flow structure. Several groups have had success in
developing WENO-based schemes in application to relativistic
astrophysics~\citep{rah05,zha06} and cosmology~\citep{fen04,
qiu_weno_rad_transfer06}. The high accuracy of WENO schemes enables them to be used
in studies of high Mach number astrophysical flows~\citep{ha05}.

\subsection{Weighted essentially non-oscillatory schemes}
\label{sec_weno}

Because of the advantages of the WENO method, we have implemented this scheme
in our code. Consider the one-dimensional grid along the $x$ axis that was
defined in section~\ref{sec_1drecon}. Its cell centres are located at points
$x_i = x_0 + h \, i$, where $h$ is the grid cell size. The right interface of
the $i$th cell is located at $x_{i+1/2} =x_0 + h \, (i+1/2-0) $.

We first consider the \emph{cell centre to right interface} reconstruction.
Given the values of a piecewise
smooth function $v(x)$ at the cell centres, $v_i = v(x_i)$, we aim to
reconstruct the function value $v_{i+1/2}$ at the interface $x = x_{i+1/2}$.
Consider $k$ candidate stencils, each of length $k$:
\begin{equation}
s_r(i) = \{x_{i-r},\dots, x_{i-r + k -1}\}, \quad r = 0, \dots, k - 1.
\end{equation}
Each of these stencils produces its own reconstruction of the function value
$v_{i+1/2}$  \citep{shu97}:
\begin{equation}  \label{eq_recon_onestencil}
v^{(s_r)}_{i+1/2} = \sum\limits_{j = 0}^{k-1} c_{rj}\, v_{i-r+j}, \quad r = 0,\dots, k - 1.
\end{equation}
with truncation error of $\Order(h^{k})$ (see Appendix~\ref{sec_optimalweights}
for numerical values of the coefficients $c_{ij}$).

Within the WENO scheme, the reconstructed value of $v(x_{i+1/2})$ is written
as a linear combination of the reconstructions due to each of the stencils
$s_r$:
\begin{equation}
v_{i+1/2} = \sum\limits_{r = 0}^{k-1} \omega_r \, v^{(s_r)}_{i+1/2};
\label{eq_weno_approx}
\end{equation}
the weights $\omega_r$ are all nonnegative and sum up to unity.
The heart of the scheme is the proper choice of these weights.

In the regions where $v(x)$ is smooth, we define a \mbox{WENO-$n$} scheme as one that
delivers a reconstruction of order $n = 2k-1$, the same order as that of
the reconstruction $v^{(S)
}_{i+1/2}$ due to the combined stencil $S =
\bigcup\limits_{r=0}^{k-1} s_r(i)$:
\begin{equation}
v^{(S)}_{i+1/2} = \sum\limits_{j = i-k+1}^{i+k-1} c_j \, v_{j},
\label{eq_total_stencil}
\end{equation}
with truncation error $\Order(h^n)$.
Usually, such constants $d_r$ can be found that for $\omega_r = d_r$ the
reconstructions given by equations \eqref{eq_weno_approx} and
\eqref{eq_total_stencil} are exactly equivalent,~\ie~\hbox{$v_{i+1/2} \equiv
v^{(S)}_{i+1/2}$}. We refer to the constants $d_r$ as the \textit{optimal
weights}; their values are given in Appendix~\ref{sec_optimalweights}.

However, the exact identity is a too strict restriction on the weights. In fact,
the weights $\omega_r$ can deviate from the optimal weights $d_r$ in smooth regions
and still produce a truncation error of the same order: by
\eqref{eq_weno_approx}, requiring that
\begin{equation} \omega_r = d_r +
\Order(h^{k-1}), \quad r = 0, \dots, k -1, \label{eq_weights_requirement}
\end{equation}
guarantees the same order of truncation error \citep[cf.\hbox{}
formula (2.57) in][]{shu97}. Equation \eqref{eq_weights_requirement} is the
requirement on the weights in regions where $v(x)$ is smooth. When there are
shocks, the stencils that contain discontinuities should be given extremely
small weight.

\subsection{Types of reconstruction}
\label{sec_typesofreconstruction}

We have just outlined the cell centre to right interface reconstruction.
It inputs a set of point values at the centres of grid
cells, $\{v_i\}$, and outputs a set of values at the grid cell interfaces,
$\{v_{i+1/2}\}$.

Analogously, we introduce the following types of reconstruction: centre to left
interface, $\{v_i\} \rightarrow \{v_{i-1/2}\}$, centre to average, $\{v_i\}
\rightarrow \{\langle v\rangle_i\}$, and average to centre, $\{\langle
v\rangle_i\} \rightarrow \{v_i\}$. Here $\langle v\rangle_i = h^{-1}
\int_{\Delta_i} v(x)\, dx$, $\Delta_i \equiv (x_{i-1/2},x_{i+1/2})$, is the
average of $v(x)$ over the $i$th grid cell. Apart from the trivial change in
the input and/or output notation,\footnote{For instance, for the \ctoa\
reconstruction one uses $\avg{v}_i$ for the symbol of the output quantity
instead of $v_{i+1/2}$.} formulae \eqref{eq_recon_onestencil} --
\eqref{eq_weights_requirement} hold for all these types of reconstruction as well.
Appendix~\ref{sec_optimalweights} gives the values of the coefficients
$c_{ij}$ and $d_r$ for each reconstruction type used by the scheme.

\subsection{WENO prescription for weights}
\label{sec_weightsprescription}

In the WENO scheme~\citep{shu96,shu97}, the weights are computed based on
\textit{smoothness indicators}, $\beta_r$. These indicators give a measure of
the variation of $p_r(x)$~-- the reconstruction polynomial due to the stencil
$s_r(i)$~-- within the grid cell $\Delta_i$. The smoothness indicators
are defined as:
\begin{equation}
\beta_r = \frac{1}{h} \int_{x_{i-1/2}}^{x_{i+1/2}}
\left[\sum_{n=1}^{k-1}\left(\frac{\partial^n p_r(x)}{\partial x^n} h^n\right)^2
\right] dx,
\label{eq_smoothness_indicator_definition}
\end{equation}
where the factors of $h$ are included to remove any grid cell size dependence
of $\beta_r$. According to the definition
\eqref{eq_smoothness_indicator_definition}, a smoothness indicator is the
average over the cell of interest of the sum of the squares of all
derivatives of the interpolating polynomial. Therefore, a smoothness indicator
measures how deviant from a constant the reconstruction is.
The smaller the smoothness indicator, the flatter
the reconstructed profile, \ie~the smaller the change of the
interpolating polynomial $p_r(x)$ over the cell of interest.  The smoothness indicators
defined by \eqref{eq_smoothness_indicator_definition} are
specified for WENO-5 by equation~\eqref{eq_bn} in appendix~\ref{sec_weno5order}.

\citet{shu96} define the unnormalized weights in the following way:
\begin{equation}
\widetilde\omega_r = \frac{d_r}{(\epsilon + \beta_r)^2},
\label{eq_weno_unnorm_weights}
\end{equation}
which, when normalized, become:
\begin{equation}
\omega_r = {\widetilde\omega_r}/{\Omega},
\label{eq_weno_norm_weights}
\end{equation} where
$\Omega =\sum_{r=0}^{k-1}\widetilde\omega_r$ is the sum of unnormalized weights and
$\epsilon$ is a small positive parameter.

The~$\epsilon$ parameter in~\eqref{eq_weno_unnorm_weights} is introduced to avoid
division by zero. Values used by workers vary from $10^{-4} - 10^{-6}$
\citep{shu96,shu97,tit04,zha06} to $10^{-10}$ \citep{bal00}.
However, setting this parameter in any problem independent
way introduces an artificial scale into the problem since discontinuities
smaller than this scale are considered to be part of the smooth flow.
It is therefore preferable to dynamically choose $\epsilon$ such that
it is large enough to avoid the triggering of the scheme on
machine level noise and yet small enough not to influence the weights
calculation in other cases~\citep[see, however,][]{shu96}. We describe the
procedure we use in Appendix \ref{sec_epsilon}.

Note that for \ctoa\ and \atoc\ reconstructions some of the optimal
weights become negative, and using negative weights can lead to an instability~\citep{shi02}.
This problem can be
avoided by a simple technique that keeps the sum of the
absolute values of the weights bounded~\citep{shi02}.  We use this technique for both
\atoc\ and \ctoa\ reconstructions.

Equation \eqref{eq_weno_norm_weights} assigns larger weights to the stencils
with smaller smoothness indicators, \ie, with flatter reconstruction profiles
and smaller degree of oscillation. These weights become extremely small for a
stencil that contains a discontinuity; this is referred to as \emph{adaptive
stencil choice} and provides the nonoscillatory property of the scheme. At the
same time, weights~\eqref{eq_weno_norm_weights} are designed to be
close to the optimal weights according to~\eqref{eq_weights_requirement}
for a smooth flow locally well-approximated by a parabola.
Therefore, for such flows, the WENO-$5$ scheme is fifth order in space. However,
there are other flows for which the scheme becomes third order.

\subsection{Convergence properties of WENO-type schemes in smooth flows}
\label{sec_conv_weno}

Preserving a high order interpolation near extrema and at the same time
maintaining the nonoscillatory property of the solution is a challenging
problem \citep{har87}. For instance, all schemes that satisfy the
monotonicity constraint reduce to first order near extrema.

WENO-$5$ reconstruction is very appealing in that it maintains high order near
extrema. For a common type of extremum, one with a nonvanishing second
derivative, it is claimed to be fifth order by \citet{shu96,shu97}. However, we
have found that in general the WENO-$5$ scheme becomes fourth order near such
extrema. Further, for a smooth flow whose Taylor series expansion is dominated
by third or higher order terms, the WENO-5 scheme reduces to third order.
Appendix~\ref{sec_weno5order} discusses the convergence properties of the
WENO-5 scheme in more detail and gives expressions for the smoothness
indicators. To help minimize the excessive reduction of order by the
standard WENO-5 scheme, our scheme uses the full
stencil~\eqref{eq_total_stencil} in regions where this stencil gives
monotonic values for the function and all of its derivatives (see Appendix
\ref{sec_jmono}).

One can construct WENO-type schemes of an arbitrarily high order (\eg\ seventh,
ninth, \etc) that deliver an even higher order when the
lower-order derivatives vanish. Such schemes provide a much better resolution
of contact discontinuities and smooth features of the flow than their
lower-order counterparts, given the same grid cell size~\citep{lat06}. However,
they also require more computational effort \citep[see, \eg, ][]{qiu02}.
Appendix~\ref{sec_higherorderindicators} proves some properties of higher order
WENO-type schemes, discusses their benefits, and points out their limitations
for handling critical points in smooth flows.

The use of WENO weights alone is
known to be insufficient in avoiding spurious oscillations near discontinuities,
and a known remedy is to use eigenvector decomposition.  In the next section, we
discuss our alternative approach that does not require computing the eigenvectors.

\subsection{Oscillation-free reconstruction}
\label{sec_oscillatoryfreerecon}

Interpolating the primitive, conserved, or any other arbitrary quantities can
in general lead to spurious oscillations. Such oscillations can be introduced
because the interpolations allow arbitrary mixing between different eigenmodes.

One can avoid significant spurious oscillations by reconstructing the so-called
wavestrengths, the wave amplitudes in the local characteristic
fields~\citep{qiu02}. We generically refer to this procedure as the eigenvector
decomposition approach. Interpolating each wavestrength individually in most
cases leads to only small mixing between different eigenmodes~\citep{har87}. There
are several types of reconstructions based on this idea: characteristic-wise
reconstruction, field-by-field reconstruction, \etc~\citep{har87,shu88,shu89,
shu97}. All are computationally intensive and require sophisticated coding that
becomes complicated for the GRMHD case.

We adopt an alternative approach that involves locally reducing the order of
the scheme near discontinuities~\citep{col84, liu98, mig05}. The Riemann solver
is fed interface values that are close to first order near shocks. This procedure
reduces the mixing between different eigenmodes and makes the scheme more robust.
Further, since this approach does not require the computation of any
eigenvectors, it can be used to solve weakly hyperbolic systems. It is
computationally more efficient than eigenvector decomposition, and is easier to
implement~\citep{liu98}. See Appendix~\ref{sec_reduction} for a discussion.

\section{Numerical tests} \label{sec_numtests}
\begin{table*}
\begin{center}
\caption{Parameter values selected for the one-dimensional nonrelativistic
Riemann problems, shock - entropy wave interaction
problem, and relativistic Riemann problems.  All problems are run on the
interval $(0,1)$, except the shock - entropy wave
interaction problem on $(-5, 5)$,
Sod's problem on $(-1.5, 1.5)$, and moving Sod's problem on $(-1.5, 25.5)$.
The initial position of the discontinuity is located at $x = x_0$,
the tests are run until the final
time $t_F$, with gas adiabatic index of $\polyconst$, with $N$ grid cells.
Here $\rho$ is the density, $v_x$ and $v_y$ the $3$-velocity components,
and $p_g$ the pressure.
`L' denotes the left state and `R' the right state. }
\begin{minipage}{\textwidth}
\begin{tabular}{@{\,}lc@{\quad}c@{\quad}c@{\quad}c@{\quad}c@{\quad}c@{\quad}c@{\quad}c@{\quad}c@{\quad}c@{\quad}c@{\quad}c@{\quad}c@{\,}}
\hline
Test & $\rho_L$ & $v_{x,L}$  & $v_{y,L}$    &    $p_{g,L}$  & $\rho_R$ & $v_{x,R}$  & $v_{y,R}$    & $p_{g,R}$   & $\polyconst$  & $x_0$ & $t_F$ & N & Sec.\\
\hline
Sod                   &  $1$   &  $0$   & $0$  & $1$  & $0.2$ &  $0$   & $0$    & $0.01$ & $5/3$ & $0.0$ & $0.77$ & $150$ & \ref{sec_test_1d_sod}\\
Moving Sod            &  $1$   & $28.87$& $0$  & $1$  & $0.2$ &  $28.87$& $0$   & $0.01$ & $5/3$ & $0.0$ & $0.77$ & $1350$ & \ref{sec_test_1d_sod}\\
Noh                   &  $1$   &  $1$     & $0$  & $0$ &    $1$   & $-1$   & $0$    & $0$ & $5/3$ & $0.5$ & $1$ & $100$ & \ref{sec_test_1d_noh}\\
Moving contact        &  $1.4$   &   $0.1$ & $0$  &   $1$     &    $1$ & $0$   &   $0.1$   & $1$ & $1.4$    & $0.5$ & $2$ & $100$& \ref{sec_test_1d_contact} \\
Test 4                &$5.99924$ & $19.5975$ & $0$ & $460.894$ & $5.99242$ & $-6.19633$ & $0$ & $46.095$ & $1.4$ & $0.4$ & $0.035$ & $200$& \ref{sec_test_1d_test4}\\
Shock - entropy  & $3.857143$ &  $2.629369$ & $0$ & $10.33333$ & $1+0.2 \sin(5x)$ & $0$ & $0$ & $1$ & $1.4$ & $-4$ & $1.8$ & $400$ & \ref{sec_test_1d_shockentropy}\\
\hline
Rel. problem 1 &  $10.0$  &   $0$ & $0$  &   $13.33$     &    $1.0$ & $0$   &   $0$   & $10^{-8}$    & $5/3$    & $0.5$ & $0.4$ & $400$  & \ref{sec_test_1d_rel_101} \\
Rel. problem 2 &  $1.0$   &   $0$ & $0$  &   $1000$     &    $1.0$ & $0$     &   $0$   & $10^{-2}$    & $5/3$    & $0.5$ & $0.4$ & $400$   & \ref{sec_test_1d_rel_102}\\
Rel. problem 3 &  $1.0$   &   $0.9$ & $0$  &   $1$      &    $1.0$ & $0$   &   $0$   & $10$         & $4/3$    & $0.5$ & $0.4$ & $400$     & \ref{sec_test_1d_rel_103}\\
Rel. problem 4 &  $1.0$   &   $0$ & $0$  &   $1000$     &    $1.0$ & $0$   &   $0.99$& $10^{-2}$    & $5/3$    & $0.5$ & $0.4$ & $400$     & \ref{sec_test_1d_rel_104}\\
Rel. problem 5 &  $1.0$   &   $0$ & $0.9$  &   $1000$     &  $1.0$ & $0$   &   $0.9$   & $10^{-2}$    & $5/3$    & $0.5$ & $0.6$ & $400$   & \ref{sec_test_1d_rel_106}\\
Rel. problem 6 &  $1.0$   & $1-10^{-10}$&$0$ & $0.001$ & \multicolumn{4}{@{\qquad\quad}l@{\qquad\quad}}{(uses reflecting boundary at $x=1$)} & $4/3$    & $1.0$ & $2$   & $100$ & \ref{sec_test_1d_rel_105}\\
\hline
\label{tab_riemtests}
\end{tabular}
\end{minipage}
\end{center}
\end{table*}

We have confirmed the accuracy and robustness of our code by running an extensive series of
tests.  We begin the discussion with four tests that show the importance of
distinguishing between the average and point values of conserved quantities.
We then discuss a number of standard tests in the literature.  Our
scheme successfully evolves all nonrelativistic hydrodynamic tests
from~\citet[\lwt]{lis03} and all chosen tests from~\citet[\zhat]{zha06}, but we
only discuss those tests we found to be most challenging.  Finally, we discuss
some general relativistic tests from~\citet{gam03}.

Table~\ref{tab_riemtests} provides detailed
information on some of the one-dimensional test problems.  Unless otherwise indicated, we use the
exact Riemann solvers by~\citet{tor97} and~\citet{gia06} to obtain the exact solutions
for nonrelativistic and relativistic one-dimensional Riemann test problems, respectively.

For nonrelativistic problems we set the value of the speed of light in the scheme to $10^{10}$
so that any velocity on the order of unity is then nonrelativistic to machine accuracy
(i.e. $\gamma-1=0$).  All problems use the ideal gas equation of state, $p_g = (\Gamma - 1) u_g$, where $\Gamma$ is
the adiabatic gas index.

For all tests, we use a \emph{constant} set of numerical parameters that control the
behavior of the scheme.  This is as opposed to many other works that fine-tune their numerical
parameters in order to make some tests work.  The fact that we are able to run all the tests
with a single set of parameters proves the robustness of our
method as a single scheme to study a vast array of problems.  We use \SWENOuf{5} scheme
with a Courant factor of $0.5$ and the approximate HLL Riemann solver for all tests.

\subsection{Smooth high Mach number flow: Hubble-type flow}
\label{sec_test_1d_hubble}

This is a simple high Mach number flow problem that tests the ability of
a numerical scheme to handle flows with disparate energy scales.
The problem is defined in section \ref{sec_hubble1d}.  For the test described
here the flow parameters are set as follows:
\begin{align}
\rho_0 &= 1, \\
\vp &= 1,\\
u_0 &= 4 \times10^{-8},
\end{align}
with $\Gamma = 1.4$.  We employ a resolution of $64$ cells on the
interval $(-1, 1)$, so that the Mach number varies from
$M_\mathrm{min} \approx 104.4$ to $M_\mathrm{max} \approx 6577.1$ at the grid cell centres.
For the values in the boundary cells we use linearly extrapolated values of
the primitive quantities.

The Hubble-type flow problem illustrates how crucial it is to account for the difference
between cell averaged and cell centred conserved quantities in high Mach
number flows.
\begin{table}
\begin{center}
\caption{
Relative \Lonenorm-error and convergence order in internal energy
at the final time \mbox{$t_\text{F} = 1$} for the Hubble-type flow
problem (section~\ref{sec_test_1d_hubble}).
The Courant factor is $C$, and the resolution is $N$. The \SWENO{5},
HARM~\citep{gam03}, and Athena~\citep{sto06} schemes converge at second order in space and at
zeroth order in time in agreement with equation~\eqref{eq_hubble_frac_error}. One has to
increase the resolution by a factor of more than $100$ to match the performance
of the WHAM scheme. WHAM converges at $4$th order in space and
time, see the note after equation \eqref{eq_conservediscr_full}.
}
\setlength{\fboxsep}{2pt}
\begin{minipage}{\textwidth}
\begin{tabular}{@{\,}ccccc@{\,}}
\hline
Scheme & $C$ & $N$ & \delone{u_g} & Order\\
\hline
WHAM        & $0.5$ & \framebox{$64$} & \framebox{$5.2\Exp{-}{2}$}  \\ 
            & $0.5$ &  $128$ & $4.3\Exp{-}{3}$ & 3.6 \\ 
            & $0.5$ &  $256$ & $3.1\Exp{-}{4}$ & 3.8 \\ 
            & $0.5$ &  $512$ & $2.1\Exp{-}{5}$ & 3.9 \\ 
            & $0.5$ & $1024$ & $1.4\Exp{-}{6}$ & 3.9 \\ 
            & $0.05$ & $64$  & $5.2\Exp{-}{6}$ & 4.0 \\ 
\hline
\SWENO{5}   & $0.5$  & $64$  & $1278$ \\ 
            & $0.5$ & $640$  & $12.78$ & 2.0 \\ 
            & $0.5$ & $6400$  & $12.78\Exp{-}{2}$ & 2.0\\ 
            & $0.5$ & \framebox{$9600$}  & \framebox{$5.7\Exp{-}{2}$} & 2.0\\ 
            & $0.5$ & $25600$  & $0.8\Exp{-}{2}$ & 2.0 \\ 
            & $0.05$ & $64$  & $1278$ & 0.0 \\ 
\hline
HARM        & $0.5$  & $64$    & $773.3$ \\ 
            & $0.5$  & $640$   & $7.8$ & 2.0 \\ 
            & $0.5$  & $6400$  & $7.8\Exp{-}{2}$ & 2.0 \\ 
            & $0.5$ & \framebox{$7680$}  & \framebox{$5.4\Exp{-}{2}$} & 2.0 \\ 
            & $0.5$  & $9600$  & $3.4\Exp{-2}$ & 2.0 \\ 
            & $0.05$  & $64$   & $1273$ & -0.2 \\ 
\hline
Athena      & $0.5$  & $64$    & $2527$ & \\ 
            & $0.5$  & $640$   & $30.43$& 1.9\\ 
            & $0.5$  & $6400$  & $0.14$ & 2.3\\ 
            & $0.5$  & \framebox{$12000$} & \framebox{$4.2\Exp{-}{2}$} & 1.9 \\ 
            & $0.05$ & $64$    & $2459$ & 0.01\\ 
\hline
\label{tab_hubble_uerr}
\end{tabular}
\end{minipage}
\setlength{\fboxsep}{0pt}
\end{center}
\end{table}
Table~\ref{tab_hubble_uerr} shows the relative
\Lonenorm-er\-ror in the internal energy and other quantities at the
characteristic time of evolution, $t_\mathrm{F} = 1$, for the WHAM,
\SWENO{5}, HARM,\footnote{The HARM scheme converges at
second order in space and time and does not account for the difference between
cell averaged and cell centred values of conserved quantities, see~\citet{gam03}.}
and Athena\footnote{We use Athena 2.0~\citep{sto06} with third order spatial interpolation
and the Roe solver. We
use a Courant factor of $0.5$. For setting the values in the boundary grid cells we use
quadratically extrapolated values of conserved quantities.} schemes.  We define the absolute
$\Lonenorm$-error norm of any quantity $u$ as
\begin{equation} \label{eq_Lonenorm}
\delonea{u} \equiv (\Delta u)_\Lonenorm = \simplefrac{ \sum_j {\abs{u^\text{numerical}_j - u^\text{exact}_j}}}{N},
\end{equation}
where $N$ is the number of elements, and the relative $\Lonenorm$-error norm as
\begin{equation}\label{eq_rel_Lonenorm}
\delone{u} \equiv (\Delta u/u_\text{max})_\Lonenorm = \delonea{u}/{\max_i\abs{u^\text{exact}_i}}.
\end{equation}
Since WHAM performs the
proper conversion between cell averaged and cell centred conserved
quantities before the reconstruction step, it is $4$th order accurate in time
and gets the final distribution of internal energy with a relative
\Lonenorm-error of only $5$\% at the default resolution of $N = 64$.\footnote{$4$th order accuracy in time at
one time step implies an error term of $\Order(\Delta t^5)$.
Therefore, the error due to evolution from $t=0$ to $t=t_\mathrm{F}$ is
$n \times \Order(\Delta t^5) \sim \Order(\Delta t^4) \sim \Order(C^4)$,
where $n = t_\mathrm{F}/\Delta t$ is the number of time steps to reach the final time $t_\mathrm{F}$ and
$C = \Order(\Delta t)$ is the Courant factor.}

\begin{figure}
\subfigure[WHAM]{
  \epsfig{figure=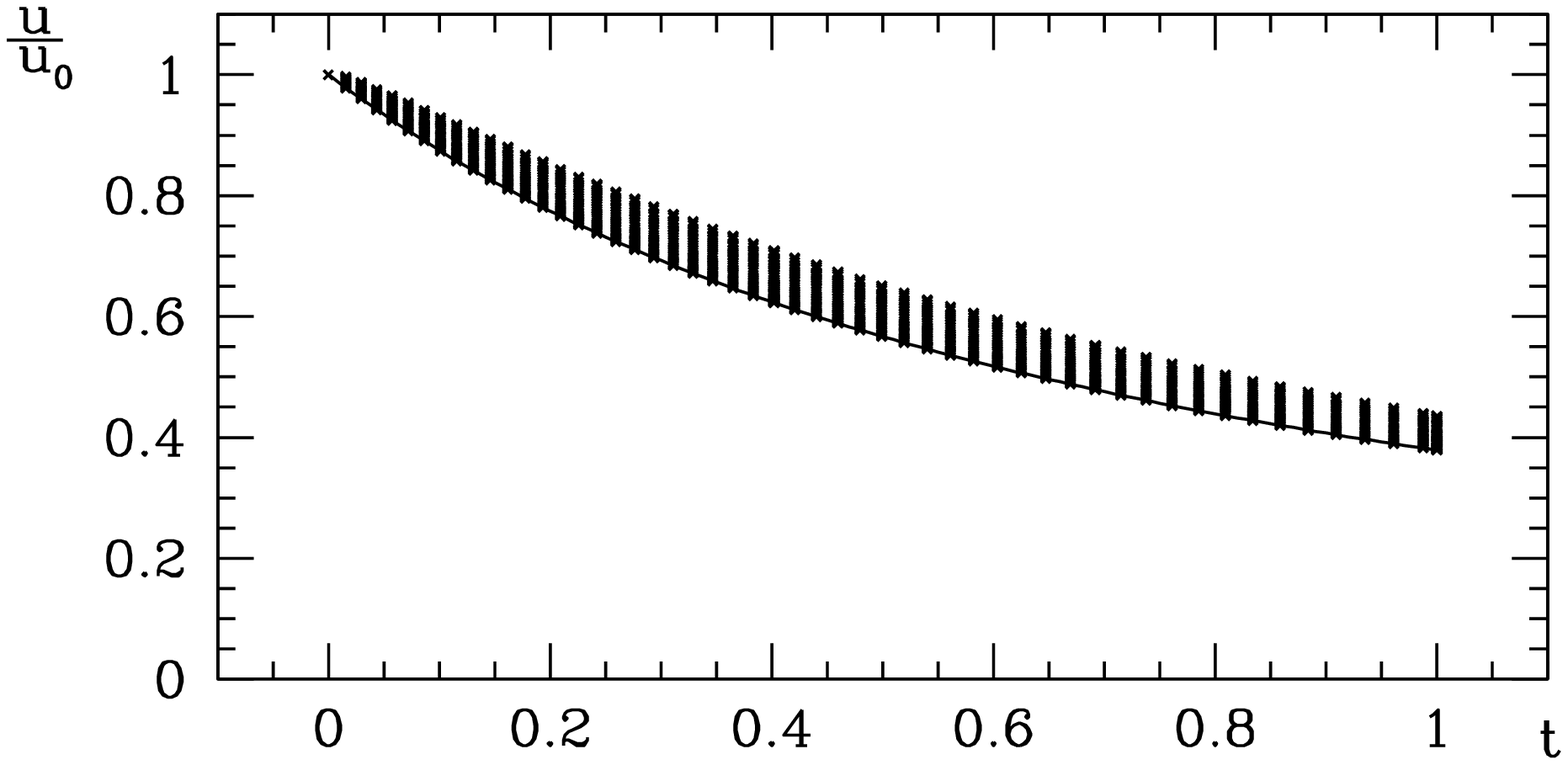,width= \columnwidth}
  }
\subfigure[HARM]{
  \epsfig{figure=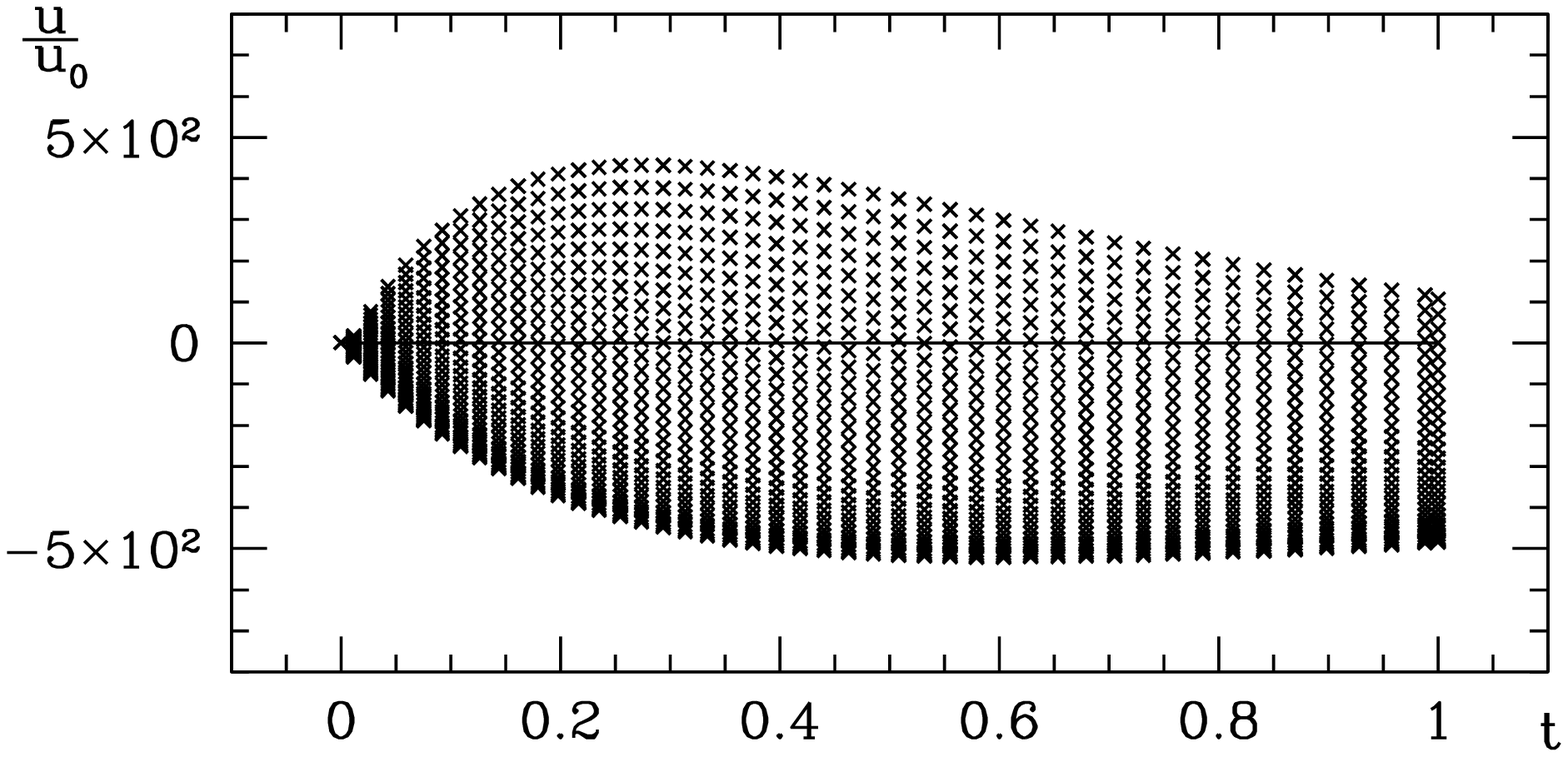,width= \columnwidth}
}
\caption{Analytical and numerical solutions for the Hubble-type flow problem,
section~\ref{sec_test_1d_hubble},  at a resolution of $N = 64$.
At each time step a scatter plot of dimensionless values of internal energy is shown
for WHAM (a) and HARM (b).
The analytical solution~\eqref{eq_hubble} for internal energy
is shown with a solid line on each panel.
Note that on panel~(b) the analytic solution is compressed into a
horizontal line because the vertical
scale has been changed by a factor of about $1000$.
WHAM accurately captures the dependence of
the internal energy in this highly supersonic flow, while
other schemes (including HARM and Athena) that do not differentiate between the
average and point values of conserved quantities makes a large
error in internal energy.
}
\label{fig_hubble1d_u}
\end{figure}

In contrast, the \SWENO{5}, HARM, and Athena codes, which
do not perform the \atoc~conversion, produce large errors. In fact,
they obtain a negative value of the internal energy already
after the first time step.\footnote{We note that a first order
Godunov scheme with an approximate HLL Riemann solver
produces inaccurate but physical values of the internal energy.}
When the internal energy is negative, for the \SWENO{5}
and HARM schemes we set the sound speed to $0$; for Athena, we use its default settings.

Figure~\ref{fig_hubble1d_u} visualizes the time dependence of the internal
energy for the WHAM and HARM schemes. The solution for internal energy given
by WHAM deviates only slightly from the analytic solution. The internal energy given
by HARM becomes negative after the first time step and is very nonuniform in
space. Any scheme (including HARM and Athena) that does not differentiate between the average and point
values of conserved quantities will make a similar error.
This error in the internal
energy is second order in space, with a large coefficient proportional to
the square of the Mach number (see eq.~\ref{eq_hubble_frac_error}).
This error does not decrease if the time step is
decreased.  For these schemes one has to use a resolution more than $100$ times larger than
the default one to match the accuracy of the WHAM
scheme at the default resolution (see table~\ref{tab_hubble_uerr}). In other words, for
high Mach number problems, high order schemes are much more effective than
lower order ones.

\subsection{Discontinuous high Mach number flow: Sod's shock tube}
\label{sec_test_1d_sod}
\begin{figure}
    \centering%
    \subfigure[]{%
      \epsfig{figure=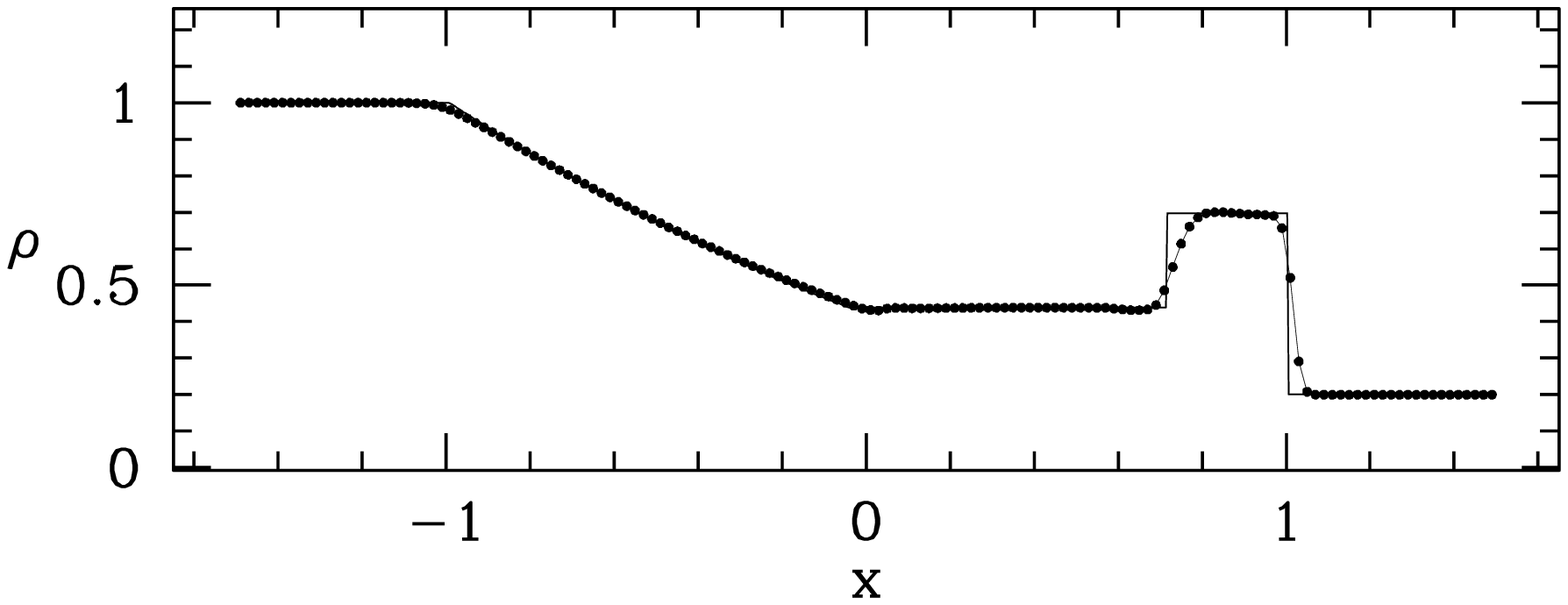,width=\columnwidth}
      \label{fig_stationarysod}
    }
    \subfigure[]{%
      \epsfig{figure=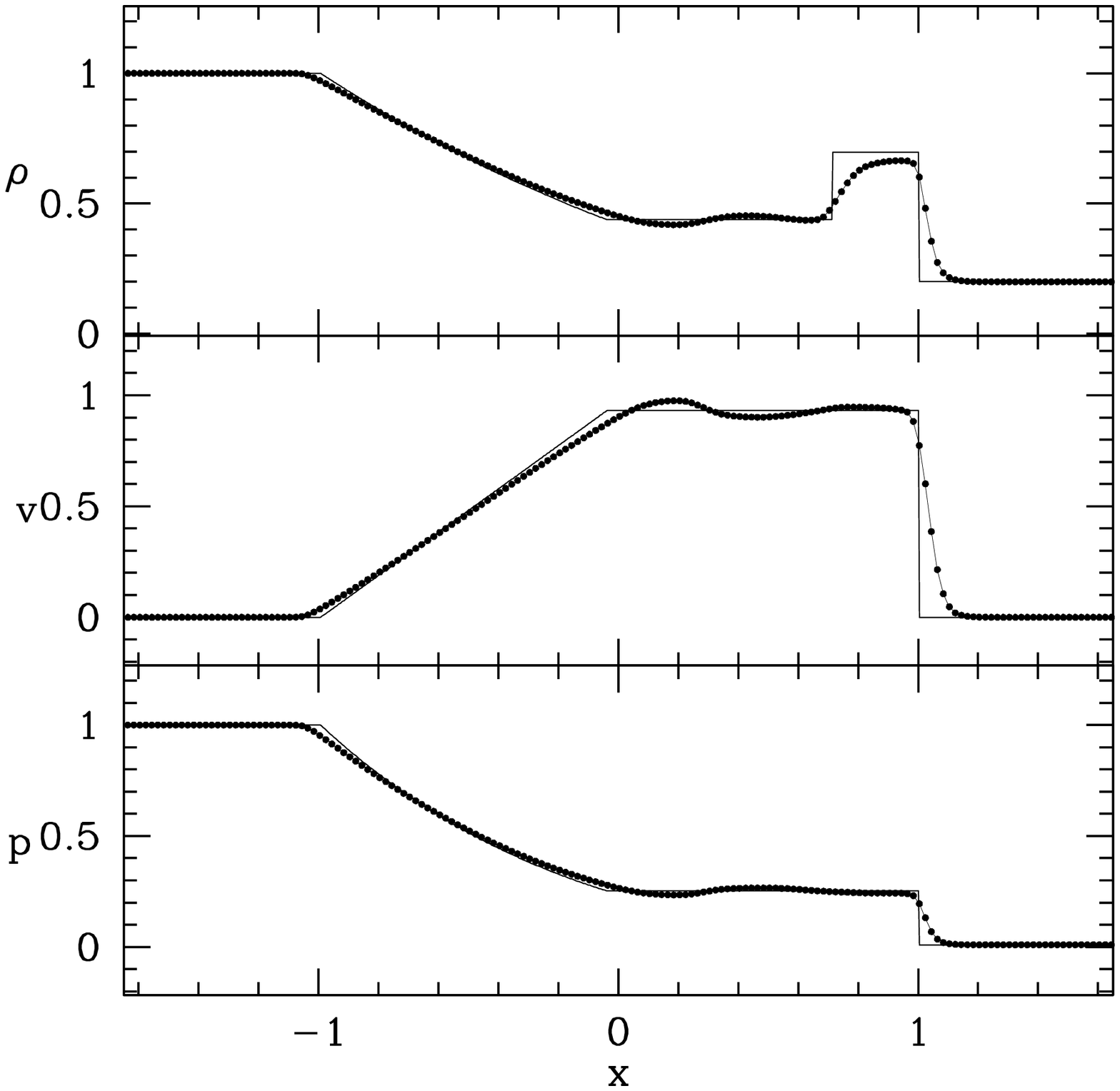,width=\columnwidth}
      \label{fig_movingsod}
    }%
  \caption{Panel~(a) shows the standard Sod's problem.
  Panel~(b) shows the `moving Sod's problem' with the pre-shock
  state moving supersonically through the grid at a Mach number of $100$.
  The numerical solution by the WHAM scheme
  is shown with connected dots and the analytic solution is shown with a solid line.
  For the lower panel~(b) the $x$-coordinate has been remapped to correspond to the
  same range as in panel~(a).
  WHAM is able to accurately
  treat shocks when both the pre-shock and post-shock regions are moving
  supersonically.
    \label{fig_sod}
  }
\end{figure}
In this section we consider two versions of Sod's problem~\citep{tra04}.
The first `stationary' version is a standard, simple shock tube problem, a variation of
the original problem by \citet{sod78}.  Table~\ref{tab_riemtests} shows the
initial conditions for this problem and
figure~\ref{fig_stationarysod}~\vpageref[]{fig_stationarysod} shows
the density distribution given by the WHAM scheme overplotted on
the exact solution at the final time.

The second `moving' version is the
`stationary' problem boosted to a supersonic speed such that the pre-shock Mach number of the
initial right state of the problem is equal to $100$. This verifies the
code's ability to handle shocks with both pre-shocked and post-shocked
gas moving supersonically with respect to the grid.
Godunov-type schemes that operate on a fixed Eulerian grid are
claimed to have serious difficulties, \eg, producing the wrong
height of the shock and generating spurious post-shock
oscillations~\citep{tra04}.

In studying the `moving' version of Sod's problem with WHAM, we keep the same
grid cell spacing and the final time as in the `stationary' version but extend the
grid so that the wave structure does not go off the grid
until the final time of the test (see table~\ref{tab_riemtests}). We have
performed this problem with both the WHAM scheme, see
figure~\ref{fig_movingsod}, and with HARM~\citep{gam03}. We find that neither
scheme shows the suggested violent post-shock oscillations claimed
by~\citeauthor{tra04}. It could be that their numerical scheme has difficulties
because it reconstructs the conserved quantities, not primitive quantities. The
lower order HARM scheme gets an incorrect height of the shock for the
`moving' problem (and produces the correct shock height in the `stationary'
version of the problem). This phenomenon also occurs with the Eulerian scheme of~\citet{tra04}.

Without the moving grid technique of~\citet{tra04}, WHAM is able to obtain
a comparable resolution of the contact discontinuity and value of the shock height.
This shows that accounting for the difference between the point and average
values of conserved quantities is an alternative to the moving grid scheme
of~\citet{tra04}.  WHAM converges uniformly at first order for this test.

\subsection{One-dimensional hydrodynamic caustics}
\label{sec_test_1d_caustics}

\begin{figure*}
\epsfig{figure=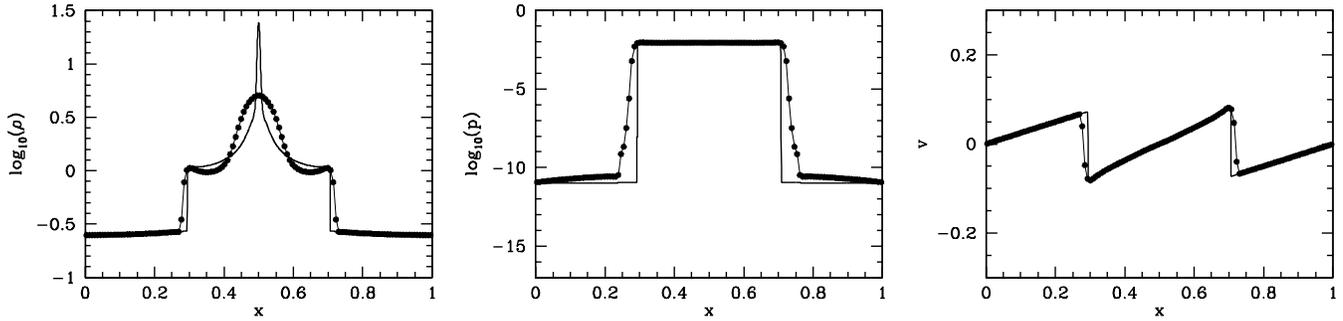,width= \textwidth}
\caption{Density, pressure, and velocity for the
one-dimensional hydrodynamic caustics problem at time $t_\text{F} = 3$.
Connected dots show the result of WHAM at the resolution of $N=128$ grid cells
with the standard Courant factor of $C=0.5$ and the solid line shows the converged solution obtained with $N=2048$ grid cells.
With $N=128$ and $C=0.5$, WHAM reproduces the low pressure pre-shock region with
an error of $250$\% that is several orders of magnitude smaller than with the energy conserving
scheme of~\citet{ryu93}.  With $N=180$ and $C=0.2$, the pre-shock pressure error is only $10\%$ with WHAM,
and with $N=256$ and $C=0.5$ the error is $25$\%.
Highly accurate, higher-order energy conserving schemes like WHAM are capable
of obtaining acceptable solutions for extremely high Mach number flows without
resorting to the dual-energy formalism, recommended by~\citet{ryu93}}
\label{fig_test_1d_caustics}
\end{figure*}

In this problem (see figure~\ref{fig_test_1d_caustics}) a highly supersonic
smooth flow steepens into a pair of shocks posing challenges for numerical
schemes: the ability to accurately evolve a smooth high Mach number flow and
handle its interaction with strong shocks. Initially the density and the
pressure are uniform, $\rho = 1$ and $p = 10^{-10}$, and the velocity is a sine
wave, $v(x) = (2\pi) ^{-1} \sin(2\pi x)$. This corresponds to Mach number
reaching $1.2\times10^4$ and the internal energy accounting only for about
$10^{-8}$ of the kinetic energy. We run the problem on an interval $(0,1)$
until time $t_\mathrm{F} = 3$ with periodic boundary conditions and we use
a resolution of $128$ grid cells as chosen by \citet{ryu93}.

For this problem \citet{ryu93} find that schemes evolving the total energy
make unacceptably large errors in the internal energy even at very high
resolutions. \citet{ryu93} therefore argue for the `dual-energy formalism' in
which one switches to evolving the entropy equation in the high Mach number
regions of the flow. Although such an approach does not suffer from the high Mach number
problem in smooth supersonic flows, it cannot account for dissipation in
weak shocks (see section~\ref{sec_intro}).

WHAM correctly captures shocks in highly supersonic flows (see the previous
section) and for smooth flows quickly converges to the correct solution as the
resolution is increased. Figure~\ref{fig_test_1d_caustics} shows WHAM getting
the final distribution of pressure in the high Mach number
region within a factor of $3$ from the converged solution. This is $4$ orders
of magnitude better than the energy-conserving scheme of \citet{ryu93}.  The shock is always resolved with
about four points. Using a resolution of $180$ and a Courant factor of $0.2$ gives a pre-shock pressure error of only $10\%$.
With a resolution of $256$ and the standard Courant factor of $0.5$,
the error is reduced by a factor of $10\approx 2^{3.5}$ to $25\%$, \ie\ our scheme
converges in pointwise sense at $3.5$th order in the pre-shock
region.   HARM or Athena, which are strictly second order in supersonic
flows and do not improve their error for smaller Courant factors, obtain an error
$\sim 25\%$ only at a resolution of $\sim 45,000$ grid cells.

\subsection{Linear wave advection}
\label{sec_test_2d_advection}

To verify the order of convergence of WHAM, we perform the advection of smooth
sound and density waves in Cartesian coordinates in two dimensions. This verifies the
order and accuracy of the multidimensional reconstruction and of the
Runge-Kutta time stepping.

We have set up this test in the same way as in the Athena code~\citep{sto06}. For
both the density and sound wave advection problems, we choose a wave with a
wave vector~$\mathbf k = (k_x, k_y) = (2\pi/\sin\alpha,2\pi/\cos\alpha)$
directed at an angle~$\alpha=\tan^{-1}(2)\approx64.4$ degrees \wrt\ the
$x$-axis. We use a computational box of size $(0,
\sin\alpha)\times(0,\cos\alpha)$ such that periodic boundary conditions can be
applied. The number of grid cells in the $x$-direction is twice that in the
$y$-direction so that each grid cell is a square even though the computational
box is a rectangle. Therefore the wave does not travel along the diagonal of
grid cells guaranteeing that the test is truly multidimensional.

We use an equation of state with $\polyconst = 5/3$ and choose a uniform
background state with a unit density $\rho_0 = 1$ and a unit sound speed $c_{s, 0}
= 1$, which corresponds to the background internal energy $u_0 = 0.9$.
For the sound wave test we choose the background velocity to be zero
($\mathbf v_0 = \mathbf 0$) and for the density wave test we choose the
background velocity to be unity along the direction of the wave
vector ($\mathbf v_0 = \mathbf k / \abs{\mathbf k}$). We run both tests
for one temporal period $T$ of the wave, \ie\ until $t_\text{F} = T =
0.4$. Relative perturbations for the
sound wave include perturbations in density, velocity, and internal energy,
\begin{align}
\left(%
\begin{array}{c}
\delta\rho/\rho_0 \\
\delta\mathbf v/c_{s,0} \\
\delta u/u_0
\end{array}
\right)
&= A \cos{(\mathbf k \cdot \mathbf r)}
\left(
\begin{array}{c}
1 \\ \mathbf k/\abs{\mathbf k} \\ \polyconst
\end{array}
\right),
\intertext{whereas for the density wave the perturbations are only in density,}
\left(%
\begin{array}{c}
\delta\rho/\rho_0 \\
\delta\mathbf v/c_{s,0} \\
\delta u/u_0
\end{array}
\right)
&= A \cos{(\mathbf k \cdot \mathbf r)}
\left(
\begin{array}{c}
1 \\ 0 \\ 0
\end{array}
\right).
\end{align}
We choose the relative magnitude $A$ of the perturbation such that nonlinear wave effects
for the sound wave at the final time lead to relative nonlinear corrections $\delta_\text{NL}$
to the solution on the order of the numerical precision for our
machine, $\delta_\text{NL} \sim A^2 t_\text{F}/T = \epsilon_\text{machine} \approx 10^{-15}$, so
we set $A = 3.2\times10^{-8}$.

We perform the calculations at different resolutions.
Table~\ref{tab_test_2d_nonrelconv} shows that the WHAM scheme indeed converges
at fifth order. For comparison, we have also run the density and sound wave
tests using the \SWENO{5}\ scheme, HARM, and Athena. These three schemes
converge only at second order despite the \ctoe\ reconstruction used in the
\SWENO{5}\ scheme being fifth order. This shows that not making a distinction
between the average and point values of conserved quantities and fluxes can
significantly impact the performance of schemes even for problems at a low Mach
number. We have also performed a smooth density advection test
described by \lwt\ and found similar convergence results.

\begin{table}
\begin{center}
\caption{Relative $\Lonenorm$-error in density,  $\delone{\rho}$,
and the order of convergence (base $2$ logarithm of the error) of WHAM,
\SWENO{5}\ and HARM schemes for the two-dimensional
density and sound wave problems (section~\ref{sec_test_2d_advection}).
The WHAM scheme converges to the analytic solution at fifth order for both
problems, while the \SWENO{5} scheme and the HARM scheme converge only at second order.}
\begin{tabular}{@{\,}lcccc@{\,}}
\hline
Resolution  & $16\times8$ & $32\times16$ & $64\times32$ & $128\times64$ \\
\hline
\multicolumn{5}{c}{Entropy wave}\\
WHAM &  $3.9\Exp{-}{09}$ & $9.2\Exp{-}{11}$ & $2.6\Exp{-}{12}$ & $8.0\Exp{-}{14}$ \\
Order & --- & $5.5$ & $5.2$ & $5.0$\\
\hline
\SWENO{5} &  $5.5\Exp{-}{09}$ & $7.1\Exp{-}{10}$ & $1.7\Exp{-}{10}$ & $4.4\Exp{-}{11}$ \\
Order & --- & $3.0$ & $2.0$ & $2.0$\\
\hline
HARM  &   $7.7\Exp{-}{09}$ & $1.6\Exp{-}{09}$ & $5.9\Exp{-}{10}$ & $1.7\Exp{-}{10}$ \\
Order & --- & $2.2$ & $1.5$ & $1.7$\\
\hline
Athena  &   $2.9\Exp{-}{09}$ & $1.1\Exp{-}{09}$ & $2.7\Exp{-}{10}$ & $6.9\Exp{-}{11}$ \\
Order & --- & $1.4$ & $2.1$ & $2.0$\\
\hline
\multicolumn{5}{c}{Sound wave}\\
WHAM &  $4.0\Exp{-}{09}$ & $9.2\Exp{-}{11}$ & $2.6\Exp{-}{12}$ &  $8.0\Exp{-}{14}$  \\
Order & --- & $5.4$ & $5.1$ & $5.0$\\
\hline
\SWENO{5} & $5.5\Exp{-}{09}$ & $7.1\Exp{-}{10}$ &  $1.7\Exp{-}{10}$ & $4.4\Exp{-}{11}$ \\
Order & --- & $3.0$ & $2.0$ & $2.0$\\
\hline
HARM &  $7.9\Exp{-}{09}$ & $1.8\Exp{-}{09}$ & $6.3\Exp{-}{10}$ & $1.9\Exp{-}{10}$ \\
Order & --- & $2.2$ & $1.5$ & $1.8$\\
\hline
Athena   &  $2.5\Exp{-}{09}$ & $6.5\Exp{-}{10}$ & $1.8\Exp{-}{10}$ & $4.2\Exp{-}{11}$ \\
Order & --- & $1.9$ & $1.9$ & $2.1$\\
\hline
\label{tab_test_2d_nonrelconv}
\end{tabular}
\end{center}
\end{table}

\subsection{Noh}
\label{sec_test_1d_noh}

\begin{figure}
\epsfig{figure=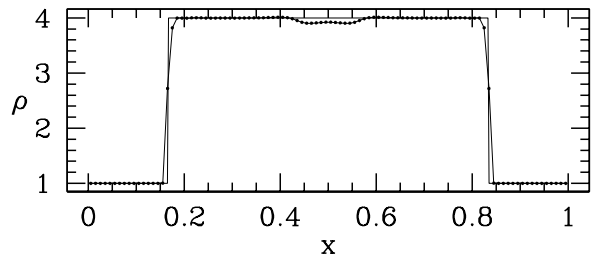,height=0.4 \columnwidth}
\caption{One-dimensional Noh problem. The analytic solution is shown with a solid line,
and the numerical solution is shown with connected dots.
WHAM does not exhibit any Gibbs phenomenon near the two strong shocks
and has a very small dip in density at the centre.}
\label{fig_test_1d_noh}
\end{figure}

This is a one-dimensional Riemann problem that tests the ability of codes to
handle infinite-strength shocks and tests the faithfulness of reproduction of
shock-shock interaction. Initially, two streams are plunging into each other
symmetrically with a constant velocity. As a result of their interaction, two
shocks develop that travel away from each other and leave matter behind at rest with
a constant density and pressure (see figure~\ref{fig_test_1d_noh}).
This test problem shows the importance of reducing to lower order in strong shocks.
If we run this problem without any stencil reduction or eigenvector decomposition
(see section~\ref{sec_fvdiscr}), the result exhibits significant Gibbs phenomenon.
Even the second order HARM scheme, which uses the MC limiter for spatial reconstruction,
produces post-shock oscillations in density with an amplitude of about~$5$\%.
With WHAM, such post-shock oscillations are absent and the
dip in density at the centre is also small. Our scheme's result is comparable to
that of the WENO-$5$ scheme and is superior to that of PPM from the
study of~\citetalias{lis03}. The \mbox{WENO-$5$} scheme uses field-by-field decomposition
that helps it avoid oscillations (see section~\ref{sec_fvdiscr}); the PPM
scheme uses a form of stencil reduction~\citep[so-called flattening, see][]{col84} that locally
lowers the order of the scheme to first order near shocks. For the
latter scheme the dip at the centre is large and some post-shock
oscillations are visible.

\subsection{Moving contact}
\label{sec_test_1d_contact}

\begin{figure}
\epsfig{figure=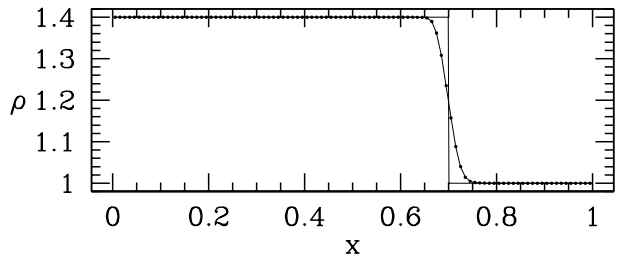,height=0.4 \columnwidth}
\caption{Moving contact problem. The analytic solution is shown with a solid line,
and the numerical solution is shown with connected dots. WHAM exhibits only a moderate
diffusion of the contact discontinuity despite avoiding eigenvector decomposition.}
\label{fig_test_1d_contact}
\end{figure}

Figure~\ref{fig_test_1d_contact} shows a test that measures the amount of
diffusion of contact discontinuities in the numerical scheme. In the exact solution the initially
sharp contact discontinuity remains sharp throughout the evolution. Used alone,
the component-wise WENO-type reconstruction would lead to excessive smearing of
the contact discontinuity because it does not use the full stencil inside the
steeply changing density profile: it chooses the one-sided stencil inside of
the discontinuity. To avoid this excessive smearing, we force the use of the
full stencil in the regions where the polynomial fit due to the full stencil
and all of its derivatives are monotonic (see Appendix \ref{sec_jmono}). Our
result is comparable to that of WENO-5 scheme, which uses eigenvector decomposition, from~\citetalias{lis03}.

\subsection{Test 4 problem from \citetalias{lis03}}
\label{sec_test_1d_test4}

\begin{figure}
\epsfig{figure=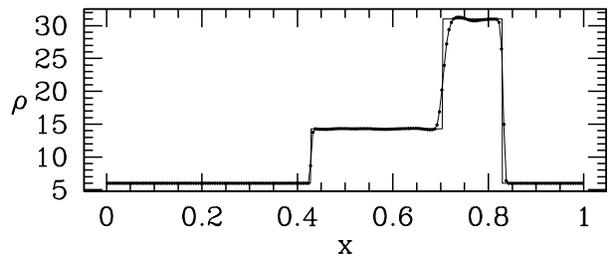,height=0.4 \columnwidth}
\caption{Test $4$ from \citetalias{lis03}.  The analytic solution is shown with a solid line,
and the numerical solution is shown with connected dots. The figure shows that WHAM
accurately resolves the structure of this complicated Riemann problem. In particular,
it provides good resolution of the moving contact discontinuity.}
\label{fig_test_1d_test4}
\end{figure}

In this problem the high resolution (small width) of the contact discontinuity,
located in figure~\ref{fig_test_1d_test4} at $x \approx 0.7$, and the absence
of oscillations in the two high density regions are important. The numerical
solution is complicated by the fact that both of these regions are `built-up'
as a result of the decay of an initial discontinuity.
The HARM scheme exhibits post-shock
oscillations at the level of about $5$\% in the left built-up state. WHAM
does not exhibit such Gibbs phenomenon. The resolution of the
contact discontinuity in our scheme
is as good as that of the WENO-5 scheme from~\citetalias{lis03},
which requires eigenvector decomposition.  The resolution of the contact by
WHAM is the same or better than that of other schemes studied by~\lwt; except the
PPM scheme, which uses an artificial contact discontinuity
sharpening technique~\citep{fry00}, a Lagrangian-remap version of PPM called VH1~\citep{vh1},
and the WAFT scheme that uses the HLLC Riemann solver.

\subsection{Shock -- entropy wave interaction test problem}
\label{sec_test_1d_shockentropy}

\begin{figure}
\epsfig{figure=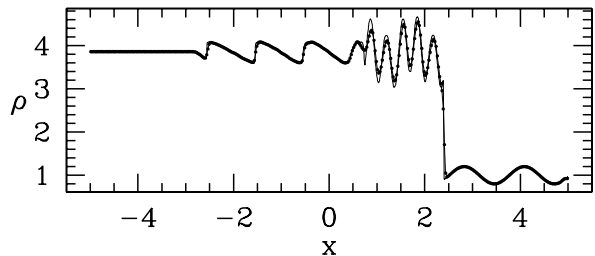,height=0.4 \columnwidth}
\caption{Snapshot of the density distribution for the shock -- entropy wave interaction
problem at the final time.
The converged
solution (at the resolution of $2000$ grid cells) is shown with the solid line,
and the numerical solution is shown with connected dots.
WHAM accurately resolves the interaction of the shock and the smooth flow.
}
\label{fig_test_1d_shockentropy}
\end{figure}

This test (figure \ref{fig_test_1d_shockentropy}) challenges the ability of the
code to handle the interaction of a shock and a smoothly varying flow.
Most second order schemes \citep{lis03short} including HARM provide
inadequate resolution of the wave structure resulting from the interaction of
the shock with the stationary density wave. WHAM is
able to accurately resolve the high-frequency waves that develop
behind the shock and provides, without the use of
eigenvector decomposition, a comparable result to that of the \mbox{CWENO-5}~\citep{qiu02}
and \mbox{WENO-5}~\citep{lis03short} schemes that use this decomposition.

\subsection{Nonrelativistic $2$D Riemann problem}
\label{sec_test_2d_case4}

\begin{figure} %
  \centering %
  \makebox[0pt][l]{\subfigure{\fbox{\epsfig{figure=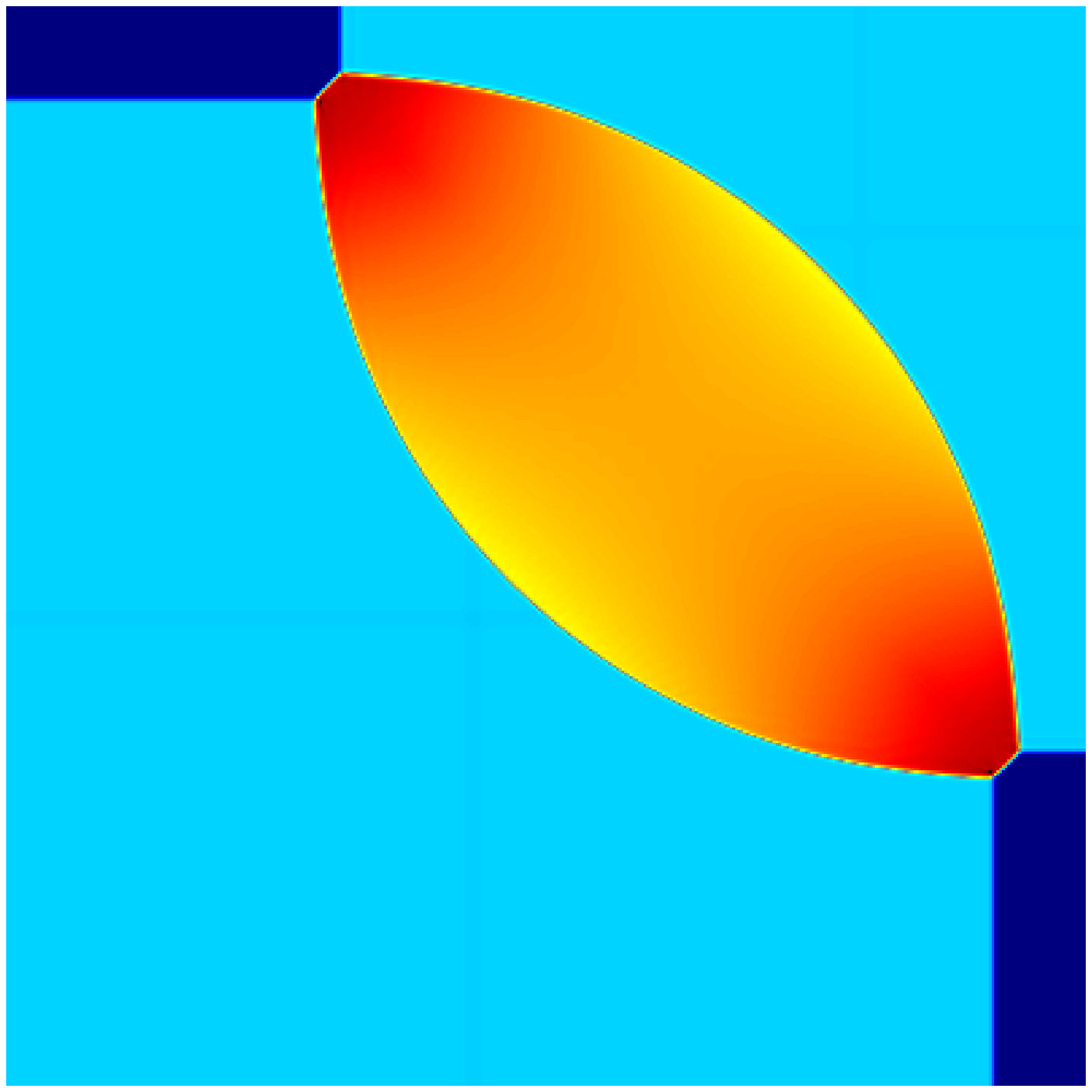,width=0.9\columnwidth,height=0.9\columnwidth,clip=}}}}%
  \subfigure{\fbox{\epsfig{figure=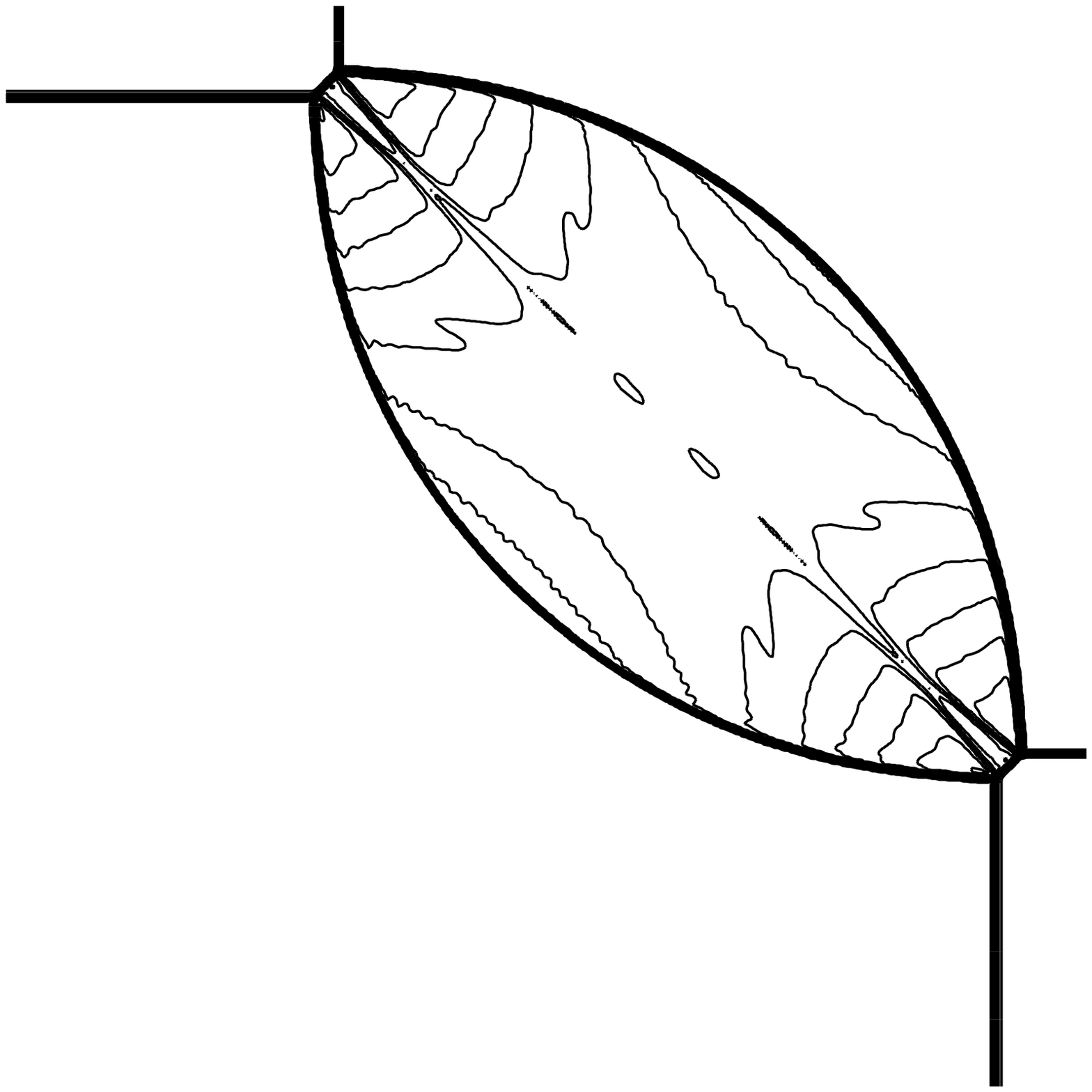,width=0.9\columnwidth,height=0.9\columnwidth}}}%
\caption{$2$D nonrelativistic Riemann problem, case $4$ from~\citetalias{lis03},
see section~\ref{sec_test_2d_case4}.
Pressure is shown in color (red denotes high values and blue low values) overplotted by a set of density contours
going from $0.52$ to $1.92$ with a step of $0.05$, the same as in~\citetalias{lis03}
for easier comparison. WHAM accurately resolves the density structure
in the eye-shaped area.  The minor oscillations are comparable to other published results.
}
\label{fig_2dcase4}
\end{figure}

We have run all two-dimensional test problems from~\citetalias{lis03} and the
results of our code are comparable to those of other codes presented there. We
picked the particular test problem~$4$ to show here because it is the only one that
exhibited any noticeable Gibbs phenomenon. The initial conditions for this test
problem are shown in table~\ref{tab_initcase4}. The problem is computed at a resolution
of $400\times400$ and uses $\polyconst = 1.4$. This test
problem initially contains $4$ planar shocks. Figure~\ref{fig_2dcase4} shows the
final state of the problem where a high-density eye-shaped area develops,
bounded by two shocks. Even though stationary
contacts are slightly less resolved in our code than in other schemes, for
moving types of contact we are the same or more accurate.
\begin{table}
\begin{center}
\caption{Initial conditions of the nonrelativistic two-dimensional Riemann problem, test case $4$
from~\citetalias{lis03}.
The final time for this test problem is $t_F = 0.25$, and $\polyconst = 1.4$.
The upper row of the table lists the initial state of the upper left and right
corners of the Riemann problem: the `L' and the `R' indices stand for the left
and right states, respectively.  The lower row lists the lower two states.
Otherwise the notation is the same as in table~\ref{tab_riemtests}.}
\begin{minipage}{\textwidth}
\begin{tabular}{@{\,}cccccccc@{\,}}
\hline
$\rho_L$ & $v_{x,L}$ & $v_{y,L}$ & $p_L$ & $\rho_R$ & $v_{x,R}$ & $v_{y,R}$ & $p_R$ \\
\hline
$0.5065$ & $0.8939$ & $0.0$ & $0.35$ & $1.1$ & $0.0$ & $0.0$ & $1.1$ \\
$1.1$ & $0.8939$ & $0.8939$ & $1.1$ & $0.5065$ & $0.0$ & $0.8939$ & $0.35$ \\
\hline
\label{tab_initcase4}
\end{tabular}
\end{minipage}
\end{center}
\end{table}

\subsection{2d Noh}
\label{sec_test_2d_noh}

\begin{figure*} %
  \centering %
  \makebox[0pt][l]{\subfigure{\fbox{\epsfig{figure=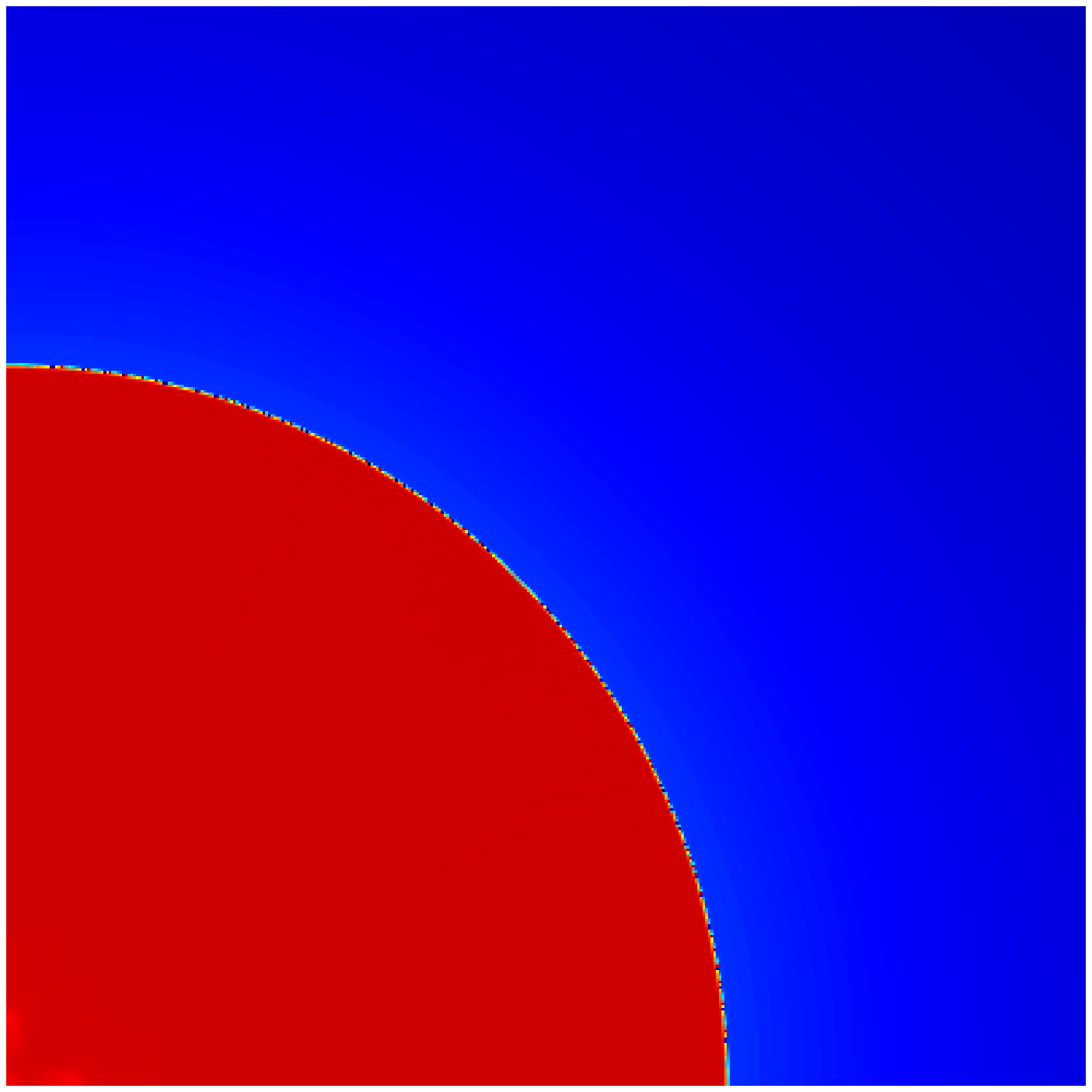,width=0.9\columnwidth,height=0.9\columnwidth}}}}%
  \subfigure{\fbox{\epsfig{figure=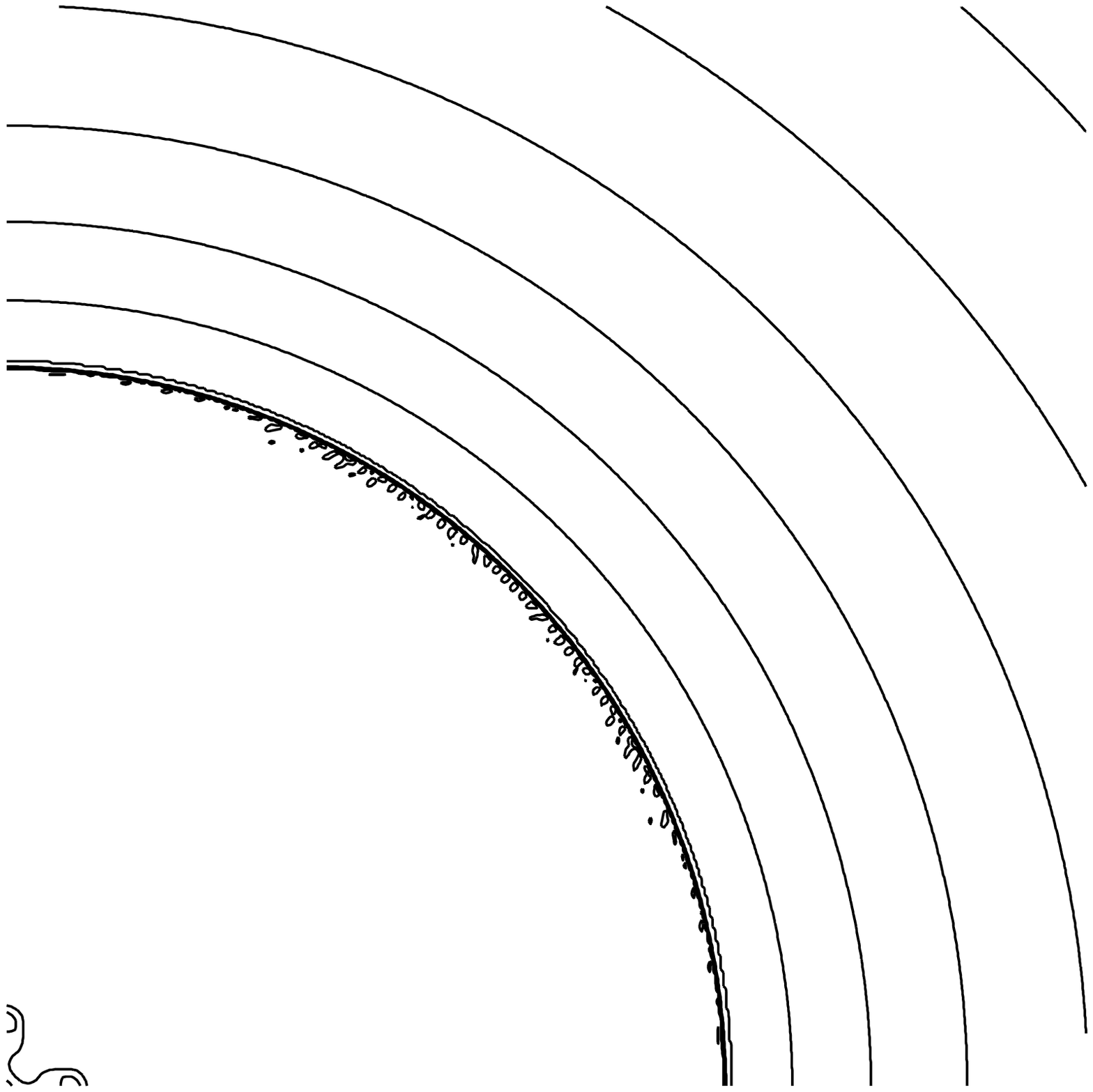,width=0.9\columnwidth,height=0.9\columnwidth}}}\label{fig_test_2d_noha} \qquad
  \subfigure{\epsfig{figure=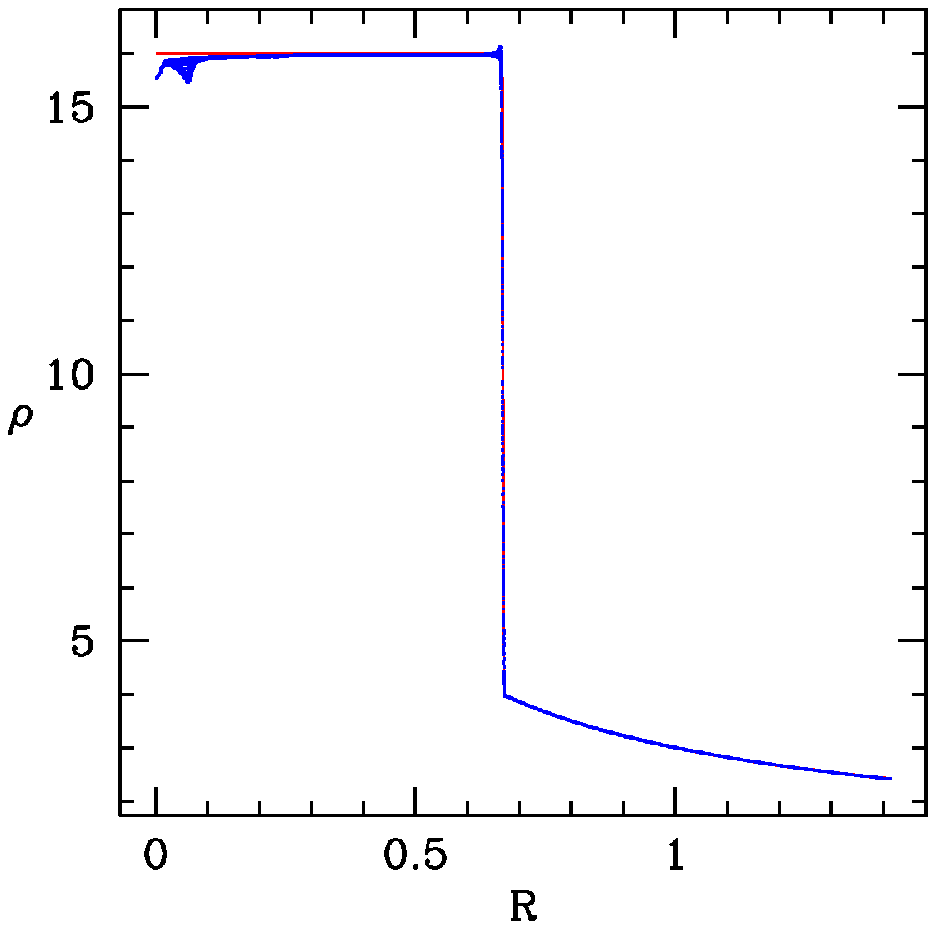,width=0.9\columnwidth,height=0.9\columnwidth}}%
\caption{Density distribution in the two-dimensional Noh problem at the final time $t_\mathrm{F} = 2$.
The left panel shows density both in color (red denotes high values and blue low values) and by a set of contours
going from $2.5$ to $4$ in steps of $0.25$ and
from $14$ to $17$ in steps of $0.2$, the same as in~\citetalias{lis03}.
The right panel shows the scatter plot of density
\wrt~radius (blue dots) and the analytic solution (thin red line).
WHAM provides a very accurate solution in the pre-shock region and shows significantly
fewer oscillations behind the shock compared to other schemes.  For the left panel,
the noise near the shock appears because one of the contours is chosen to the
analytic value.
}
\label{fig_2dnoh}
\end{figure*}

This is a $2$-dimensional version of the Noh problem for an ideal gas with
$\polyconst = 5/3$ set up in Cartesian coordinates. Initially, the flow is
cylindrically symmetric and converging on to the origin with a
constant radial velocity, $v_r = 1$. The density distribution is uniform, $\rho
= 1$, and the pressure is zero. Since the flow converges to a point, an
outgoing shock develops as can be seen in
figure~\ref{fig_2dnoh}. The shock position is located at
$R = R_s = t/3$, where $R = \sqrt{x^2+y^2}$. In the post-shock region ($R <
R_s$) the density is $\rho = 16$, the velocity is zero and the pressure is $p_g
= 16/3$. In the pre-shock region ($R > R_s$) the density is $\rho = 1 + t/R$,
the velocity stays constant $\abs{\mathbf{v}} = 1$, and the pressure remains
zero ~\citep[][]{noh87,lis03}. The initial conditions are evolved until
$t_\mathrm{F} = 2$. We use a resolution of $400\times400$ cells. Further, for
consistency with~\citetalias{lis03}, we initialize the pressure with a
small value $p_g = p_0 = 10^{-6}$ so that it is dynamically unimportant.

This test problem is unusual in that most of the final state is determined
by the time-dependent boundary conditions. In contrast to the one-dimensional Noh problem,
here the boundary conditions have to be the \emph{evolved} initial conditions.
For each of the Runge-Kutta substeps we set the time
we use for the boundary conditions to correspond to the time of that trial time step.
We find that using $p_g=p_0$ for the boundary condition as is done in, \eg, \citetalias{lis03},
results in a sharp kink in pressure along the diagonal in the region of
influence of the boundary conditions.
To avoid this, instead of keeping the pressure constant at the boundary,
we use an approximate solution for the pressure $p_g = p_0 (1 + \polyconst t /R)$
in the pre-shock region. We obtain this approximate solution by solving the internal energy
advection equation,
\begin{equation}
\frac{\partial u_g}{\partial t} + \frac{1}{R} \frac{\partial}{\partial R} \left[R (u_g + p_g) v\right] = 0,
\end{equation}
with an initial condition
$p_g=p_0$ and with all other quantities determined by the above analytic
solution (\ie~we neglect the effect of the pressure force on the evolution of
density and velocity). We have not found a closed analytic solution to this
problem for a nonzero initial pressure.

The smooth pre-shocked region is a highly supersonic flow with an initial Mach
number of $M \approx 775$. For the evolution of the internal energy to be
accurate in this supersonic region, the truncation error has to be small. We
find that for this both the de-averaging of the conserved quantities and the
transversal averaging of the fluxes are important. If we do not perform either
of these two operations, the internal energy in the smooth region becomes
negative. We expect that any scheme, such as PPM used in~\citetalias{lis03},
that ignores the difference between points and averages will generate a
significant error in the internal energy in the pre-shock region.

The numerical results are shown in
figure~\ref{fig_2dnoh} and are much superior to those of HARM and
the other schemes considered by~\citetalias{lis03}. The numerical solution is very
smooth in the pre-shock region and does not show any visible oscillations for
radii far enough ($r\gtrsim 0.2$) from the origin in the post-shock region except
near the shock.
The solution for density remains accurate even if we use $p_g=p_0$ for the boundary
condition. This should be contrasted to the results of other codes, all of
which show much more significant oscillations both in the pre-shock and
post-shock regions.

\subsection{Implosion}
\label{sec_test_2d_implosion}

The implosion test problem~\lwp\ corresponds to the interaction of a
low-density, low-pressure diamond ($\rho_i = 0.125$, $p_i = 0.14$) with a
high-density, high-pressure exterior ($\rho_o = 1$, $p_o = 1$) in a square box
with reflecting boundaries, both the diamond and the box centred on the
origin. The gas adiabatic index is $\polyconst = 1.4$ The vertices of the
diamond are located at the intersections of coordinate axes with a circle of
radius $0.15$. The computational box is $(-0.3, 0.3) \times (-0.3, 0.3)$.
Initially the velocities are zero. We perform the computations only for the
upper-right quadrant of the box (as in \lwt\ we use reflecting boundary
conditions on all $4$ boundaries) at a resolution of $400\times400$. In order
to avoid grid-induced artifacts in the initial conditions, for grid cells
intersected by the discontinuity we use the average of the two states.
The initial discontinuity evolves into two shocks with a contact
discontinuity between them.

\begin{figure} %
  \centering %
  \makebox[0pt][l]{\subfigure{\fbox{\epsfig{figure=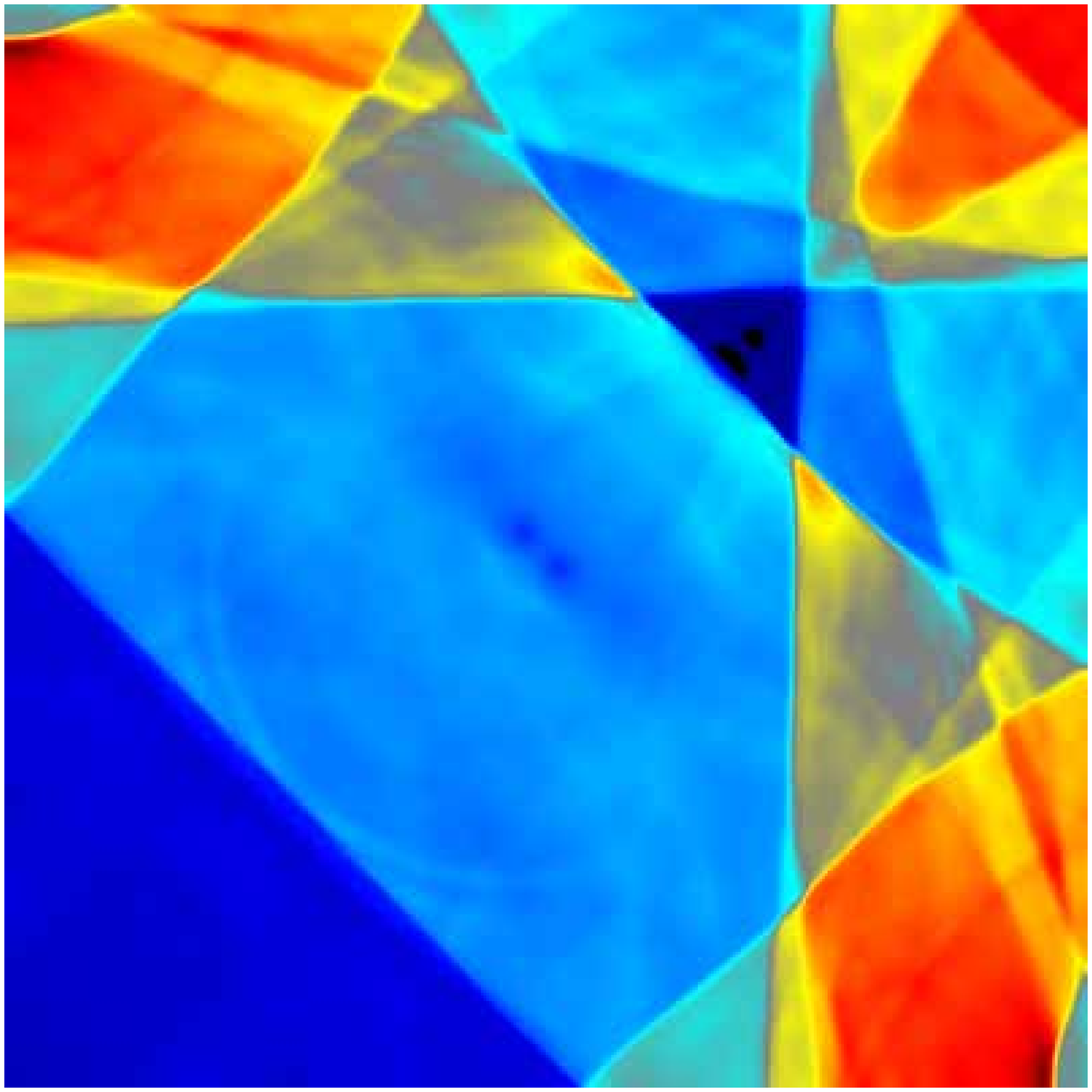,width=0.9\columnwidth,height=0.9\columnwidth,clip=}}}}%
  \subfigure{\fbox{\epsfig{figure=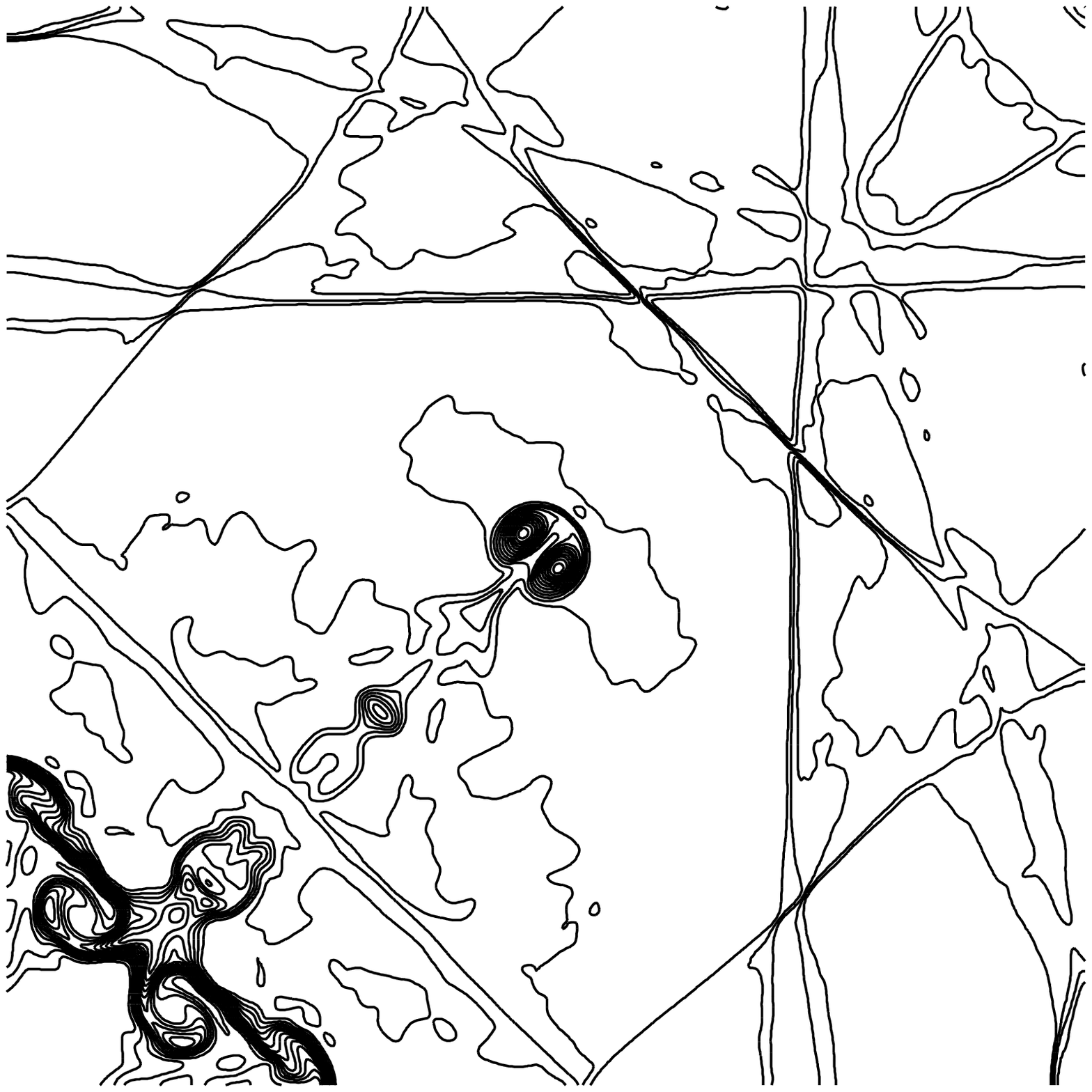,width=0.9\columnwidth,height=0.9\columnwidth}}}%
  \caption{Implosion problem~\citepalias[][]{lis03}.
  Pressure is shown in color (red denotes high values and blue low values) overplotted by a set of density contours
  going from $0.35$ to $1.1$ in steps of $0.025$, the same as in~\citetalias{lis03}.
  The figure shows that WHAM is able to
  maintain perfect symmetry of the solution and has low diffusivity that is required
  for producing the narrow jet.
    \label{fig_2dimplosion}
  }
\end{figure}

The interaction of the contact discontinuity with
the shocks generates streams that travel toward the origin and, after colliding there,
form an outflow in the form of a narrow jet. Figure~\ref{fig_2dimplosion}
shows the configuration at the final time $t_\mathrm{F} = 2.5$.
This problem tests the ability of the code to resolve contact
discontinuities and maintain symmetry in two dimensions:
if a code does not preserve symmetry, the jet will
not be produced or will be distorted.
Lower order schemes significantly diffuse the
jet at this resolution.  Unlike some of the schemes
from \lwt, our code gives exactly symmetric results. Further, the resolution
of the jet provided by our code is comparable to that of the WENO-5 scheme from
\lwt\ and is superior to all other schemes discussed in that paper and HARM.
Since Athena has a lower dissipation than WHAM, their jet head travels about $10$\% further~\citep{sto06}.
The result obtained with WHAM could be improved if we use an approximate HLLC Riemann solver,
which provides a better resolution for contact discontinuities.

\subsection{Explosion}
\label{sec_test_2d_explosion}

The explosion problem verifies the ability of the
code to evolve unstable contacts. In particular this problem studies how sensitive the
code is to numerical perturbations, which arise from the discreteness of the grid.

\begin{figure} %
  \centering %
  \makebox[0pt][l]{\subfigure{\fbox{\epsfig{figure=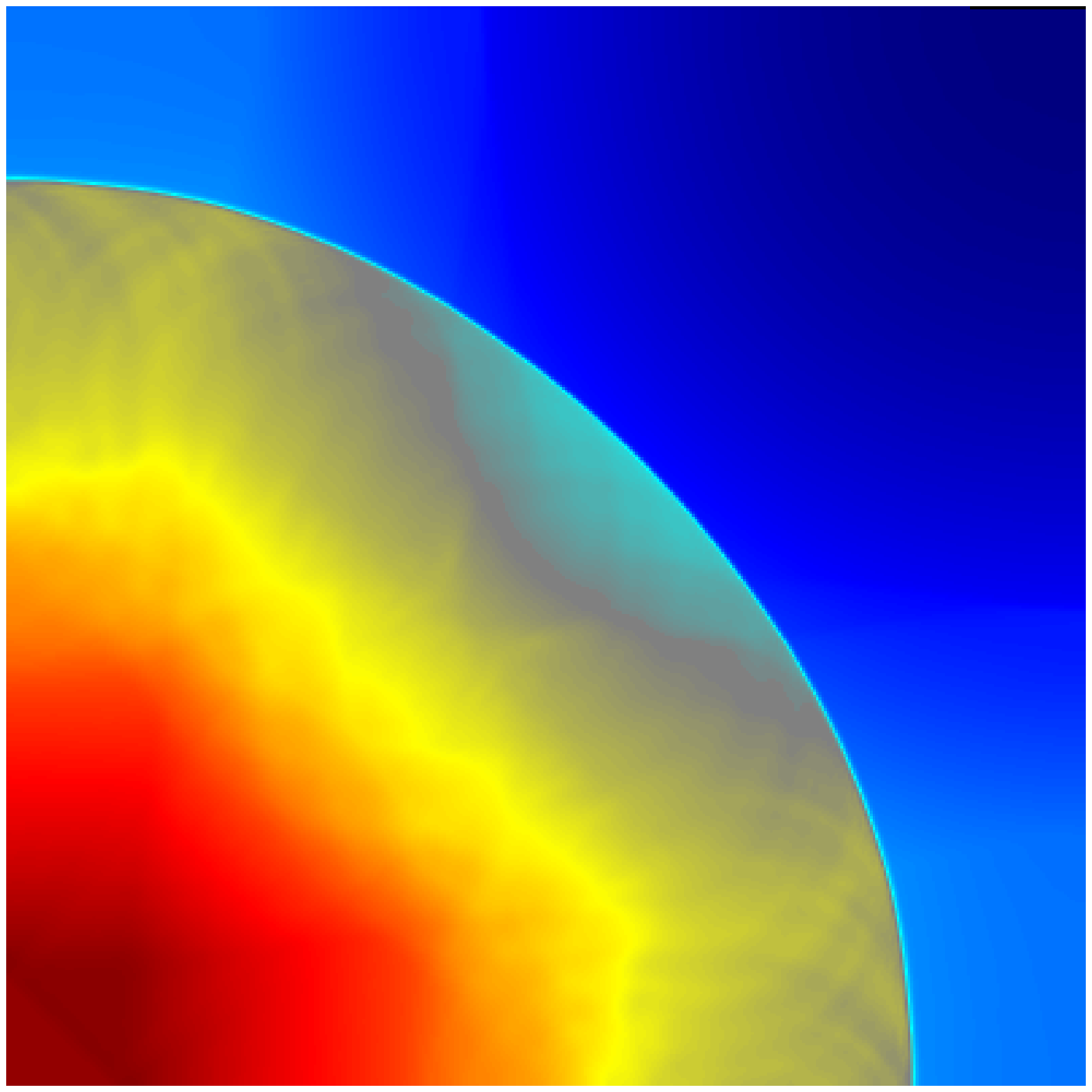,width=0.9\columnwidth,height=0.9\columnwidth,clip=}}}}%
  \subfigure{\fbox{\epsfig{figure=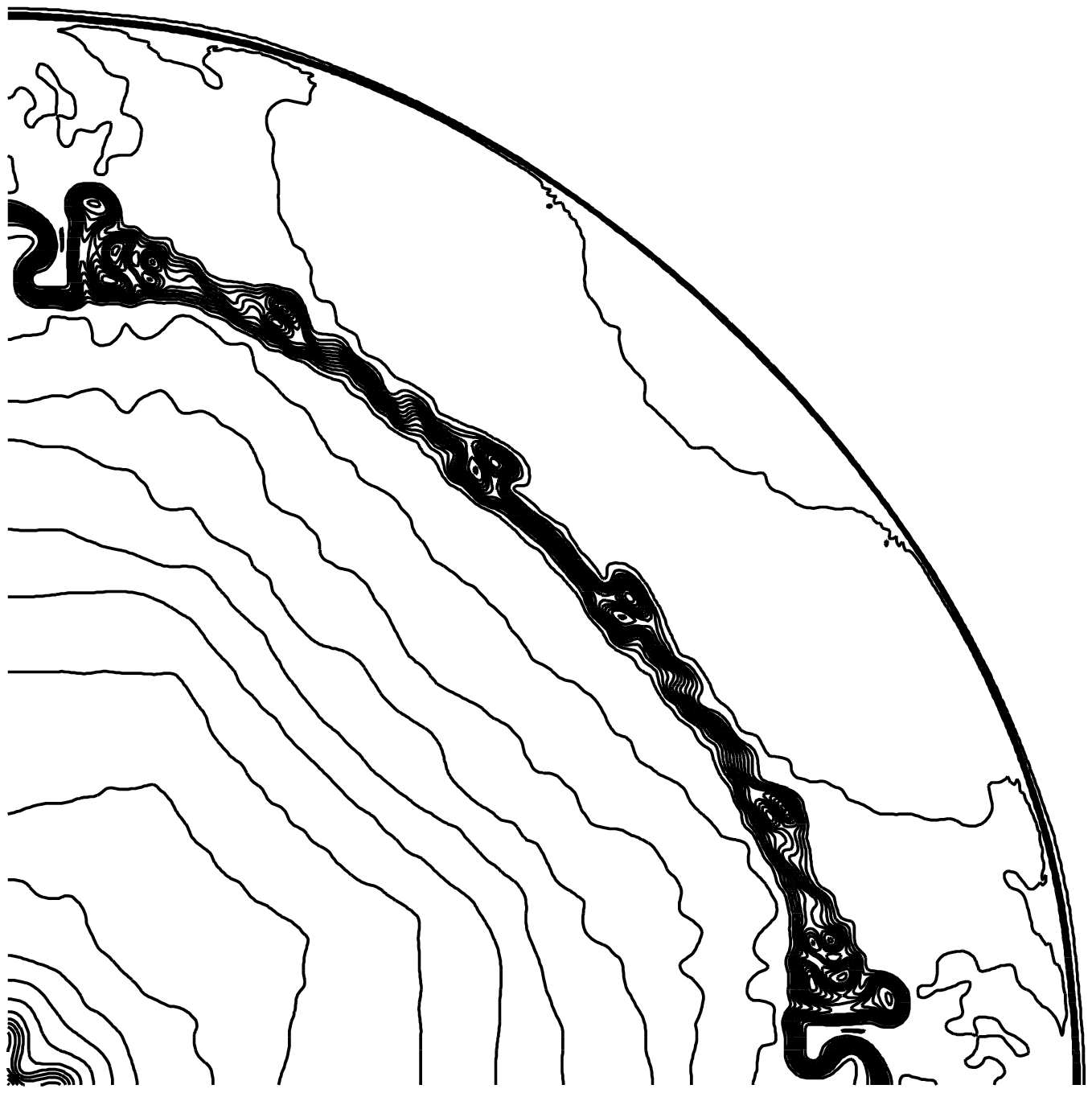,width=0.9\columnwidth,height=0.9\columnwidth}}}%
  \caption{Explosion problem~\citepalias[see, \eg,][]{lis03}.
  Pressure is shown in color (red denotes high values and blue low values) overplotted by a set of density contours
  going from $0.08$ to $0.21$ in steps of $0.005$, the same as in~\citetalias{lis03}.
  WHAM shows the same or smaller degree of breakup of the
  unstable contact discontinuity compared to the WENO-5 scheme from \lwt.
  }
\label{fig_2dexplosion}
\end{figure}

The initial conditions are cylindrically symmetric, with a high-density, high-pressure
cylinder, $\rho_i = 1$, $p_i = 1$, with a radius of $0.4$,
embedded in a low-density, low-pressure
medium, $\rho_o = 0.125$, $p_o = 0.1$. The gas constant is $\polyconst = 1.4$.
As in~\lwt, we perform the computations in Cartesian coordinates
in a square box $(0,1.5)\times(0,1.5)$ at
a resolution of $400\times400$ grid cells, with reflecting boundary conditions at the
left and bottom boundaries and zero-derivative conditions at the upper and right
boundaries.
The initial discontinuity evolves into two shocks with a contact discontinuity
in between.  Figure~\ref{fig_2dexplosion} shows a snapshot of the problem at
a final time of $t_\mathrm{F} = 3.2$ after
the outgoing shock leaves the computational domain through the
upper and right boundaries and the ingoing shock bounces off the origin and passes
through the contact discontinuity.

The interface of the contact discontinuity in this test is unstable to
perturbations, so the initial conditions have to be properly averaged for grid
cells that are intersected by the discontinuity to avoid seed perturbations in
the contact discontinuity coming from the grid structure in the initial conditions.
However, one cannot fully avoid perturbations to the contact discontinuity,
and the discontinuity breaks up similarly to the way it happens for other high order
schemes~\citepalias{lis03}.  As pointed by~\citetalias{lis03}, this problem
is sensitive to the conditions that are applied at the upper
and right boundaries. Therefore, the interaction of these
boundary conditions with the flow and, in particular, the contact discontinuity
could seed additional perturbations to the contact discontinuity.
The width and degree of break up of the contact discontinuity is the same as or
smaller than that of the WENO-5 scheme in \lwt. However, unlike
the WENO-5 scheme, our result is obtained without the use of eigenvector
decomposition.

\subsection{Moving Gresho problem}
\label{sec_test_2d_mgresho}

\begin{figure}
  \makebox[\linewidth][c]{
  \subfigure[Initial]{%
     \fbox{\epsfig{figure=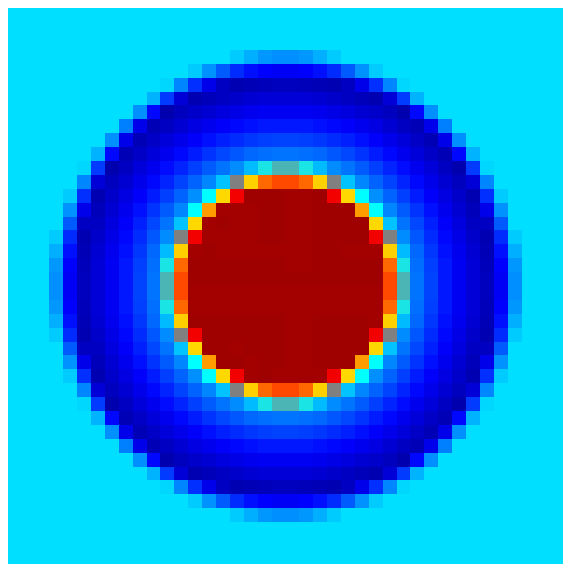,width=0.45 \columnwidth,clip=}}
  }
  }
  \linebreak%
      \subfigure[Final WHAM]{%
        \makebox[0pt][l]{ %
          \fbox{\epsfig{figure=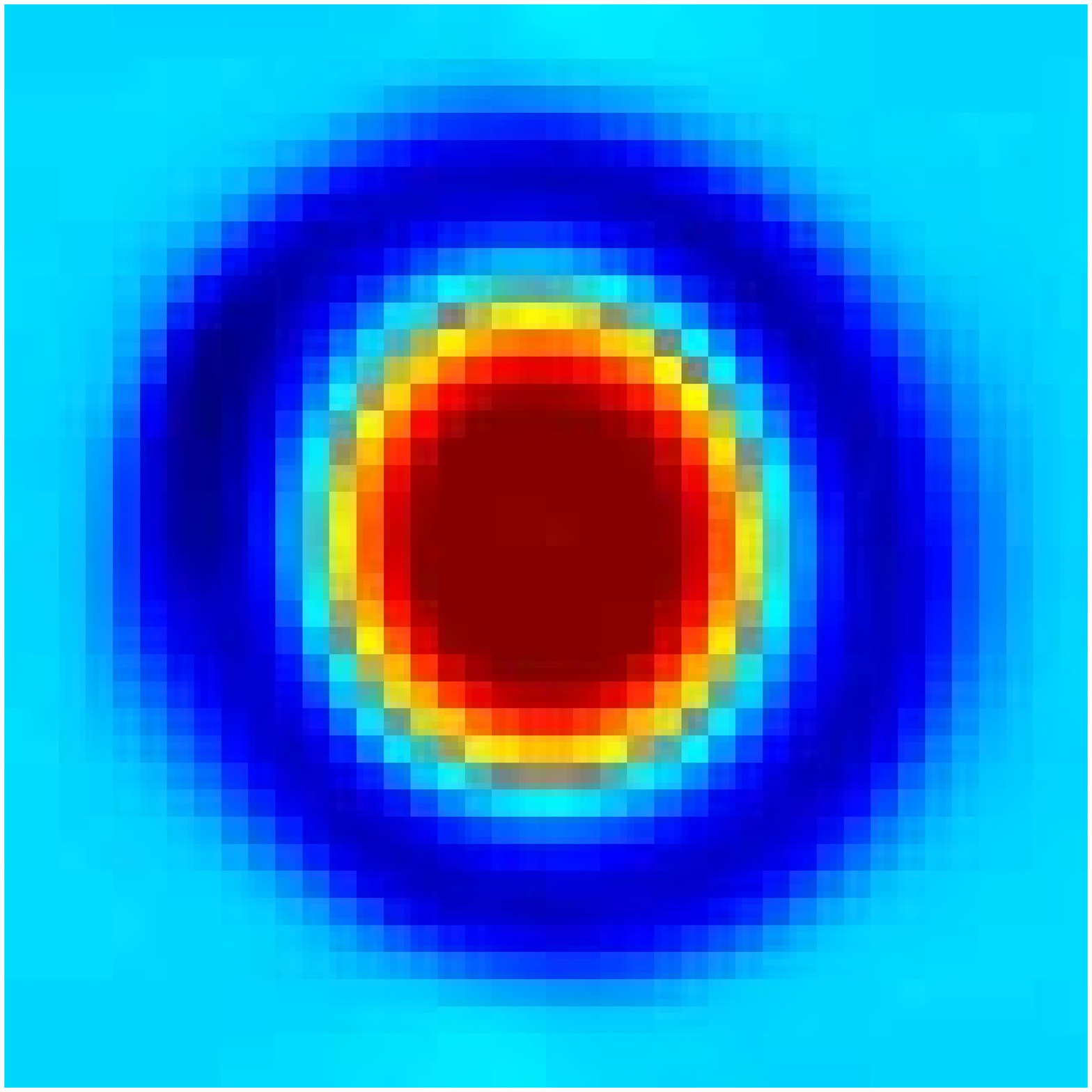,width=0.45 \columnwidth,clip=}}
          }
          \fbox{\epsfig{figure=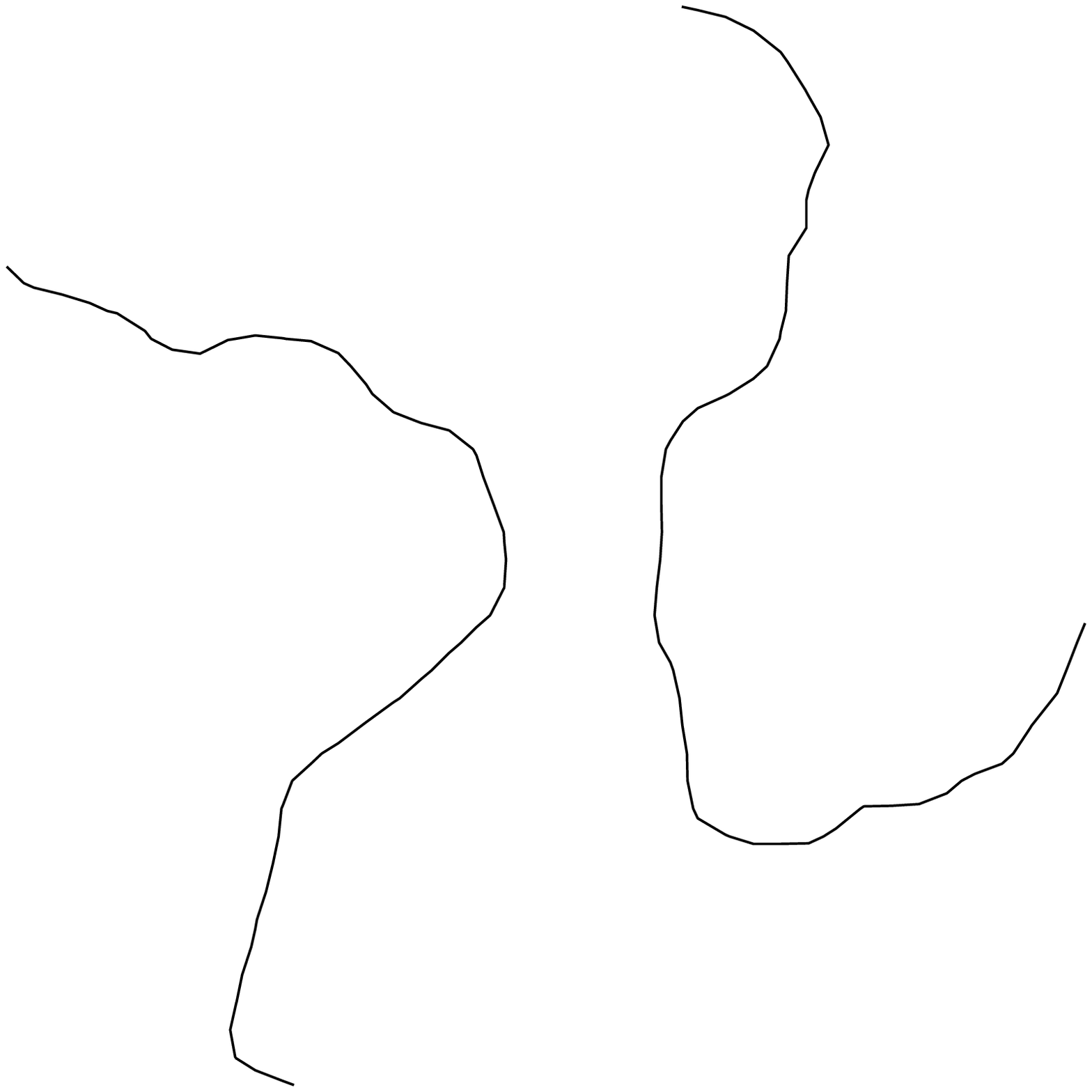,width=0.45 \columnwidth}}
    }%
  \hfill
  \subfigure[Final HARM]{%
      \makebox[0pt][l]{ %
        \fbox{\epsfig{figure=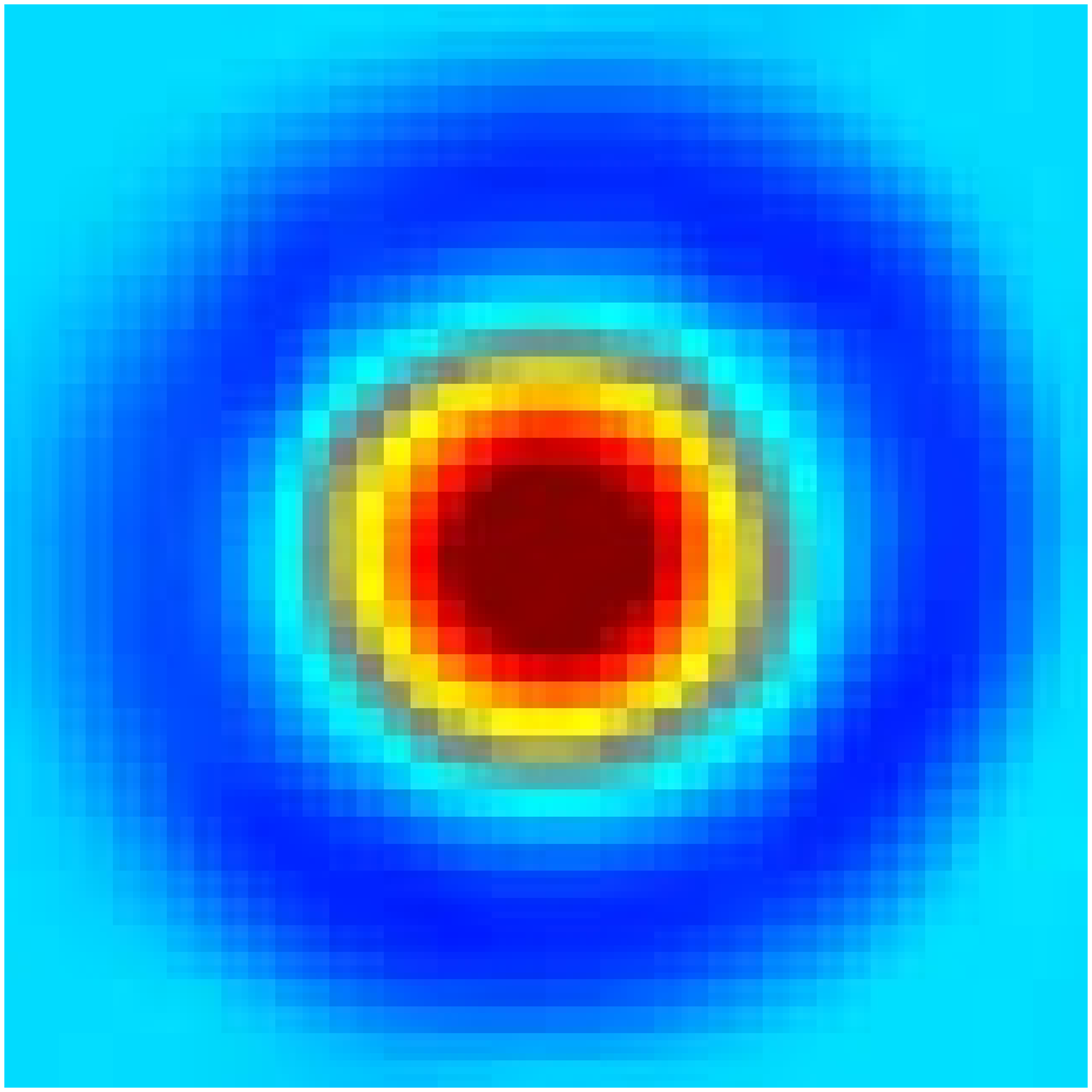,width=0.45 \columnwidth,clip=}}
        }
      \fbox{\epsfig{figure=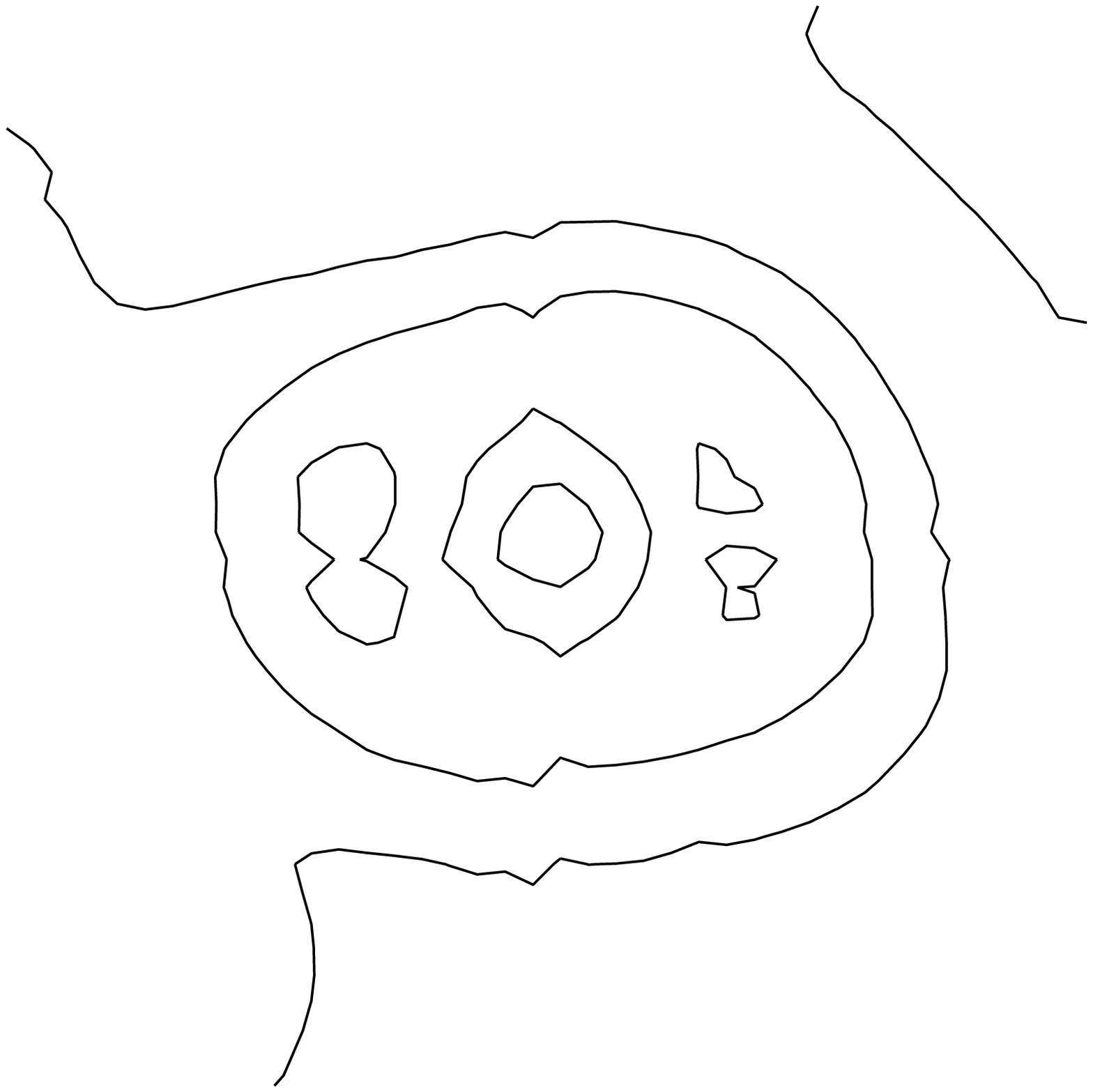,width=0.45 \columnwidth}}
  }%
  \caption{Moving Gresho problem~\citepalias[see, \eg,][]{lis03}.
    Initial~(a) and final~(b) \& (c) distributions of vorticity
    are shown in color (red denotes high values and blue low values) overplotted by a set of density contours
    going from $0.97$ to $1.03$ in steps of $0.006$, the same as in~\citetalias{lis03}.
    The resolution of the fragments shown is $40\times40$,
    the resolution of the full grid is $160\times40$.  The WHAM scheme (panel b)
    preserves the structure of the vortex in density and velocity
    exceptionally well compared to HARM (panel c) and the various schemes considered in~\lwt.
    \label{fig_mgresho}
  }
\end{figure}

This problem tests how well a numerical scheme is able to
advect a smooth vortex supported by the balance of pressure and rotation.
A non-moving vortex centred at $(0,0)$
has the following distribution of velocity and pressure in polar coordinates $(r,\varphi)$
with the origin at its centre:
\begin{alignat*}{3}
v_\varphi &= 5r,    &\enspace p_g &= 5 + \myfrac{25}{2}r^2, &\enspace 0.0 &\le r < 0.2, \\
v_\varphi &= 2- 5r, &\enspace p_g &= 9 + \myfrac{25}{2}r^2 - 20r + 4\ln 5r, \enspace&0.2 &\le r < 0.4, \\
v_\varphi &= 0,     &\enspace p_g &= 3 + 4 \ln 2, &\enspace 0.4 &\le r,
\end{alignat*}
the radial velocity $v_r$ is zero, the density is unity, and the polytropic
index of the gas is $\polyconst = 1.4$.
In the test, this vortex is initially imparted unit velocity in the $x$-direction.
Figure~\ref{fig_mgresho} shows snapshots of the vortex at the initial time and
the final time $t_\mathrm{F} = 3$ when the vortex centre has moved to $(3,0)$.
We perform the computation in cartesian coordinates in
the rectangle $(-0.5, 3.5)\times(-0.5, 0.5)$
at a resolution of $160\times40$ and use zero-derivative outflow boundary conditions
at all boundaries. This is the same as used by \lwt.

WHAM does extremely well in preserving the structure of the vortex in both
velocity and density. Our code maintains the vorticity and makes a very small
error in density so that it ends up with much fewer density contours compared to
HARM and all schemes in~\citetalias{lis03}. HARM as well as the codes considered
in~\lwt\ such as WENO-5, PPM, VH1, \etc, destroy the shape of the vortex,
making it ellipsoidal and/or significantly diffuse in the radial profile. This
shows the superiority of our high accuracy finite-volume approach in applications
involving smooth problems.  Relative $\Lonenorm$-errors of the final solution
at $t_\text{F} = 3$ \wrt\ to the analytic solution for WHAM and HARM
are shown in table~\ref{tab_test_mgresho}.

\begin{table}
\begin{center}
\caption{Relative $\Lonenorm$-errors for the Moving Gresho problem (sec.~\ref{sec_test_2d_mgresho}).
The errors are computed over the core of the vortex within radius $0.2$
from the vortex centre, at the final time $t_\text{F} = 3$.
WHAM makes about a factor of $5$ smaller error in baryon density and
internal energy than HARM at the resolution of $160\times40$.
}
\begin{minipage}{\textwidth}
\begin{tabular}{@{\,}lcccc@{\,}}
\hline
Scheme &            \delone{\rho}&
                    \delone{u_g} &
                    \delone{v_x} &
                    \delone{v_y} \\
\hline
WHAM &  $1.68\Exp{-}{03}$ & $3.90\Exp{-}{03}$ & $2.10\Exp{-}{02}$  &  $1.09\Exp{-}{02}$ \\
HARM &  $8.36\Exp{-}{03}$ & $2.61\Exp{-}{02}$ & $3.87\Exp{-}{02}$  &  $2.06\Exp{-}{02}$ \\
\hline
\label{tab_test_mgresho}
\end{tabular}
\end{minipage}
\end{center}
\end{table}

\subsection{$1$D relativistic Riemann problem 1}
\label{sec_test_1d_rel_101}

\begin{figure}
\centering\epsfig{figure=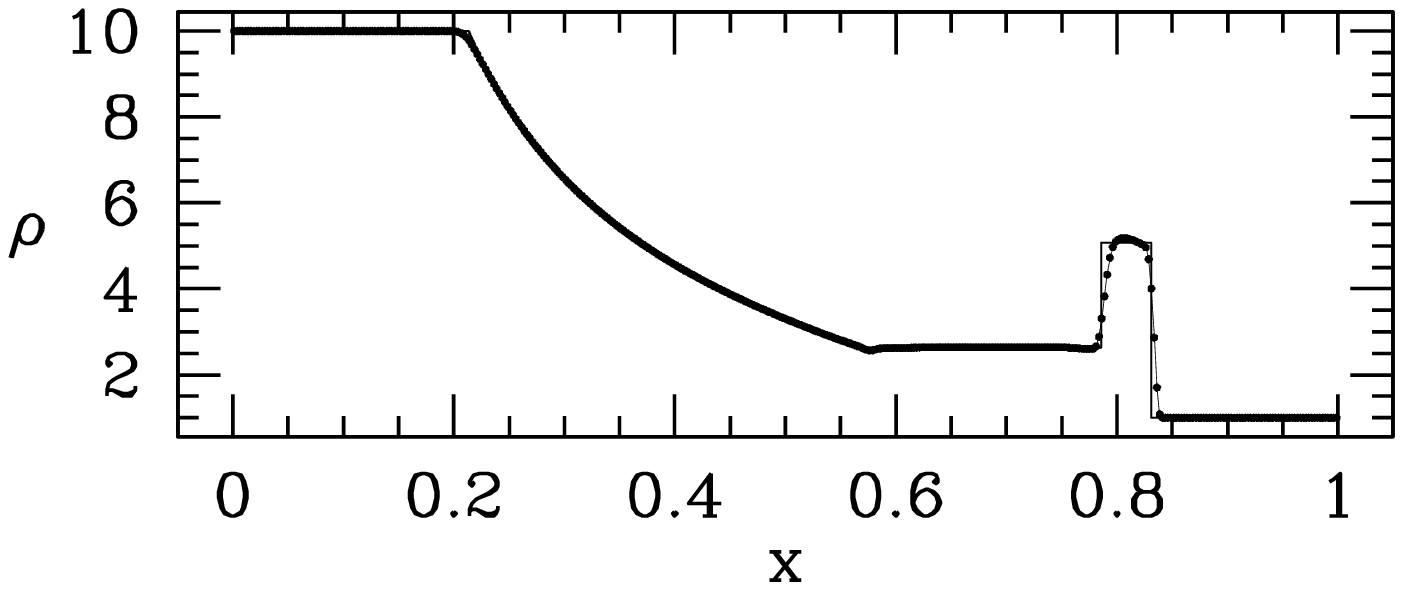,width=1.0 \columnwidth}
\caption{The distribution of density at the final time of the one-dimensional relativistic
Riemann problem 1 from~\citetalias{zha06}. \plotnotation.
WHAM correctly reproduces all ingredients of flow:
the rarefaction wave, the contact discontinuity, and the shock.
}
\label{fig_1d_rel_101}
\end{figure}

The initial conditions are given in table~\ref{tab_riemtests}. This problem
verifies the ability of a numerical scheme to treat basic relativistic
problems. As seen in figure~\ref{fig_1d_rel_101}, the initial discontinuity
decays into a rarefaction wave, a contact discontinuity, and a shock. The
resolution of the contact discontinuity and the amount of post-shock
oscillation is comparable to that of the WENO scheme in~\citetalias{zha06}.
Note that our code achieves a comparable result despite not using the
characteristic decomposition that certainly helps the resolution of Riemann
problems. Table~\ref{tab_rel_conv} shows the $\Lonenorm$-errors of the solution
and the order of convergence of WHAM for
this problem as well as the following five relativistic Riemann problems.

\begin{table}
\begin{center}
\caption{Absolute and relative $\Lonenorm$ errors in density and the order of convergence of WHAM
for relativistic Riemann problems 1 -- 6.  For a unit interval, at which these problems
are set up, the definition of the absolute $\Lonenorm$ error~\eqref{eq_Lonenorm} is
equivalent to the definition used by \zhat.  WHAM converges at approximately first
order for all tests, where the errors and convergence rates are similar to
other published work. }
\begin{minipage}{\textwidth}
\begin{tabular}{@{\,}rrcclc@{\,}}
\hline
Problem  & Resolution & $\delonea{\rho}$ & $\delone{\rho}$ & Order & Section\\
\hline
$1$            &   $100$      &  $1.39\Exp{-}{01}$  &  $1.39\Exp{-}{02}$  &   --  &  \ref{sec_test_1d_rel_101} \\
               &   $200$      &  $7.26\Exp{-}{02}$  &  $7.26\Exp{-}{03}$  &  $0.94$ \\
               &   $400$      &  $3.57\Exp{-}{02}$  &  $3.57\Exp{-}{03}$  &  $1.00$ \\
               &   $800$      &  $2.08\Exp{-}{02}$  &  $2.08\Exp{-}{03}$  &  $0.78$ \\
               &  $1600$      &  $1.06\Exp{-}{02}$  &  $1.06\Exp{-}{03}$  &  $0.97$ \\
               &  $3200$      &  $5.59\Exp{-}{03}$  &  $5.59\Exp{-}{04}$  &  $0.93$ \\
\hline
$2$            &   $100$      &  $2.12\Exp{-}{01}$  &  $2.04\Exp{-}{02}$  &  --   &  \ref{sec_test_1d_rel_102} \\
               &   $200$      &  $1.48\Exp{-}{01}$  &  $1.42\Exp{-}{02}$  &  $0.52$ \\
               &   $400$      &  $9.40\Exp{-}{02}$  &  $9.03\Exp{-}{03}$  &  $0.65$ \\
               &   $800$      &  $5.17\Exp{-}{02}$  &  $4.96\Exp{-}{03}$  &  $0.86$ \\
               &  $1600$      &  $2.57\Exp{-}{02}$  &  $2.47\Exp{-}{03}$  &  $1.00$ \\
               &  $3200$      &  $1.52\Exp{-}{02}$  &  $1.46\Exp{-}{03}$  &  $0.76$ \\

\hline
$3$            &   $100$      &  $9.32\Exp{-}{02}$  &  $1.41\Exp{-}{02}$  &   --  &  \ref{sec_test_1d_rel_103} \\
               &   $200$      &  $5.91\Exp{-}{02}$  &  $8.97\Exp{-}{03}$  &  $0.66$ \\
               &   $400$      &  $3.46\Exp{-}{02}$  &  $5.24\Exp{-}{03}$  &  $0.78$ \\
               &   $800$      &  $1.93\Exp{-}{02}$  &  $2.93\Exp{-}{03}$  &  $0.84$ \\
               &  $1600$      &  $1.04\Exp{-}{02}$  &  $1.58\Exp{-}{03}$  &  $0.89$ \\
               &  $3200$      &  $5.56\Exp{-}{03}$  &  $8.43\Exp{-}{04}$  &  $0.91$ \\
\hline
$4$            &   $100$      &  $6.20\Exp{-}{01}$  &  $2.63\Exp{-}{02}$  &  --   &  \ref{sec_test_1d_rel_104} \\
               &   $200$      &  $3.60\Exp{-}{01}$  &  $1.53\Exp{-}{02}$  &  $0.79$ \\
               &   $400$      &  $2.01\Exp{-}{01}$  &  $8.55\Exp{-}{03}$  &  $0.84$ \\
               &   $800$      &  $1.02\Exp{-}{01}$  &  $4.34\Exp{-}{03}$  &  $0.98$ \\
               &  $1600$      &  $6.79\Exp{-}{02}$  &  $2.88\Exp{-}{03}$  &  $0.59$ \\
               &  $3200$      &  $3.57\Exp{-}{02}$  &  $1.51\Exp{-}{03}$  &  $0.93$ \\
\hline
$5$            &   $100$      &  $6.88\Exp{-}{01}$  &  $1.54\Exp{-}{01}$  &  --   &  \ref{sec_test_1d_rel_106} \\
               &   $200$      &  $5.96\Exp{-}{01}$  &  $1.34\Exp{-}{01}$  &  $0.21$ \\
               &   $400$      &  $4.13\Exp{-}{01}$  &  $9.24\Exp{-}{02}$  &  $0.53$ \\
               &   $800$      &  $2.68\Exp{-}{01}$  &  $6.00\Exp{-}{02}$  &  $0.62$ \\
               &  $1600$      &  $1.47\Exp{-}{01}$  &  $3.30\Exp{-}{02}$  &  $0.86$ \\
               &  $3200$      &  $7.83\Exp{-}{02}$  &  $1.75\Exp{-}{02}$  &  $0.91$ \\
\hline
$6$            &   $25$      &  $7.18\Exp{}{03}$  &  $2.54\Exp{-}{02}$  &  --   &  \ref{sec_test_1d_rel_105} \\
               &   $50$      &  $4.57\Exp{}{03}$  &  $1.62\Exp{-}{02}$  &  $0.65$ \\
               &  $100$      &  $1.73\Exp{}{03}$  &  $6.11\Exp{-}{03}$  &  $1.4$ \\
               &  $200$      &  $1.14\Exp{}{03}$  &  $4.04\Exp{-}{03}$  &  $0.6$ \\
               &  $400$      &  $4.24\Exp{}{02}$  &  $1.50\Exp{-}{03}$  &  $1.4$ \\
               &  $800$      &  $2.82\Exp{}{02}$  &  $9.96\Exp{-}{04}$  &  $0.59$ \\
               & $1600$      &  $1.04\Exp{}{02}$  &  $3.67\Exp{-}{04}$  &  $1.4$ \\
               & $3200$      &  $6.91\Exp{}{01}$  &  $2.44\Exp{-}{04}$  &  $0.59$ \\
\hline
\label{tab_rel_conv}
\end{tabular}
\end{minipage}
\end{center}
\end{table}

\subsection{$1$D relativistic Riemann problem 2}
\label{sec_test_1d_rel_102}

\begin{figure}
\centering\epsfig{figure=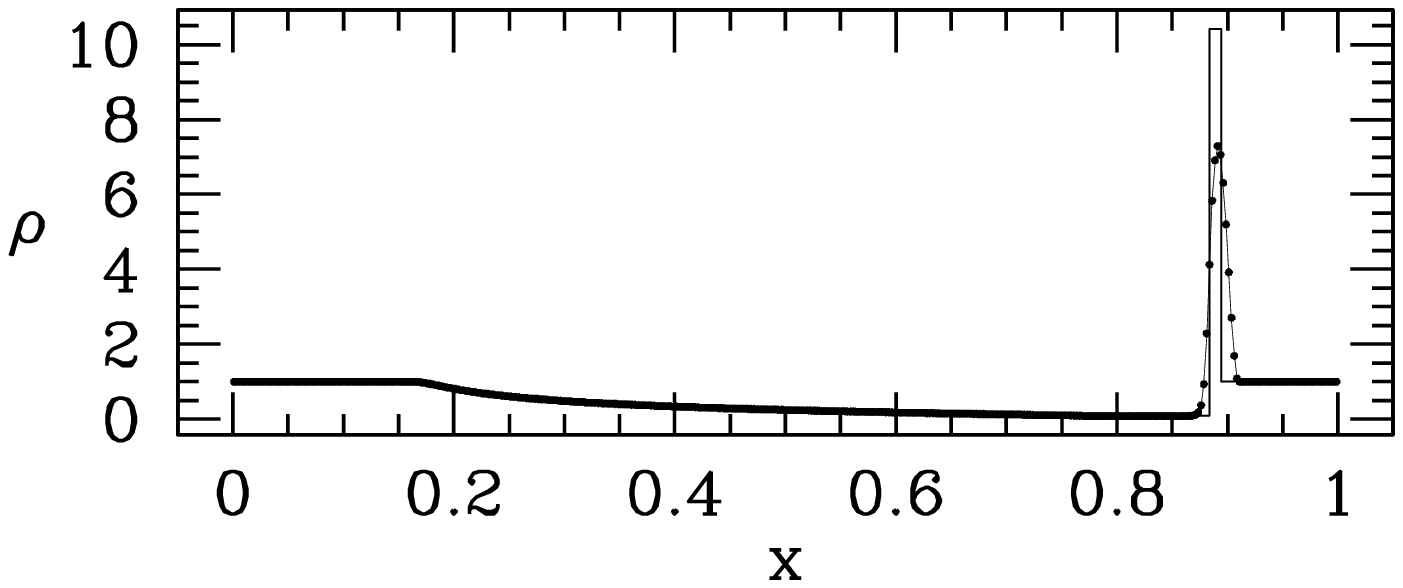,width=1.0 \columnwidth}
\caption{Density distribution for the one-dimensional relativistic Riemann problem 2 from~\citetalias{zha06}.
\plotnotation.
WHAM performs well by getting the height of the built-up state
to about $70$\% of the analytic value, a height
comparable to other codes~\citep{zha06,mor06}.
}
\label{fig_1d_rel_102}
\end{figure}

The data for this test are given in table~\ref{tab_riemtests}. This is a
challenging test with sparse resolution.  It is very hard to get the correct height of
the built-up state due to the smearing of the contact discontinuity and the
proximity of the shock (figure~\ref{fig_1d_rel_102}).
Comparable to that of other codes~\citep{alo99,zha06,mor06}, WHAM gets
the height of the built-up state to about $70$\% of the analytic value.

\subsection{$1$D relativistic Riemann problem 3}
\label{sec_test_1d_rel_103}

\begin{figure}
\centering\epsfig{figure=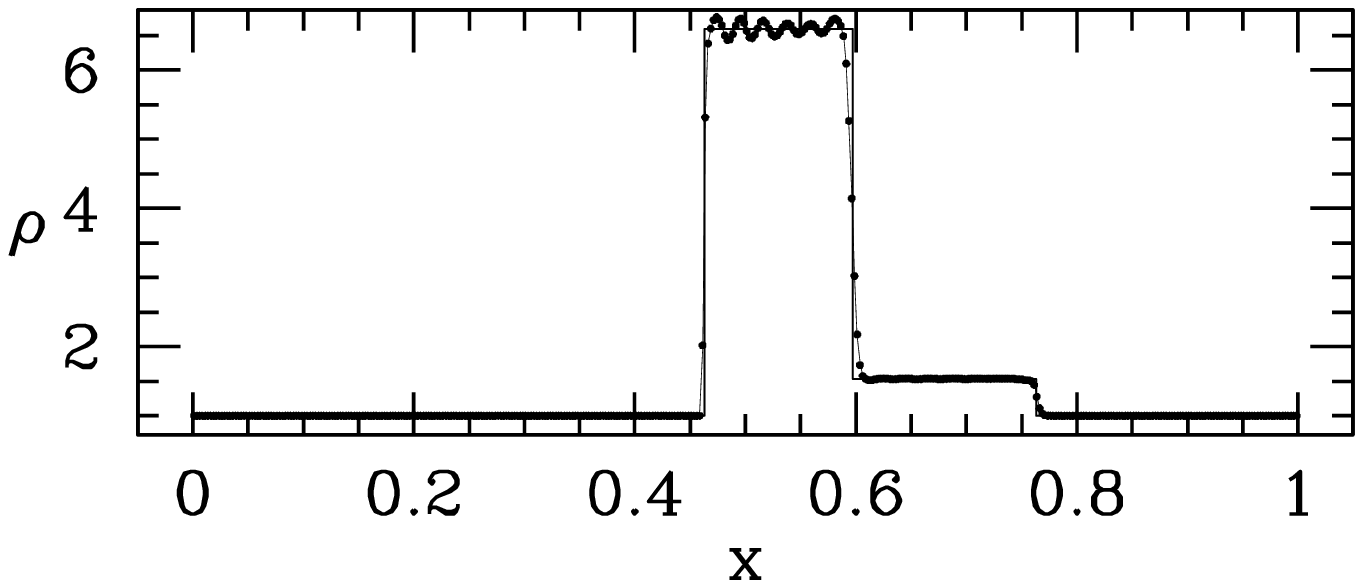,width=1.0 \columnwidth}
\caption{Density distribution of the one-dimensional relativistic Riemann problem 3
from~\citetalias{zha06}. \plotnotation. This figure shows that WHAM gets the
positions of discontinuities correctly but sometimes produces
small amplitude oscillations near slowly moving shocks. Similar
oscillations are present at some level in most codes
that do not use eigenvector decomposition~\citepalias{zha06}.
}
\label{fig_1d_rel_103}
\end{figure}

The initial discontinuity in this problem generates two shocks and a contact
discontinuity, see figure~\ref{fig_1d_rel_103}.
Table~\ref{tab_riemtests} lists the problem parameters. The
slow moving reverse shock appears to be hard for most codes to handle.
Even the most dissipative second order schemes like $\MINMOD$ produce
oscillations at a noticeable level. Our code exhibits them too
at about twice the level of second order
codes that do not use eigenvector decomposition.
We note that the relativistic version of the FLASH code, which uses a
flattening procedure
to decrease the order of the scheme to first order in shocks,
does not produce post-shock oscillations for this test~\citep{mor06}.  However, this flattening
procedure
requires parameter tuning for strong shocks (see
section~\ref{sec_test_1d_rel_105}).

\subsection{$1$D relativistic Riemann problem 4}
\label{sec_test_1d_rel_104}

\begin{figure}
\centering\epsfig{figure=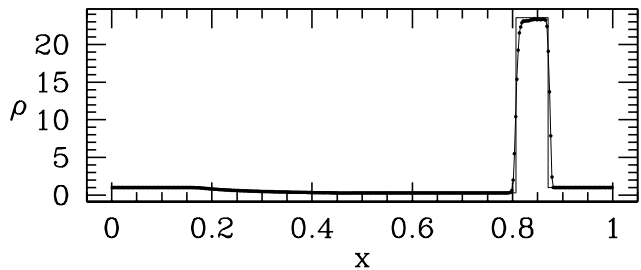,width=1.0 \columnwidth}
\caption{One-dimensional relativistic Riemann problem 4 from~\citetalias{zha06}.
WHAM is able to accurately resolve the contact
discontinuity and the height of the built-up state for this flow that has a large
shearing velocity.
}
\label{fig_1d_rel_104}
\end{figure}

The two initial states in this test have a large shearing velocity
(table~\ref{tab_riemtests}). Shearing flows are intrinsically hard in the
relativistic case because of the coupling between the directions through the
Lorentz factor. This configuration, however, does not lead to problems.
Figure~\ref{fig_1d_rel_104} shows that the width of the contact discontinuity
is the same as that of the WENO code from~\citetalias{zha06}; in our case,
without the use of eigenvector decomposition, the built-up state near
the contact discontinuity slightly undershoots compared to the analytic value.

\subsection{$1$D relativistic Riemann problem 5}
\label{sec_test_1d_rel_106}

\begin{figure}
\centering\epsfig{figure=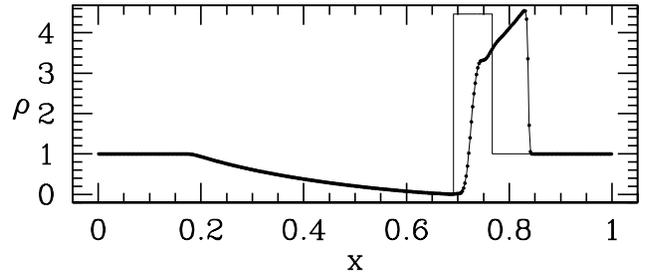,width=1.0 \columnwidth}
\caption{Density distribution for the one-dimensional relativistic Riemann problem 5 from~\citetalias{zha06}.
For large transverse velocities as in this problem, one has to use an increased resolution
in order to obtain the correct positions of discontinuities.
}
\label{fig_1d_rel_106}
\end{figure}

This test is very similar to the previous one in terms of the initial
conditions (table~\ref{tab_riemtests}). In this case, however, there is no
shear; rather, both states are moving at a relativistic velocity in the
transverse direction. Figure~\ref{fig_1d_rel_106} shows that at a moderate
resolution of $400$ cells WHAM gets the positions of the discontinuities
incorrectly. In order to get reasonable convergence to the analytic solution,
one has to use an increased resolution for this test. Our code's result agrees
with that of other codes at the same resolution~\citep{zha06,mor06}.

\subsection{$1$D relativistic Riemann problem 6}
\label{sec_test_1d_rel_105}

\begin{figure}
\centering\epsfig{figure=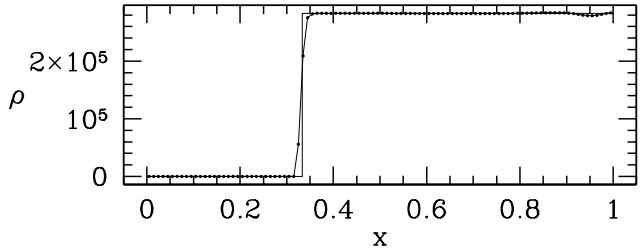,width=1.0 \columnwidth}
\caption{Density distribution for the one-dimensional relativistic Riemann problem 6 from~\citetalias{zha06}.
The numerical solution is shown with connected dots, and the analytical
solution~\citepalias{zha06} is shown with a solid line.
WHAM is able to evolve extremely relativistic
supersonic flows with strong shocks with
minimal Gibbs phenomenon.
}
\label{fig_1d_rel_105}
\end{figure}

This is a very stringent test that probes the ability of a code to handle
flows at extremely large Lorentz factors with strong shocks. It is a
generalization of the nonrelativistic one-dimensional Noh problem
(section~\ref{sec_test_1d_noh}) in which a highly relativistic flow with
$\gamma \approx 7 \times 10^4$ hits a reflecting boundary (see
table~\ref{tab_riemtests}). Figure~\ref{fig_1d_rel_105} shows that a reverse
shock forms at the boundary $x = 1$ and moves to the left leaving the matter
behind at rest.

WHAM produces minimal post-shock oscillations comparable to other schemes, such as
a WENO code that uses eigenvector decomposition~\citepalias{zha06}.
In the post-shock region the density reaches a maximum relative error of $1.5$\% in the vicinity of
the reflecting wall. This is twice as small as with WENO \zhap\ and
HARM. Unlike
the PPM algorithm~\citepalias{zha06}, we do not have to perform any fine-tuning
of our code parameters for this test.

A similar yet more extreme test is discussed by~\citet{alo99} for which they
chose $\gamma=2.24\times 10^5$, $p=7.63\times 10^{-6}$, and a resolution of
$200$ grid cells. As shown in Appendix~\ref{sec_cancellatonrel}, their test problem can be
barely resolved by a computer with a machine accuracy of $\sim 10^{-16}$ (i.e.
double precision). In order to minimize post-shock oscillations, they chose a
Courant factor of $0.1$ and tuned their reconstruction parameters. Using WHAM's
standard numerical settings (\eg\ Courant factor of $0.5$), we obtain an error
of $\lesssim 0.5$\% in all quantities for this test, an error similar to
that of~\citet{alo99} (here we use their definition of relative error).
The errors are dominated by the numerical solution having a shock width of
a few points.

\subsection{$2$D relativistic shock-tube problem}
\label{sec_test_2d_relriem}

\begin{figure}
\centering\epsfig{figure=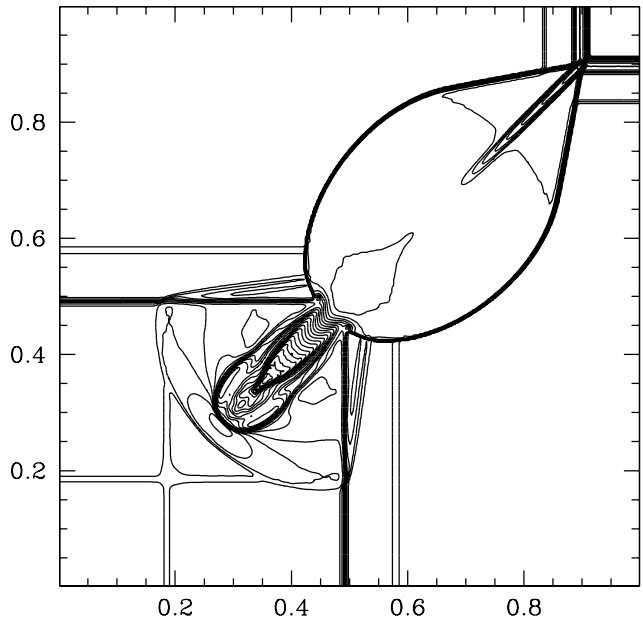,width=\columnwidth}
\caption{$2$D relativistic Riemann problem~\citep{del02,zha06} at the final time $t_\text{F} = 0.4$, see
section~\ref{sec_test_2d_relriem}.
The plot shows $30$ equally spaced contours of $\log_{10}\rho$ that go from
$-2.241$ to $0.8243$ in steps of $0.1057$.
WHAM is able to handle highly relativistic multidimensional flows
with minimal Gibbs phenomenon.
}
\label{fig_2drelriem}
\end{figure}

The initial conditions for this problem are shown in table~\ref{tab_2drelriem}.
The test is run at a resolution of $400\times400$ until $t_\mathrm{F} = 0.4$.
The contour plot of density is shown in figure~\ref{fig_2drelriem}. This is a
difficult highly relativistic two-dimensional Riemann problem~\citep{del02,
luc04, zha06,mor06}. The Lorentz factor reaches values larger than $25$ inside
the sharp diagonal jet-like feature in the lower left quadrant. The amount of
oscillation in the lower-left quadrant is less than with the relativistic FLASH
code~\citep{mor06} and the RAM code~\citepalias{zha06}, which uses
finite-difference fifth order WENO with field-by-field decomposition. WHAM does
not use such decomposition. The resolution of the diagonal feature
in the upper-right quadrant is comparable to RAM.

\begin{table}
\begin{center}
\caption{Initial conditions for the two-dimensional relativistic shock-tube
problem~\citep{del02}, see section~\ref{sec_test_2d_relriem}.
The notation is the same as in table~\ref{tab_initcase4}. The problem
is computed on the domain $(0,1)\times(0,1)$ at a resolution of $400\times400$
until time $t_\text{F}=0.4$ with $\polyconst = 5/3$. }
\begin{minipage}{\textwidth}
\begin{tabular}{@{\,}cccc@{\qquad}cccc@{\,}}
\hline
$\rho_L$ & $v_{x,L}$ & $v_{y,L}$ & $p_L$ & $\rho_R$ & $v_{x,R}$ & $v_{y,R}$ & $p_R$ \\
\hline
$0.1$ & $0.99$ & $0.0$ & $1.0$ & $0.1$ & $0.0$ & $0.0$ & $0.01$ \\
$0.5$ & $0.0$ & $0.0$ & $1.0$ & $0.1$ & $0.0$ & $0.99$ & $1.0$ \\
\hline
\label{tab_2drelriem}
\end{tabular}
\end{minipage}
\end{center}
\end{table}

\subsection{$2$D relativistic jet in cylindrical geometry}
\label{sec_test_2d_jet}

We study a two-dimensional relativistic jet in cylindrical geometry. This
problem couples many elements, which were separately tested in the individual
problems described so far, in a real astrophysical setting:
relativistic, highly supersonic flow, containing strong relativistic shocks,
shear flows, and instabilities. Along with being of astrophysical significance,
this test allows us to benchmark our numerical scheme against others. For the
sake of comparison, we have used the same conditions for the test as
\citet{zha06} and \citet[model C2]{mar97}. The size of the computational domain
is $(0,15)\times(0,45)$ in the $(R,z)$ plane, and we use a resolution of $384
\times 1152$. The ambient medium is initially at zero velocity, with density
$\rho_m = 1$ and pressure $p_m = 0.000170305$. The jet is injected along the
$z$-axis from the lower boundary with a density of $\rho_b = 0.01$, pressure
equal to the ambient pressure, $p_b = p_m$, and a speed of $v_b = 0.99 c$, which
corresponds to a Lorentz factor of $\gamma \approx 7$. Numerically, we implement the injection by assigning the values of the
boundary grid cells within $R < 1$, $z < 0$ to the state of the jet material.
We use the zero-derivative outflow boundary conditions at the low-$z$ ($R\geq1$), high-$z$, and
high-$R$ boundaries and use the appropriate boundary condition on the axis $R =
0$.  We evolve the system until $t_\text{F} = 100$.

\begin{figure}%
\begin{minipage}[t]{\columnwidth}
  \begin{minipage}[t]{0.67\linewidth}
    \par\vspace*{0mm}
    \subfigure{\fbox{\epsfig{figure=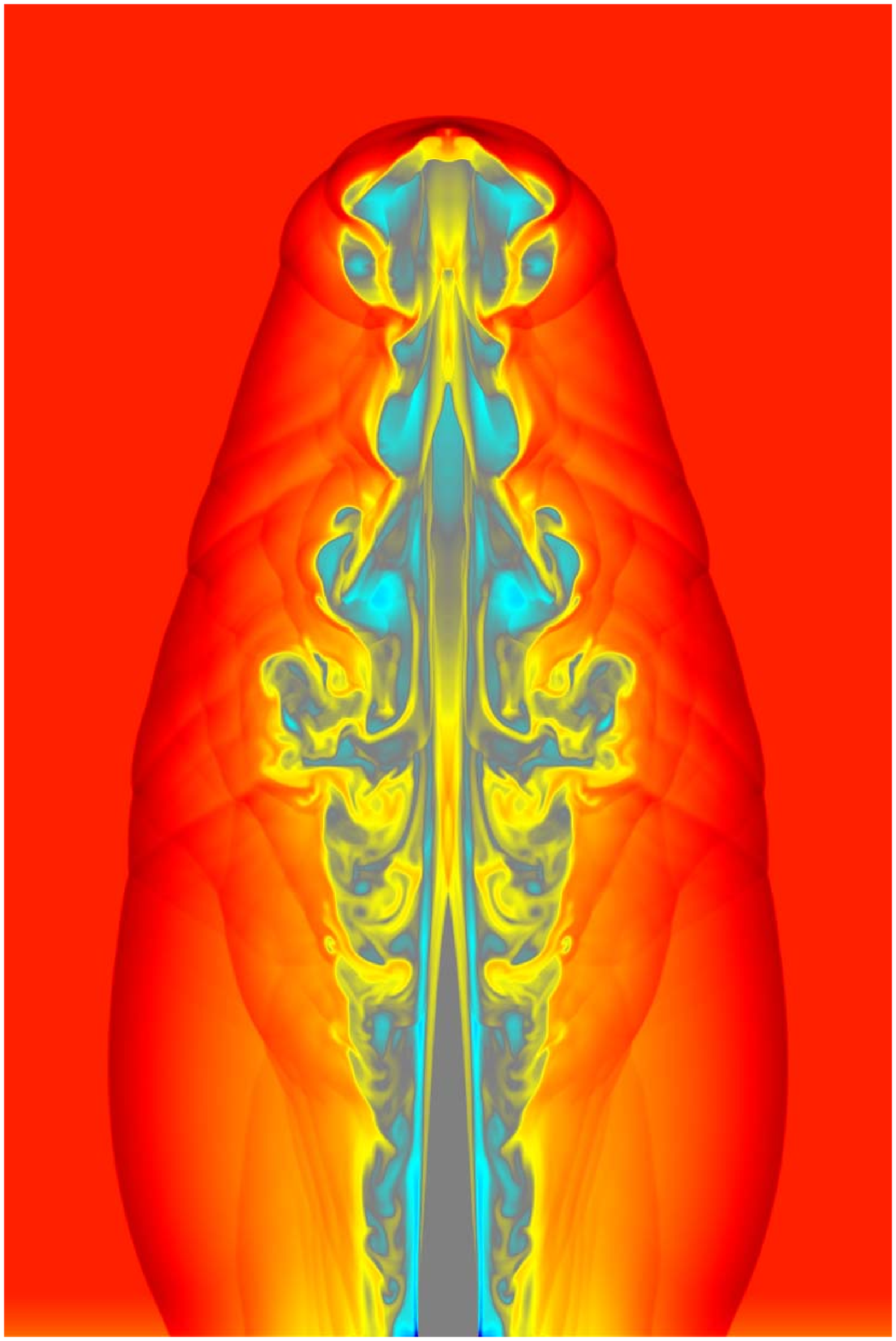,width=\linewidth,clip=}}}%
  \end{minipage}
  \hfill
  \begin{minipage}[t]{0.31\linewidth}
   \par\vspace*{0mm}
   \subfigure{\begin{sideways}\epsfig{figure=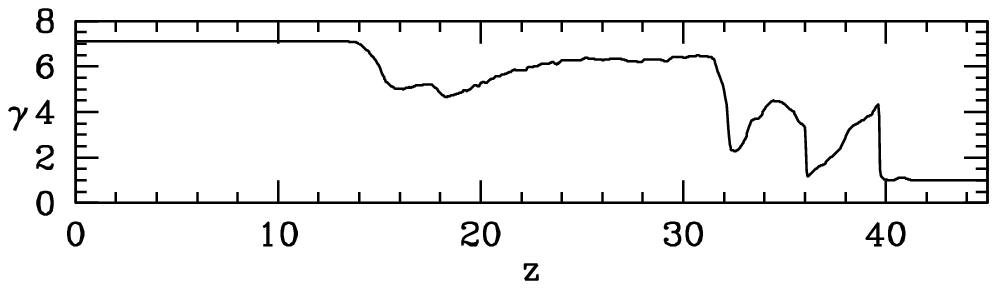,height=\linewidth}\end{sideways}}%
  \end{minipage}
\end{minipage}
\caption{Relativistic two-dimensional jet problem in cylindrical geometry:
 the left panel shows the color-coded distribution of the logarithm of
 the fluid frame rest-mass density (red denotes high values and blue low values), and the right panel shows
 the $z$-dependence of $\gamma$ along the axis of the jet.
 The simulation was run at a resolution of $384\times1152$,
 the same as in~\citetalias{zha06},
 and the result is mirrored across the axis in order to obtain the figure in the left panel.
 For more detail, see section~\ref{sec_test_2d_jet}.
 WHAM reproduces all essential elements of the jet structure for this
 astrophysically relevant problem.
}
\label{fig_test_2d_jet}
\end{figure}

The interaction of the jet material with the ambient medium forms an expanding
bow shock and a Mach shock with a contact discontinuity in between that
goes Kelvin-Helmholtz
unstable, see figure~\ref{fig_test_2d_jet}. The jet
contains internal shocks, backflows, and shear flows.  The
jumps in the $\gamma$-factor on the right panel of the figure correspond to
relativistic shocks crossing the axis of the jet. The essential structure of
the jet, its head position, the shape of the bow shock, and the development of
the Kelvin-Helmholtz instability agree with other codes' results~\citep{mar97,
zha06}.

\subsection{Bondi flow in Schwarzschild geometry}
\label{sec_test_bondi}

We study the order of convergence of our the scheme for stationary spherically
symmetric accretion on to a non-spinning black hole~\citep{bon52}. Even though
the exact solution is spherically symmetric, for WHAM in 2D the Bondi problem
is in fact two-dimensional: in the $\theta$-momentum equation, a gradient of
momentum flux ({}$\p_\theta(p_g \sin\theta)$ in the nonrelativistic limit) is
balanced by a source term ($p_g\cos\theta$) that numerically cancel out to
within the truncation error~\citep{gam03}. We find that in order to obtain an
accurate solution, one has to properly average the source terms on the \rhs\ of
the equations of motion~\eqref{eq_conserve}. Further, for this test WHAM uses
Kerr-Schild coordinates that have nonzero space-time mixing even for the case
of a non-spinning black hole: $7$ out of $10$ components of the metric are nonzero,
so this problem involves many of the terms that appear in the general equations
of motion. For this test problem we set the source term analytically in order to test
convergence near machine precision accuracy.

We follow the setup of the problem as discussed by~\citet{haw84,gam03}.
We fix the radius of the sonic surface, $R_\text{s} = 8$, the adiabatic
index of the gas, $\polyconst = 4/3$,
the gas adiabat, $K = 1 \,(= p_g/\rho^\polyconst)$,
and choose the mass of the black hole, $M = 1$.
These determine the critical solution and the three integrals of the problem:
the radial flux of mass ($= \gdet\rho u^r$), the flux of energy
($=\tensor{T}{r}{t}$),
and the entropy of the flow.
The solution can be found
semi-analytically by solving an $8$th order polynomial equation for
the temperature of the flow~\citep{haw84}.

We treat the boundary conditions as follows. We use the value of the `baryon
flux' $C_1$ from the outermost active grid cell together with the critical
values of $K$ and the `energy flux' $C_2$ to set the state of the outer-$r$
boundary cells \citep[for definitions of the integrals $C_1$ and
$C_2$, see][]{haw84}. At the inner-$r$ boundary we choose the states of the
cells to correspond to the three integrals of the flow determined from the
numerical solution at the innermost active grid cell in the active grid. At the
$\theta$-boundaries we use the usual polar axis boundary conditions. As the
system approaches steady state, the mass flux is determined by the
critical condition at the sonic surface. Since in steady state the fluxes
become uniform in space within the truncation error, extrapolating the flux
allows us to approximate the boundary conditions very accurately, compared to
extrapolating, \eg, the nontrivially varying density. This also avoids the order
of the extrapolation limiting the order of convergence of the scheme.

\begin{table}
\begin{center}
\caption{Relative $\Lonenorm$-errors in density and the order of convergence
of the WHAM, HARM, and \SWENOnos\ schemes
for the 1D and 2D spherical accretion problems (section~\ref{sec_test_bondi}),
at $N\times N$ resolution.
At the same resolution, the higher order WHAM scheme provides a much smaller
error than the other two schemes.
The averaging of the source terms is crucial for preserving the high order accuracy.
Radial velocity and internal energy similarly converge to the analytic solution.}
\begin{tabular}{@{\,}l@{\quad}c@{\quad}c@{\quad}c@{\quad}c@{\quad}c@{\quad}c@{\,}}
\hline
$N$ & $16$ & $32$ & $64$ & $128$ & $256$ & $512$ \\
\hline
\multicolumn{7}{c}{1D Bondi problem}\\
WHAM &  $1.6\Exp{-}{5}$ & $1.8\Exp{-}{7}$ & $1.5\Exp{-}{9}$ & $1.6\Exp{-}{11}$ & $1.2\Exp{-}{12}$ & $5.0\Exp{-}{15}$ \\
Order & --- & $6.4$ & $6.9$ & $6.6$  & $3.7$  & $7.9$ \\
\hline
HARM &  $9.7\Exp{-}{4}$ & $2.1\Exp{-}{4}$ & $4.9\Exp{-}{5}$ & $1.2\Exp{-}{5}$ & $2.8\Exp{-}{6}$ & $6.9\Exp{-}{7}$ \\
Order & --- & $2.2$ & $2.1$ & $2.0$  & $2.0$  & $2.0$ \\
\hline
\SWENOnos\ &  $2.0\Exp{-}{3}$ & $5.2\Exp{-}{4}$ & $1.3\Exp{-}{4}$ & $3.2\Exp{-}{5}$ & $7.9\Exp{-}{6}$ & $2.0\Exp{-}{6}$ \\
Order & --- & $2.0$ & $2.0$ & $2.0$  & $2.0$  & $2.0$ \\
\hline
\multicolumn{7}{c}{2D Bondi problem}\\
WHAM &  $1.6\Exp{-}{5}$ & $1.8\Exp{-}{7}$ & $1.5\Exp{-}{9}$ & $1.6\Exp{-}{11}$ & $1.2\Exp{-}{12}$ & $4.9\Exp{-}{15}$ \\
Order & --- & $6.4$ & $7.0$ & $6.5$  & $3.7$ & $7.9$ \\
\hline
HARM  &  $6.4\Exp{-}{4}$ & $1.6\Exp{-}{4}$ & $3.9\Exp{-}{5}$ & $9.7\Exp{-}{6}$ & $2.4\Exp{-}{6}$ & $6.0\Exp{-}{7}$ \\
Order & --- & $2.0$ & $2.0$ & $2.0$  & $2.0$ & $2.0$  \\
\hline
\SWENOnos &  $1.4\Exp{-}{3}$ & $3.3\Exp{-}{4}$ & $8.2\Exp{-}{5}$ & $2.0\Exp{-}{5}$ & $5.1\Exp{-}{6}$ & $1.3\Exp{-}{6}$ \\
Order & --- & $2.0$ & $2.0$ & $2.0$  & $2.0$ & $2.0$ \\
\hline
\label{tab_test_bondi}
\end{tabular}
\end{center}
\end{table}

We have performed the problem in 1D and 2D in modified Kerr-Schild coordinates,
the same coordinate system as in~\citet{gam03}, over the domain $r\in(1.9,20)$
in units of $G M/c^2$. We initialize the problem with the analytic solution and
let the system evolve for $t = 1000$ in units of $GM/c^3$, corresponding to
several sound crossing times, by which time the system reaches steady state. We
then compute the relative $\Lonenorm$-error norm in density between the final
solution and the initial conditions over the active grid. The results are shown
in table~\ref{tab_test_bondi} for various resolutions. WHAM converges at an
order higher than $5$ and makes a much smaller error than the \SWENOnos\ scheme,
which does not average the source terms, or the HARM scheme. The latter two
schemes converge at second order as expected, since they do not properly account
for the difference between the point and average source terms and/or fluxes and
conserved quantities. The nonuniform convergence rate appears to be due to the
sensitivity of the monotonicity indicator (see section~\ref{sec_jmono}) for
particular parts of the solution.

\subsection{Equilibrium torus in Kerr geometry}
\label{sec_test_2d_torus}

Finally, we consider a rotating fluid torus surrounding a spinning
black hole that is in equilibrium under the action of pressure and
centrifugal forces~\citep{fis76}. We choose the black hole spin to be
$a/M=0.95$, and take the following parameters for the torus: location
of the inner edge $r_\text{in} = 3.7$, angular momentum per unit mass
$u^tu_\varphi = 3.85$~\citep[as in][]{mck04,gam04}, and equation of
state $p_g = K \rho^\polyconst$ with $K = 10^{-3}$, $\polyconst =
4/3$.

The exact \citeauthor{fis76} torus solution is embedded into a vacuum. In order to
avoid having zero densities that are problematic numerically, we embed the torus in a dynamically
unimportant atmosphere by introducing floors on the density and internal energy:
$\rho_\text{min} = 10^{-4} (r/r_\text{in})^{-3/2}$, $u_\text{min} = 10^{-6}
(r/r_\text{in})^{-5/2}$ \citep[as in][]{gam03}.

We use the same modified Kerr-Schild coordinates as in \citet{gam03},
concentrating the resolution toward the midplane, and run the simulation until
$\tf = 10$ in the $2$D domain $(r,\theta) \in (0.98 r_h,\, 20) \times(-\pi/2,
\pi/2)$, where $r_h$ is the radius of black hole horizon.
We compute the relative
$\Lonenorm$-error norm between the final solution and the initial conditions
over the region where $\rho > 0.02 \rho_\text{max}$ to minimize the influence
of the atmosphere on the convergence results. The results are shown in
table~\ref{tab_test_torus}. We see that WHAM asymptotically converges at fifth
order,\footnote{At low resolutions, the truncation error of WHAM
is dominated by unresolved edges of the torus where the conserved quantities
rapidly fall off to zero and the reconstruction becomes
second order. Even though the torus is resolved with about $30$ grid cells
at a resolution of $64\times64$, the drop of $U_0 = \gdet \rho u^t$ at the
torus edge is resolved with only $4$ grid cells. Because of this,
at low resolutions HARM is comparable to WHAM,
but at high resolutions WHAM is far more accurate.}
while HARM converges at second order.

This final example tests all the
general relativistic aspects of WHAM -- equations of motion, metric,
connection coefficients, and source terms -- in Kerr space-time.

\begin{table}
\begin{center}
\caption{Relative $\Lonenorm$-error in density and order of convergence in density
of the WHAM and HARM schemes for an equilibrium torus problem, section~\ref{sec_test_2d_torus},
at $N \times N$ resolution.
Similar convergence is observed in all other quantities.
For this general relativistic problem in the Kerr metric, WHAM achieves asymptotic fifth order convergence.}
\begin{minipage}{\textwidth}
\begin{tabular}{@{\,}lcccccc@{\,}}
\hline
$N$ & $16$ & $32$ & $64$ & $128$ & $256$ & $512$ \\
\hline
WHAM &  $1.6\Exp{-}{2}$ & $1.7\Exp{-}{3}$ & $2.3\Exp{-}{4}$ & $2.8\Exp{-}{5}$ & $1.4\Exp{-}{6}$ & $2.7\Exp{-}{8}$ \\
Order & --- & $3.3$ & $2.9$ & $3.1$  & $4.3$  & $5.7$ \\
\hline
HARM &  $1.5\Exp{-}{2}$ & $3.5\Exp{-}{3}$ & $6.6\Exp{-}{4}$ & $1.4\Exp{-}{4}$ & $3.3\Exp{-}{5}$ & $8.3\Exp{-}{6}$ \\
Order & --- & $2.1$ & $2.4$ & $2.2$  & $2.1$  & $2.0$ \\
\hline
\label{tab_test_torus}
\end{tabular}
\end{minipage}
\end{center}
\end{table}

\section{Limitations} \label{sec_limitations}

The hydrodynamical numerical scheme we have described actually does
include magnetic fields, with the divergence-free
constraint~\eqref{eq_nomonopoles} applied using the FLUX-CT
constrained transport method \citep{gam03}. This method averages the
fluxes of the magnetic field in such a way that the associated update
to the magnetic field is guaranteed to preserve a certain numerical
representation of magnetic field divergence. However, the magnetic
fields are treated to lower order than the hydrodynamical quantities.
In a followup paper (McKinney, Tchekhovskoy, \& Narayan in prep.), we
will describe a consistent high order scheme that generalizes our
hydrodynamical method to a full MHD method with a constrained
transport method that is a high order version of the staggered grid
method used by Athena~\citep{gar05}.
We expect that the stiff regime associated with the MHD equations in the highly
magnetized limit will significantly benefit from our higher-order
reconstruction method since the effective magnetic Mach number is $M_\text{mag}$,
which is on order of $10^3$ or larger for models of black hole or neutron star
systems.

WHAM is capable of performing simulations in three dimensions. However, a
description of our method and a suite of 3D tests will be given in a followup
paper (McKinney, Tchekhovskoy, \& Narayan in prep.).

We have not implemented any performance optimizations in our scheme
since much has been already optimized in HARM which WHAM is based upon.
Currently, for one-dimensional problems, WHAM is about $4$ times
slower than the third order method used by \citet{mck06jf}. We expect
that further optimization will lead to a factor of several speed increase and
make WHAM competitive with any scheme of similar accuracy/order.  Our
experience suggests, however, that even without optimizations the
duration of simulations will not be the bottleneck in completing
science projects.

\section{Proposed applications}
\label{sec_applications}
The numerical scheme that we have developed provides superior accuracy
in highly supersonic flows as well as flows in which the energy scales are
very different. The applications where the scheme is
particularly advantageous therefore include the study of supersonic
disk winds, jets, and pulsar magnetospheres. There have been numerical
studies of these systems within the
force-free~\citep{kom02,kom06,mck06ffcode,mck06ff,mck06pulff,spit06,mck06bp}
and full MHD~\citep{kom05,mck04,buc06,mn06a,mck06jf} limits.
\citet[][]{mck06jf} achieved high magnetization, $b^2/(\rho+u_g+p_g)
\sim 10^3$, and could handle values of up to $10^4$ but it was not
clear whether the evolution was accurate in that regime. However, even
such high magnetization values are still below the actual observed
values of $10^5-10^6$ in the case of pulsar outflows.

We plan to study jet and pulsar systems in the limit of cold GRMHD, \ie\ with
the internal energy of the plasma vanishing exactly. This limit should allow us
to achieve higher magnetisation factors more easily. Since our code does not
require the use of eigenvector decomposition, only the equations of motion
have to be changed to study systems in this limit.

Our code can also be
used to perform three-dimensional studies of thin discs around black holes that involve
highly supersonic motion in the $\varphi$-direction. Such studies would
determine how jet speed and power is related to disk thickness, address the
problem of the multidimensional structure of disks near the innermost stable circular
orbit~\citep{bes05}, and test the applicability of well-known analytic
models~\citep{sha73,nov73} there.

\section{Conclusions}
\label{sec_conclusions}

We have developed a high order accurate conservative finite volume general
relativistic hydrodynamic (GRHD) scheme called WHAM that is robust in the
vicinity of shocks and is very accurate in smooth flows even for
an arbitrary metric and coordinates. It compares very well
with its special relativistic hydrodynamic analogs and is, as far as we know,
the first GRHD code that converges at fifth order in smooth flows. In contrast
to most high order schemes~\citep{jia98,tit04,fen04,zha06}, WHAM uses
primitive quantities to reconstruct the values of quantities at interfaces.
This is a safe choice in the relativistic regime because it guarantees that the
interface values, which determine inter-cell fluxes, are physical.

WHAM avoids the use of the computationally intensive eigenvector
decomposition approach used by many high order schemes~\citep{jia98,tit04,fen04,zha06,
xin06}. Instead, we adaptively reduce the order of the scheme near
discontinuities, while, unlike the standard WENO formalism, avoid excessive
reduction in smooth monotonic flows.
The resolution of contact discontinuities we obtain is comparable to that of
WENO schemes that use eigenvector decomposition without contact steepeners.

Unlike finite difference schemes~\citep[\eg,][]{jia98,fen04,zha06}, which
conserve the integrals of motion up to truncation error, our finite volume
scheme conserves them exactly, to machine precision. Our scheme performs the
proper conversion between the average and point values of conserved quantities,
fluxes, and source terms using fifth order WENO-type reconstruction.
We find that the alternative method of quadratures is less accurate for this
purpose~\citep[\eg,][]{tit04,noe06, xin06}.

We have demonstrated that our scheme provides accurate solutions in
highly supersonic flows, which are intrinsically hard for conservative
numerical schemes. In particular, the scheme is able to accurately
evolve shocks for which both pre-shock and post-shock material moves
supersonically through the computational grid. The scheme is likely to
be particularly good for studies of relativistic and/or supersonic
flows near compact objects.  In the future we plan to extend the
formalism to include magnetic fields and study models for which the
magnetization is large to identify why present schemes have
difficulties in such regimes (McKinney, Tchekhovskoy, \& Narayan in
prep.).

\section*{Acknowledgments}

We thank Charles Gammie, John Hawley, Anatoly Spitkovsky, and Jim Stone for
useful discussions. We also thank the anonymous referee for valuable
suggestions and corrections that have helped to improve the clarity of the
presentation.

\appendix
\section{Implementation notes}

\subsection{Truncation error analysis for a smooth high Mach number flow}
\label{sec_hubble_error_analysis}
This section presents a detailed error analysis of the evolution
of the Hubble-type diverging flow, see sections~\ref{sec_hubble1d}
and~\ref{sec_test_1d_hubble}.  We are particularly interested in the
consequences of neglecting the difference between the cell average and cell centred
values of conserved quantities.  As we show below, this leads to a secular
error in the evolution of the internal energy so that it becomes negative.
The error is first order in time and second order in space, independent of
the actual order of discretisation of the scheme in space and time.  This means that
lowering the time step does not diminish the error; only by significantly increasing the
spatial resolution can the error be reduced.

Firstly, we verify that solution~\eqref{eq_hubble} satisfies the conservation laws:
\begin{align}
\label{eq_hubble_sol}
\frac{\p\rho}{\p t} &= - \frac{\vp \rho_0}{(\den)^2} = - \frac{\p (\rho v)}{\p x}, \\
\frac{\p(\rho v)}{\p t} &= - \frac{2 \rho_0 \vp^2 x}{(\den)^3} = - \frac{\p(\rho v^2)}{\p x},\\
\frac{\p u_g}{\p t} &= - \frac{ \polyconst u_0 \vp }{(\den)^{\polyconst+1}} = - \frac{\p \left[(u_g+p_g) v\right]}{\p x},
\end{align}
so the system \eqref{eq_hubble} correctly describes the time evolution of the
initial distribution described by setting $t = 0$ in the system.

Let us assume that the spatial reconstruction on
the primitive quantities is exact but one does not take into account the
difference between cell averaged and cell centred values of conserved quantities.
We use the Euler method to discretise time stepping and check if the method
converges at second order in time as one would naively expect. We only need to
consider the first time step since for all successive time steps the result of
the previous time step(s) can be considered an initial condition. We define
a uniform numerical grid consisting of grid cells $\Delta_i = [x_{i-1/2},
x_{i+1/2}]$ of size $h$. For spatial discretisation we use the simplest finite
volume scheme: we linearly interpolate the values of primitive quantities from
cell centres to cell interfaces within each of the grid cells. Since in the
above solution all primitive variables depend linearly on $x$, this linear
interpolation is exact.

Consider the $i$th grid cell, $\Delta_i$. For clarity we shall omit the
index $i$ where it does not lead to confusion. Let us integrate the equations
from $t = 0$ to $t = \dt$. The conserved variables
are the density $\rho$, specific momentum $p = \rho v$, and the
total specific energy $E = \rho v^2/2 + u_g$. The discretisation of conservation
laws then takes the form:
\begin{align}
\Delta \langle\rho\rangle = &- \dt\left[\rho_0 \vp (x+h/2) - \rho_0 \vp (x-h/2)\right]/h \notag \\
 = &- \rho_0 \vp \dt, \label{eq_rhoavg.update}\\
\Delta \langle p\rangle = &- \dt
               \left[
                     \rho_0 \vp^2(x+h/2)^2
                   - \rho_0 \vp^2(x-h/2)^2
              \right]/h \notag\\
            = &- 2 \rho_0 v_0 \vp \dt, \label{eq_pavg.update}\\
\Delta \langle E\rangle = &- \dt
                    \left[
                          \rho_0 \vp^3 (x + h/2)^3/2 - \rho_0 \vp^3 (x-h/2)^3/2
                    \right]/h \notag \\
                     &- \dt (u_g+p_g)_0 \vp
                    \notag\\
            = &- (\rho v^2/2)_0 \, 3\vp \dt \left(
                     1 + \myfrac{1}{12} \, \simplefrac{h^2}{x^2}
               \right)- (u_g+p_g)_0 \vp \dt, \label{eq_Eavg.update}
\end{align}
where angle brackets indicate averaging over the grid cell and index~`0' indicates
values taken at $t = 0$. The latter equation corresponds to the following conservation law:
\begin{equation}
\frac{\p E}{\p t} = - \ppx\left[(E+P)v\right] \equiv
                    - \ppx\left[\rho v^3/2 + (u_g + p_g) v \right].
\end{equation}
The above expressions \eqref{eq_rhoavg.update} -- \eqref{eq_Eavg.update} are so
far exact up to first order in time. Now suppose we remove the angle brackets
on the \lhs\ in the above expressions, \ie\ do not make a distinction between
cell centred and cell averaged quantities. Equations \eqref{eq_rhoavg.update}
and \eqref{eq_pavg.update} will still be exact up to first order in time for
the particular problem considered, \ie\ they will still give the updates to
cell centred quantities $\rho$ and $p$ that are first order accurate in time.
This is because $\rho$ and $p = \rho v$ are linear functions of $x$ and thus
their average value in any grid cell is the same as the point value at the cell
centre. This is not the case, however, for equation \eqref{eq_Eavg.update}
where energy $E$ is a nonlinear function of $x$: $E \ne \langle E \rangle$.
This also can be seen from the expansion of equation~\eqref{eq_hubble}, according to which the
cell centred conserved quantities evolve according to
\begin{align}
\Delta \rho &= \rho_0 \left(- \vp \dt + \vp^2 \dt^2 - \dots\right), \notag\\
\Delta p &= p_0 \left( - 2\vp \dt + 3 \vp^2 \dt^2 - \dots\right), \label{eq_cons.evol}\\
\Delta E &= (\rho v^2/2)_0 \left( - 3 \vp \dt + 6 \vp^2 \dt^2 - \dots \right) \notag\\
                &\; + (u_g + p_g)_0 \left( - \vp \dt + (\polyconst+1)\vp^2\dt^2/2-\dots\right).\notag
\end{align}

So, by treating the update to the average of the energy in the grid cell as the
update to the point value of energy, we are making an absolute error in $\Delta E$ equal to
\begin{equation}
\err(\Delta E) = \Delta \langle E \rangle - \Delta E =
        - \frac{1}{8} \rho_0 v_0^2 \vp h^2 \dt \frac{h^2}{x^2}
        \equiv - \frac{1}{8} \rho_0 \vp^3 h^2 \dt. \label{eq_E.1storder.error}
\end{equation}
Note that up to first order in time, $\err(\Delta \rho) = \err(\Delta p) = 0$.
Since the internal energy is computed from the values of conserved quantities via
\begin{equation}
u_g = E - p^2 / ( 2 \rho ),
\end{equation}
the error in the update to internal energy is
\begin{equation}
\err(\Delta u_g) = \err(\Delta E) = - \frac{1}{8} \rho_0 \vp^2 h^2 \, \vp\dt.
\label{eq_hubble_uerror}
\end{equation}
It is first order in time for every step, which means that using smaller steps
(\ie~lowering the Courant factor) does not lead to convergence.
We note that using higher order discretisation in time and/or space does not
help to avoid this error.
If we were to use a higher order time
stepping discretisation scheme, the time stepping error would remain the same
because higher order time stepping schemes assume that each simple first order
time step is accurate to first order. However, here this is not the case.  So for higher
order time stepping error~\eqref{eq_hubble_uerror} remains the same. Only by
increasing the spatial resolution can one get convergence of second order in
space, independent of the actual spatial and temporal order of approximation of
the scheme.

We can estimate the error in the internal energy at a characteristic time of the evolution
by setting $\dt = 1/\vp$,
\begin{equation}
\label{eq_hubble_sol_u}
\frac{\err(\Delta u_g)}{u_0}\Big|_{\vp \dt = 1} \sim -\frac{\rho_0 \vp^2 h^2}{u_0} \sim - M_\text{min}^2.
\end{equation}
This error is on the order of the minimum of the Mach number on the grid squared.
Therefore, for large Mach numbers, the
schemes that do not take into account the difference between the point
and average values of conserved quantities make very large errors in the
evolution of the internal energy and require the use of the
resolution proportional to the Mach number of the flow in order
to correctly capture the evolution of the internal energy.
For instance, for a Mach number $M_\text{min} = 100$, the resolution
has to be increased by about two orders of magnitude.
See section \ref{sec_test_1d_hubble} for a numerical verification of these results.

\subsection{Reconstruction matrices and optimal weights for WENO-$5$}
\label{sec_optimalweights}
In WENO-$5$ the integer $k$ in section \ref{sec_weno} is equal to three.
Thus we have $3$ stencils, each of length $3$.
The reconstruction matrices in equation \eqref{eq_recon_onestencil} for \atoc\ and \ctoa\ reconstructions,
$c^\text{(ac)}_{ij}$ and $c^\text{(ca)}_{ij}$, are the inverse of each other:
\begin{alignat*}{3}
c^\text{(ac)}_{ij} = &\left(
\begin{array}{rrr}
 \myfrac{23}{24} & \myfrac{1}{12} & -\myfrac{1}{24} \\
 -\myfrac{1}{24} & \myfrac{13}{12} & -\myfrac{1}{24} \\
 -\myfrac{1}{24} & \myfrac{1}{12} & \myfrac{23}{24}
\end{array}
\right), \\
c^\text{(ca)}_{ij}  = &\left(
\begin{array}{rrr}
 \myfrac{25}{24} & -\myfrac{1}{12} & \myfrac{1}{24} \\
 \myfrac{1}{24} & \myfrac{11}{12} & \myfrac{1}{24} \\
 \myfrac{1}{24} & -\myfrac{1}{12} & \myfrac{25}{24}
\end{array}
\right).
\intertext{The reconstruction matrices for centre to left interface and
centre to right interface
reconstructions, $c^\text{(cl)}_{ij}$ and $c^\text{(cr)}_{ij}$, are:}
c^\text{(cl)}_{ij} = &\left(
\begin{array}{rrr}
 \myfrac{15}{8} & -\myfrac{5}{4} & \myfrac{3}{8} \\
 \myfrac{3}{8} & \myfrac{3}{4} & -\myfrac{1}{8} \\
 -\myfrac{1}{8} & \myfrac{3}{4} & \myfrac{3}{8}
 \end{array}
\right),  \\
c^\text{(cr)}_{ij} = &\left(
\begin{array}{rrr}
 \myfrac{3}{8} & \myfrac{3}{4} & -\myfrac{1}{8} \\
 -\myfrac{1}{8} & \myfrac{3}{4} & \myfrac{3}{8} \\
 \myfrac{3}{8} & -\myfrac{5}{4} & \myfrac{15}{8}
\end{array}
\right).
\end{alignat*}

The optimal weights for centre to edge, average to centre, and centre to
average reconstructions can be computed by equating the expressions
\eqref{eq_weno_approx} and \eqref{eq_total_stencil}.  For completeness,
we give them here, see table~\ref{table_optimalweights}.  Note that for
average to centre and centre to average reconstructions, negative optimal weights
appear that require a special treatment.
See section \ref{sec_weightsprescription} for a discussion.

\begin{table}
\begin{center}
\caption{Optimal weights for various types of
WENO-$5$ reconstruction.  According the convention of \citet{shu97}
adopted in this paper, $d_0$, $d_1$, and $d_2$ are the right, central, and
left optimal weights.}
\begin{minipage}{\textwidth}
\begin{tabular}{@{\,}lccc@{\,}}
\hline
Reconstruction type &  $d_0$ & $d_1$ & $d_2$  \\
\hline
Centre to left face &   $\phantom{-}\simplefrac{1}{16}\phantom{-}$  & $\simplefrac{5}{8}$  & $\phantom{-}\simplefrac{5}{16}\phantom{-}$              \\
Centre to right face &   $\phantom{-}\simplefrac{5}{16}\phantom{-}$  & $\simplefrac{5}{8}$  &  $\phantom{-}\simplefrac{1}{16}\phantom{-}$             \\
Average to centre &   $-\simplefrac{9}{80}\phantom{-}$  & $\simplefrac{49}{40}$  & $-\simplefrac{9}{80}\phantom{-}$                \\
Centre to average &   $-\simplefrac{17}{240}\phantom{-}$  & $\simplefrac{137}{120}$  & $-\simplefrac{17}{240}\phantom{-}$      \\
\hline
\label{table_optimalweights}
\end{tabular}
\end{minipage}
\end{center}
\end{table}

\subsection{Choosing the value of $\epsilon$}
\label{sec_epsilon}

The standard WENO weights computation procedure uses a fixed value of
$\epsilon$ for the purpose of avoiding division by zero
in~\eqref{eq_weno_unnorm_weights} \citep{shu96,shu97}. However, setting
$\epsilon$ to a constant value independent of the function being interpolated
sets artificial, irrelevant scales for a problem.  Any discontinuity in
the interpolant, such that
smoothness indicators for the stencils that cross that discontinuity are much smaller than
$\epsilon$, is mistakenly treated by the reconstruction as part of smooth flow.

Let us consider $\epsilon \ll 1$.
Then a contact discontinuity, which is a density jump with $\Delta\rho =1$
at $x = 0$,
\begin{equation}
\rho(x) = \begin{cases}
2, & x \ge 0 \\
1, & x < 0,
\end{cases}
\end{equation}
is correctly treated by the reconstruction, \ie~the reconstruction
avoids using the stencils that pass through the discontinuity.
However, if the same density jump is recast in different density units,
\begin{equation}
\tilde\rho(x) = \begin{cases}
2 \epsilon, & x \ge 0 \\
\epsilon, & x < 0,
\end{cases}
\end{equation}
it is not avoided by the reconstruction because in \eqref{eq_weno_unnorm_weights}
$\epsilon \gg \beta_r \sim (\Delta\tilde\rho)^2 \sim \epsilon^2$.
The addition of $\epsilon$ to the smoothness indicators in this case
effectively hides the jump from the reconstruction and leads to an
oscillatory reconstruction.

We propose the following algorithm for the adaptive choice of $\epsilon$.
We leave the weights computation \eqref{eq_weno_unnorm_weights} --
\eqref{eq_weno_norm_weights} the same but modify the smoothness indicators that come into
it:
\begin{equation} \label{eq_smoothnessindnew}
\beta'_r = \beta_r+ \epsilon \norm{v^2} + \delta,
\end{equation}
where $\norm{v^2}$ is the sum of $v_i^2$ within the WENO stencil and
$\delta$ is the smallest
positive number the floating-point type used in the numerical implementation
of the method can hold; it is added
to avoid division by zero when  $\norm{v^2}$ vanishes.
We  choose an~$\epsilon$ such that it is larger than the relative machine precision
in the computation of smoothness indicators and at the same time is negligible compared to them
in shocks, $\epsilon = (c \epsilon_\mathrm{machine})^2$.
Here $\epsilon_\mathrm{machine}$ is of order of relative machine
precision for the float type used
and $c$ is a sufficiently large constant. We choose $c = 100$, so for
\emph{double precision} we have  $\epsilon_\mathrm{machine}\sim 10^{-15}$ and
 $\epsilon = 10^{-26}$.

Further, in order to extend the dynamic range of the reconstruction
and avoid division by small numbers (smoothness
indicators become small for small $\norm{v^2}$), prior to plugging $\beta'_r$'s into
the weights computation
\eqref{eq_weno_unnorm_weights} we rescale them so that they add up to unity:
\begin{equation} \label{eq_smoothnessindnorm}
\beta''_r = \frac{\beta'_r}{\norm{\beta'_r}},
\end{equation}
where
$\norm{\beta'_r} = \sum_{r=0}^{r=k-1}\beta'_r$.

Finally, we use the usual weights computation
procedure~\eqref{eq_weno_unnorm_weights}~-- \eqref{eq_weno_norm_weights}
in which we use modified smoothness indicators~$\beta''_r$~\eqref{eq_smoothnessindnorm}
and the value of $\epsilon$ determined above.

\subsection{At what order does WENO-$5$ actually converge in smooth flows?}
\label{sec_weno5order}
Assuming we can neglect~$\epsilon$ in the
definitions of the weights, then in order  to comply
with~\eqref{eq_weights_requirement} we need to have for \hbox{$r = 0, \, \dots, \, k-1$}:
\begin{equation}
\beta_r = D \, \left[1 + \Order(h^{k-1})\right], \label{eq_beta_req}
\end{equation}
where $D$ is a positive constant common to all stencils at a given resolution
(so $D$ may depend on the grid cell size, $h$).

Let us verify if for a smooth function $v(x)$ smoothness
indicators~\eqref{eq_smoothness_indicator_definition} satisfy
requirement~\eqref{eq_beta_req} for the case of WENO-$5$, \ie~$k=3$.
The smoothness indicators for \ctoe, \atoc, and \ctoa\ reconstructions
are the same for the case of WENO-5 (see appendix \ref{sec_higherorderindicators})
and are given by:
\begin{align}
\beta_0 &= \frac{1}{4} (-3 v_i+4 v_{i+1}+v_{i+2})^2+\frac{13}{12} (v_{i}-2 v_{i+1}+v_{i+2})^2, \notag\\
\beta_1 &= \frac{1}{4} (v_{i+1}-v_{i-1})^2+\frac{13}{12} (v_{i-1}-2 v_i+v_{i+1})^2,\label{eq_bn}\\
\beta_2 &= \frac{1}{4} (v_{i-2}-4 v_{i-1}+3 v_i)^2+\frac{13}{12} (v_{i-2}-2 v_{i-1}+v_i)^2. \notag
\end{align}
Using Taylor expansion, this can be cast into
\begin{align}
\beta_0 &= \frac{1}{4} (2v'h-2/3v'''h^3-1/4v^{(4)}h^4)^2 + \frac{13}{12} (v''h^2+v'''h^3)^2,\notag\\
\beta_1 &= \frac{1}{4} (2v'h+1/3v'''h^3)^2 + \frac{13}{12} (v''h^2)^2,\label{eq_be}\\
\beta_2 &= \frac{1}{4} (2v'h-2/3v'''h^3+1/4v^{(4)}h^4)^2 + \frac{13}{12} (v''h^2-v'''h^3)^2, \notag
\end{align}
with the truncation error $\Order(h^6)$. Here the derivatives are evaluated at
the cell centre $x = x_i$: $v^{(n)}\equiv v^{(n)}(x_i)$. If $v' \ne 0$,
then
\begin{equation}
\beta_r = (v'h)^2\left[1+\Order(h^2)\right], \quad r = 0, 1, 2, \label{eq_1stderk3}
\end{equation}
which satisfies \eqref{eq_beta_req} and agrees with \citet{shu96}. This means
that if the $v'$-containing terms dominate in \eqref{eq_be},
then the smoothness indicators give nearly
optimal weights that satisfy \eqref{eq_weights_requirement}, and hence the
truncation order of WENO reconstruction \eqref{eq_weno_approx} is
$\Order(h^{2k-1}) \sim \Order(h^5)$ .

However, at simple extrema with $v' = 0$ and $v'' \ne 0$, or more precisely
with
$\abs{v'}h \ll \abs{v'''}h^3 \ll \abs{v''}h^2$, we have for smoothness
indicators,
\begin{align}
\beta_r &= \frac{13}{12} (v''h^2)^2 \left[1+2\, (v'''\!\!/v'') h + \Order(h^2)\right], \quad r = 0, 1, 2,
\label{eq_2derderk3}
\end{align}
which means that neither \eqref{eq_beta_req} nor \eqref{eq_weights_requirement}
are satisfied, and WENO reconstruction \eqref{eq_weno_approx} has
a larger truncation error, $\Order(h^{2k-2}) \sim \Order(h^4)$, near the
such extrema. This reduction in the order of the scheme occurs because the second
derivative approximation in the smoothness indicators \eqref{eq_bn} is only
first order accurate.
This disagrees with the result of \citet{shu96}. They
erroneously consider the numerical approximation to the 2nd derivative to be
2nd order accurate, \ie~they are missing the term proportional to $v'''$ in
\eqref{eq_2derderk3}.
Therefore in general the approximation order of the WENO-$5$ scheme
at extrema reduces to $4$, and only in the special case of extrema with $v'''=0$ does WENO-$5$ give fifth
order approximation, \eg~for $v(x) = \sin x$.

Finally, if both first and second derivatives become small compared to higher
order ones (at a high order extremum, a critical point, or in a smooth region
of the flow dominated by high-order derivatives) the order of the WENO-$5$
scheme reduces to the third one. This happens in contact discontinuities and
leads to their excessive diffusion, this also happens near inflection points,
\eg~of the $v(x) = x^3$ function: here WENO-$5$ is unable to capture
the dependence near $x=0$ exactly even though a fifth order polynomial would be able
to.

\subsection{High order approximation in smooth monotonic flows}
\label{sec_jmono}

WENO-$5$~scheme aims to obtain the flattest reconstruction profile inside a
grid cell of interest: it favors the stencils that minimize the absolute value
of the first and second derivatives within the grid cell. In smooth flows that
are locally well approximated by a parabola these derivatives do not change
significantly between the stencils, so such a prescription leads to nearly
optimal weights given to stencils and results in the fifth order of
approximation. However, if any terms higher than second order ones are
significant in the Taylor expansion of the function, the WENO-$5$ scheme
reduces the reconstruction order down to the third one. This reduction leads to
unwanted diffusion of contact discontinuities that becomes very large without
the use of field-by-field decomposition.

To avoid such a reduction of the order of the scheme, we use a fifth order
interpolation polynomial for reconstruction if the polynomial and its
derivatives are monotonic. This minimizes the amount oscillation in the
reconstructed solution and maintains a high order of interpolation in the
monotonic regions of the flow. We use this approach for the centre to edge
reconstruction. A similar idea is used in the PPM reconstruction procedure
\citep{col84,mig05}.

Consider an exponential dependence $f(x) = e^x$ with the grid cell centres
located at $x_i = h i$, $h = 1$. In this case the WENO-$5$ weights
\eqref{eq_weno_norm_weights} are $(\omega_0,\omega_1,\omega_2) \propto (0.007
d_0, 0.09 d_1, 0.9 d_2)$ independent of the value of $i$; here $d_r$'s are the
optimal weights (see text after equation~\eqref{eq_total_stencil}). WENO-$5$ is
giving about $90\%$ of the weight to one of the stencils, therefore the
WENO-$5$ reconstruction order is effectively reduced from $5$ to $3$. In this
case, the fifth order polynomial interpolation through, \eg, points $x_{-2},
\dots,x_2$ would have no oscillations at the interval $[x_{-1},x_1]$ and could
well be used.

Therefore, prior to applying the standard WENO-$5$ reconstruction inside an
$i$th grid cell, we consider the fifth order reconstruction polynomial $f_5(x)
$. It interpolates the set of discrete values $f_{i-2},\dots,f_{i+2}$ at points
$x_{i-2},\dots,x_{i+2}$. We check if all of the polynomial's derivatives are
monotonic at the interval $[x_{i-1}, x_{i+1}]$ and if they are, we use this
polynomial for interpolation instead of the WENO-$5$ interpolation. For a
fifth order polynomial, it suffices to check that each of the first three
derivatives, $f_5^{(n)}$, $n = 1,2,3$, has the same sign at both ends of the
interval; below for compactness we shall omit the index `5'.

We now describe an indicator
that vanishes unless the function values $f_{i-2},\dots,f_{i+2}$ form a monotonic
sequence and the interpolation polynomial $f(x)$ and all of its derivatives
are monotonic at $(x_{i-1}, x_{i+1})$ (in which case it is unity). In order
to avoid switching between the schemes, we design the indicator to be
a smooth function of discrete function values. We define the
minimum value by absolute magnitude of an $n$th derivative at the interval,
$f^{(n)}_\mathrm{min} = \MINMOD\left[f^{(n)}(x_{i-1}), f^{(n)}(x_{i+1})\right]$.
In order to ensure monotonicity of discrete values, we redefine
$f^{(1)}_\mathrm{min}=\MINMOD\left[f^{(1)}_\mathrm{min},f_{i-1}-f_{i-2},f_{i+2}-f_{i+1}\right]$.
Finally, we compute the minimum absolute value of nondimensionalized derivatives of
the reconstruction polynomial $f(x)$,
$f^{(\cdot) }_\mathrm{min} = \min\limits_{n=1,\,2,\,3} \Abs{f^{(n)}_\mathrm{min}h^n}$.
We refer to the weight that we give to the fifth order polynomial reconstruction as
the monotonicity indicator,
\begin{equation}
\alpha = \max %
 \left\{
    0, \min
      \left[
      1, f^{(\cdot)}_\mathrm{min}/(\sqrt\epsilon\,\norm{f}+\delta)
      \right]
 \right\},
\end{equation}
where $\norm{f} = \sum_{j=i-2}^{i+2} \abs{f_j}$ is the norm of $f(x)$ at the
stencil and $\delta$~-- added to
avoid division by zero -- is the minimum positive
value that a floating-point variable can hold on our machine.
In the above formula $\epsilon$ controls how fast the use of the fifth order
polynomial for reconstruction kicks in when the function and its derivatives
become monotonic.  For consistency, we choose $\epsilon$ to be
the same as the one in the WENO weights computation
procedure~\eqref{eq_smoothnessindnew},
\eqref{eq_smoothnessindnorm}, \eqref{eq_weno_unnorm_weights},
\eqref{eq_weno_norm_weights}.

The reconstructed value we use is given by a linear combination of
the fifth order accurate value of the polynomial, $f_5$,
and that of the WENO-type reconstruction, $f_\mathrm{WENO}$:
\begin{equation}  \label{eq_jmonotransition}
f = \alpha f_5 + (1-\alpha) f_\mathrm{WENO},
\end{equation}
where we have returned back to the notation $f_5$ for denoting the fifth order
reconstruction polynomial.

Since the reconstruction due to the fifth order polynomial is
equivalent to the WENO-$5$ reconstruction with weights set to optimal
ones, equation \eqref{eq_jmonotransition} is equivalent to using the WENO-$5$
reconstruction with modified weights $\omega'_r$:
\begin{equation}
\omega'_r = \alpha d_r + (1-\alpha) \omega_r, \quad r = 0, \dots, k -1.
\end{equation}
This provides a smooth
transition between using the polynomial and the WENO-type reconstructions.

\subsection{Higher order smoothness indicators}
\label{sec_higherorderindicators}

High order WENO-type schemes provide better convergence properties near
critical points of smooth flows than their lower-order counterparts but have a
higher computational cost~\citep{qiu02}. In this section we summarize the properties of
higher order smoothness indicators. In particular, we concentrate on the
convergence properties of WENO-type schemes near the critical
points of smooth flows: when does a WENO-$n$ scheme provide
the maximal, $n$th, order of convergence?

One can show for any $k$ (at least up to $k = 7$ as we have checked in
Mathematica) that any smoothness indicator
\eqref{eq_smoothness_indicator_definition} can be cast in the form analogous to
that of \eqref{eq_be}: as a linear combination of the squares of discrete
approximations to the derivatives at $x = x_i$. For instance for the case of
$k=4$, we have for the smoothness indicators:
\begin{align}
\beta_0 = &-\myfrac{1}{32} (-3 v_i+4 v_{i+1}-v_{i+2})^2
  \notag\\
      &+\myfrac{1}{32} (-11 v_i+18 v_{i+1}-9 v_{i+2}+2 v_{i+3})^2
    \notag\\
    &+\myfrac{13}{12} (2 v_i-5 v_{i+1}+4 v_{i+2}-v_{i+3})^2
   \notag\\
   &+\myfrac{3169}{2880} (-v_i+3 v_{i+1}-3 v_{i+2}+v_{i+3})^2, \\
\beta_1 = &+ \myfrac{1}{16} (v_{i+1}-v_{i-1})^2 \notag\\
  &+\myfrac{1}{48} (-2 v_{i-1}-3 v_i+6 v_{i+1}-v_{i+2})^2 \notag\\
    &+\myfrac{13}{12} (v_{i-1}-2 v_i+v_{i+1})^2 \notag\\
  &+\myfrac{3109}{2880} (-v_{i-1}+3 v_i-3 v_{i+1}+v_{i+2})^2,\\
\beta_2 = &+ \myfrac{1}{16} (v_{i+1}-v_{i-1})^2 \notag\\
  &+\myfrac{1}{48} (v_{i-2}-6 v_{i-1}+3 v_i+2 v_{i+1})^2 \notag \\
    &+\myfrac{13}{12} (v_{i-1}-2 v_i+v_{i+1})^2 \notag\\
  &+\myfrac{3109}{2880}(-v_{i-2}+3 v_{i-1}-3 v_i+v_{i+1})^2, \\
\beta_3 = &-\myfrac{1}{32} (v_{i-2}-4 v_{i-1}+3 v_i)^2 \notag\\
  &+\myfrac{1}{32} (-2 v_{i-3}+9 v_{i-2}-18 v_{i-1}+11 v_i)^2 \notag\\
    &+\myfrac{13}{12} (-v_{i-3}+4 v_{i-2}-5 v_{i-1}+2 v_i)^2 \notag\\
  &+\myfrac{3169}{2880} (-v_{i-3}+3 v_{i-2}-3 v_{i-1}+v_i)^2.
\end{align}
This is a much more conceivable representation of higher order smoothness
indicators than the one used in \citet{bal00}. In fact, this form allows one to
easily prove the properties of higher order smoothness indicators the same way
as we did for the case $k=3$. Again, using Taylor expansion, we get:
\begin{align}
\beta_0 =
   &-\frac{1}{8} \left(v' h-\frac{1}{3} v^{(3)} h^3-\frac{1}{4} v^{(4)} h^4\right)^2
   +\frac{9}{8} \left(v'h+\frac{1}{4} v^{(4)} h^4\right)^2 \notag\\
   &+\frac{13}{12} \left(v''h^2-\frac{11}{12} v^{(4)} h^4\right)^2
   +\frac{3169}{2880} \left(v^{(3)} h^3+\frac{3}{2} v^{(4)} h^4\right)^2
   ,\\
\beta_1 =
   &+\frac{1}{4} \left(v' h+\frac{1}{6} v^{(3)} h^3\right)^2
   +\frac{3}{4} \left(hv'-\frac{1}{12} h^4 v^{(4)}\right)^2 \notag\\
   &+\frac{13}{12} \left(v'' h^2+\frac{1}{12} v^{(4)} h^4\right)^2
   +\frac{3109}{2280} \left(v^{(3)}h^3+\frac{1}{2}h^4 v^{(4)}\right)^2
   , \\
\beta_2 =
   &-\frac{1}{4} \left(v' h+\frac{1}{6} v^{(3)} h^3\right)^2
   +\frac{3}{4} \left(v'h+\frac{1}{12} v^{(4)} h^4\right)^2 \notag \\
   &+\frac{13}{12} \left(v'' h^2+\frac{1}{12} v^{(4)}h^4\right)^2
   +\frac{3109}{2880} \left(v^{(3)} h^3 -\frac{1}{2} h^4 v^{(4)}\right)^2
   , \\
\beta_3 =
   &-\frac{1}{8} \left(v' h-\frac{1}{3} v^{(3)} h^3+\frac{1}{4} v^{(4)} h^4\right)^2
   +\frac{9}{8} \left(v'h-\frac{1}{4} h^4 v^{(4)}\right)^2 \notag\\
   &+\frac{13}{12} \left(v''h^2 -\frac{11}{12} h^4 v^{(4)}\right)^2
   +\frac{3169}{2880} \left(v^{(3)}h^3 -\frac{3}{2} h^4 v^{(4)}\right)^2
   ,
\end{align}
with the truncation error of $\Order(v' h^6)+\Order(v'' h^7)+\Order(h^8)$.

If $v' \ne 0$, we get the analog of equation \eqref{eq_1stderk3},
\begin{equation}
\beta_r = (v'h)^2\left[1+\Order(h^2)\right], \quad r = 0, 1, 2, 3;
\label{eq_1stderk4}
\end{equation}
further, for $v' = 0$, $v'' \ne 0$ we obtain the analog of equation \eqref{eq_2derderk3},
\begin{equation}
\beta_r = 13/12(v''h^2)^2\left[1+\Order(h^2)\right], \quad r = 0, 1, 2, 3.
\label{eq_2ndderk4}
\end{equation}
This shows that the smoothness indicators for WENO-7 (with $k=4$) are capable of
handling the extrema, where $v'=0$, without any reduction in order: the order of the scheme is then
$\Order(h^{2k-1}) \sim \Order(h^7)$. However, when both $v'=0$ and $v''=0$,
\begin{equation}
\beta_r = {1043}/{960}(v^{(3)}h^3)^2[1+\Order(h)], \quad r = 0, 1, 2, 3,
\label{eq_3rdderk4}
\end{equation}
so the order of the scheme reduces to $\Order(h^{2k-2}) \sim
\Order(h^6)$ for this case.

More generally, consider a WENO-$(2k-1)$ scheme, which uses stencils of length $k$. As
can be verified for each particular $k$, the numerical approximations to $v'$,
\dots, $v^{(k-2)}$, that the smoothness indicators are comprised of, have
truncation errors that do not contain any terms proportional to $v'$, \dots,
$v^{(k-2)}$. Therefore, in analogy with the above discussion of the $k=3$ and
$k = 4$ cases, the WENO scheme that uses stencils of length $k$ can be shown to
be able to handle the maxima without reducing the reconstruction order if at
least one of the $1$st, $2$nd, \dots, $(k-2)$th derivatives at $x = x_i$ is
nonzero.

Note that for $k \le 3$ the smoothness indicators are the same for all
reconstruction types: \atoe, $\langle v\rangle_i \rightarrow v_{i+1/2}$, \ctoe,
$v_i \rightarrow v_{i+1/2}$, \atoc, $\avg{v}_i \rightarrow v_{i}$, and \ctoa,
$v_{i} \rightarrow \avg{v}_i$, reconstructions and are given by equations
\eqref{eq_bn}. This is because for $ k \le 3$ the difference between
cell centre and cell average values is the same for all points in the stencil.
However, for larger $k$, this difference starts to vary from point to point
within the stencil, and for $k \ge 4$ the smoothness indicators become
different for these types of reconstruction. So, the $4$th order smoothness
indicators presented here for \ctoe, $v_i \rightarrow v_{i+1/2}$,
reconstruction are different from the smoothness indicators given by
\citet{bal00} for the \atoe, $\avg v_i \rightarrow v_{i+1/2}$, reconstruction.

\subsection{Algorithm for reducing the stencil size} \label{sec_reduction}

WENO schemes only operate on stencils of a fixed
length and this length is the smallest the stencils can get (\eg, it is~$3$ for
the case of WENO-$5$). There are several cases when there may not be large
enough smooth region for a stencil of a large size to fit in, \eg\ (1)~for a
reconstruction between two shocks that are about to interact, (2)~in the case
of a state that has to be `built-up', and (3)~for the grid cells
inside unresolved discontinuities \citep{har87}.

We have developed an algorithm of adaptive reduction of the stencil size: the
algorithm locates the grid cells inside sharp discontinuities in the flow, and
a lower-order \mbox{WENO-$3$} reconstruction~\citep{shu97} is used there. In this
algorithm we do not rely on the pressure jumps \citep[as is done in the PPM
algorithm, see, \eg,][]{col84} as indicators of shocks because in the
supersonic regime, where the pressure is negligible and, possibly, noisy, this
approach may lead to excessive reduction of the order of the scheme.

The information about the discontinuities in the flow is actually encoded in the
weights~\eqref{eq_weno_unnorm_weights} used by the WENO reconstruction. For the
purposes of the stencil reduction algorithm described below, we compute the
unoptimized weights that are obtained by setting $d_r =1$ in~\eqref{eq_weno_unnorm_weights},
which, given the smoothness indicators, is a fast computation.
We want to find out if we need to reduce
the order of reconstruction for the $i$th grid cell and we are using the WENO-$5$
scheme with $k = 3$. Suppose, the weight $\omega_0^i$ is very large compared to $\omega_2^{i+2}$,
\ie~the flow around the $i$th cell has a steeper profile than that near the
$(i+2)$nd cell. This is an indication that the $i$th cell is located inside a discontinuity,
so we use the lower-order WENO-$3$ reconstruction there~\citep{shu97}.
Namely, for $i$th grid cell we define the ratio
\begin{equation}
\mathcal{R} = \mathcal{F}\left(\frac{\omega_0^i}{\omega_2^{i+2}},\frac{\omega_2^i}{\omega_0^{i-2}}\right),
\end{equation}
where we choose function $\mathcal{F}$ below depending on the reconstruction type, and define
the fraction of the WENO-$3$ reconstruction that we use in that grid cell as
\begin{equation}
\alpha_3 = \max\biggl[0,\min\Bigl(\frac{\mathcal{R}-\mathcal{R}_\text{min}}{\mathcal{R}_\text{max}-\mathcal{R}_\text{min}}, 1\Bigr)\biggr],
\end{equation}
so the fraction of the WENO-5 reconstruction is $(1 - \alpha_3)$.
This means that for
the ratios of weights $\mathcal{R}< \mathcal{R}_\mathrm{min}$ we fully use the WENO-5
reconstruction procedure, for $\mathcal{R} > \mathcal{R}_\mathrm{max}$ we fully use the WENO-3 one,
and linearly transition between them for intermediate $\mathcal{R}$.
For the \ctoe\ reconstruction we use $\mathcal{F}(x,y) = \max(x,y)$,
$\mathcal{R}_\text{min} = 1.3$, and $\mathcal{R}_\text{max} = 1.6$. This
provides reduction in the order of reconstruction in unresolved discontinuities whose width is smaller than about the
stencil size ($5$ points for WENO-$5$) and avoids reduction in the kinks of the flow (discontinuities
in first derivative).  For \ctoa\ and
\atoc\ reconstructions we use $\mathcal{F}(x,y) = \min(x,y)$, $\mathcal{R}_\text{min} = 10$, and
$\mathcal{R}_\text{max} = 15$.  This allows reduction in unresolved discontinuities
and in the kinks of the flow.  Lastly, the stiff relativistic regime where the
effective Lorentz factor $\gamma_\text{eff}\equiv \gamma(\rho+u_g+p_g)/\rho \geq10$
stresses higher-order methods since,
unlike with TVD methods, the interpolated value can lie outside of the range given by the surrounding values.  To avoid
mild oscillations in the ultrarelativistic test in section~\ref{sec_test_1d_rel_105}, we use the latter
reduction method for all reconstructions if $\gamma_\text{eff}\geq 10$.  This helps by keeping
the interpolations consistent with each other and only makes a mild difference near
the shock in this single ultrarelativistic test.

In smooth flows where the first and second derivatives do not vary much from
grid cell to grid cell WENO-5 maintains high order accuracy: all weights
are on the order of $1/3$, and the reduction does not operate. However,
if there is a discontinuity in the flow, the weights for the points that are near
it will vanish for the stencils crossing the discontinuity, so a lower order
reconstruction
will be used for the points inside of the discontinuity.  Similar to what is done in the
flattening algorithm used by PPM \citep{col84,mig05}, we additionally reduce the order near
shocks but avoid such a reduction near contact discontinuities.  For each grid cell
we modify $\alpha_3$ as follows:
\begin{align}
\alpha_3 &= \max\left(\alpha_3, \, {\mathcal{S}}_i\,\alpha_{3,i-1}, \, {\mathcal{S}}_i\,\alpha_{3,i+1}\right),
\end{align}
where $\alpha_{3,i\pm1}$ are the unmodified values of $\alpha_3$ at the grid cells adjacent
to the $i$th one and $\mathcal{S}_i$ is an indicator of the shock strength that
vanishes for no shock and becomes unity for a strong shock:
\begin{align}
\mathcal{S}_i &=  \max\left[0,\min%
      (
      1, \, 4\,\tilde{\mathcal{S}}_i-1
      )\right],\\
\tilde{\mathcal{S}}_i &= \frac{\rho_i\Delta\Abs{u^t(1+u_t)}+\Delta\Abs{(p_g+u_g)u^tu_t+p_g}}{\left[\rho u^t(1+u_t)+(p_g+u_g)u^tu_t+p_g\right]_i}.
\end{align}
$\tilde{\mathcal{S}}_i$ has the meaning of the relative jump in the energy $(-\tensor{T}{t}{t}-\rho u^t)$, eq.~\eqref{eq_integrals},
and $\Delta q = q_{i+1}-q_{i-1}$.

We introduce \emph{reduced} WENO-5 weights,
$\tilde\omega_r = (1-\alpha_3) \omega_r$, and rewrite the reconstruction result,
\begin{equation}
f_\mathrm{WENO} = \tilde f_5 + \tilde\alpha_3 f_3,
\end{equation}
where $\tilde f_5$ is the reconstruction due to WENO-$5$ with the reduced weights,
$f_3$ is the reconstructed value due to the WENO-$3$,
and $\tilde\alpha_3 = 1 - \sum_r\tilde\omega_r$.  We will modify the reduced
WENO-$5$ weights in the following paragraph.

We perform the integration of source terms in the usual component-by-component way.
However, we have to be more careful for \atoc\ and \ctoa\ reconstructions
since they operate on the conserved quantities. The difference between cell averaged and
cell centred values cannot be treated in a pure component-by-component way
because of the nonlinear coupling between the conserved quantities.
For example, we would
like to avoid using different stencils in \atoc~reconstructions for total
conserved energy and components of the total conserved momentum.
For the conserved quantities  we separately
perform the reconstruction of the energy component in the way described in the previous paragraph.
For each of the other components we modify the
reduced WENO-5 weights to be equal to the minimum of the reduced WENO-5 weights
for the energy component and
ten times the WENO-5 weights for that component. This way in smooth regions for the reconstruction of
all conserved quantities we use a common set of weights computed for the total
conserved energy.
Similarly, for the \ctoa\ reconstruction of fluxes in $i$th direction,
we use the same procedure as described above for the conserved quantities, but
instead of using a common set of weights computed for the energy component,
we use a common set of weights computed
for the flux of $i$th component of momentum in the $i$ the direction.

We make the reconstruction even more robust by reducing its order
in non-smooth features of the flow that we refer to as cusps (defined below).
Further, if there is a cusp in the~$\gamma$-factor of the flow
in a grid cell and cell to cell change of $\gamma$ is larger than $1$\%, 
we do not perform the \ctoa\ reconstruction of the fluxes
whose values depend on the
reconstruction of $\gamma$ that has the cusp (since in this case the
fluxes, which are steep functions of~$\gamma$, are not smooth). We now explain
what we mean by a cusp. Define $f'_{i+\myfrac{1}{2}} = f_{i+1}-f_i$, $f''_i =
f'_{i+\myfrac{1}{2}} - f'_{i-\myfrac{1}{2}}$. First, consider the case of an
increasing function, $f'_{i-\myfrac{1}{2}} > \sqrt\varepsilon \norm{f}$; we
later will generalize the procedure. We then define a point $i$ to be in a cusp
if it is located between an inflection point (\ie, $f''_{i-1} >
\sqrt\varepsilon \norm{f}$ and $f''_{i} < -\sqrt\varepsilon \norm{f}$) and a
local maximum with one-sided derivatives different by no more than $50$\% by
absolute value (\ie~either
$4\abs{f'_{i+\myfrac{1}{2}}+f'_{i-\myfrac{1}{2}}}<\abs{f'_{i+\myfrac{1}{2}}}+\abs{f'_{i-\myfrac{1}{2}}}$
or
$4\abs{f'_{i+\myfrac{3}{2}}+f'_{i-\myfrac{1}{2}}}<\abs{f'_{i+\myfrac{3}{2}}}+\abs{f'_{i-\myfrac{1}{2}}}$).
In general, we define the function to contain the cusp at the point
$i$ if either the above is true or if it becomes true after the flip of the
function sign, grid direction, or both. For consistency, we choose $\epsilon$ to be
the same as the one in the WENO weights computation
procedure~\eqref{eq_smoothnessindnew},
\eqref{eq_smoothnessindnorm}, \eqref{eq_weno_unnorm_weights},
\eqref{eq_weno_norm_weights}.

\subsection{Failsafe integration}
\label{sec_failsafeint}

As in other codes~\citep{zha06,gam03}, we implement certain safety features
that enable our code to succeed in especially stiff regimes. In relativistic
flows, there are states of conserved quantities~\eqref{eq_conserve} for which
there exists no corresponding set of primitive quantities~\eqref{eq_primitives}
at all. Further, a change in conserved quantities of less than~$1$\% may
correspond to an \emph{infinite} change in primitive quantities. Therefore,
any reconstruction operations directly performed on the conserved quantities or
on the fluxes may in principle lead to unphysical states or instabilities. This
is the reason why we carry out the crucial \ctoe~reconstruction
\ref{enum_c2e} (see section~\ref{sec_algorithmoutline}) on the primitive quantities instead of conserved quantities or
fluxes: this guarantees that we always obtain a physically meaningful state at
the interface and, therefore, a physically meaningful flux. However, the update
to the conserved quantities due to this flux may sometimes lead to an
unphysical state with a negative internal energy or density, or, as explained
above, the corresponding state of primitive quantities may not at all exist.
Here we describe the procedure that we follow in such cases in order to make
the integration failsafe.

In our scheme the de-averaging and  averaging reconstructions,~\ref{enum_c2a} and
\ref{enum_a2c} (see section~\ref{sec_algorithmoutline}),
change the representation of conserved quantities but do not
alter the locations of those. We avoid using the~\atoc~reconstruction if it makes a
large difference for the total energy density in the fluid frame, $(\rho+u_g)$
in the hydrodynamic case we consider in this paper.  In particular, we make sure that the
primitive quantities~$\mathbf{p}_c$, that correspond to cell centred conserved quantities,
are not very different from~$\mathbf{p}_a$, that correspond to
cell averaged conserved quantities: we use $\mathbf{p}_c$ if its difference
from~$\mathbf{p}_a$ in the value of energy density in the fluid frame  is smaller than $5$\%,
use~$\mathbf{p}_a$ if that difference is larger than $10$\%, and linearly transition
between these two values for an intermediate difference.  If no
primitive state $\mathbf{p}_c$ has been found by the inversion, we use
$\mathbf{p}_a$.  This locally lowers the order of the scheme
in stiff regimes where operations on the conserved quantities are not safe, so
there our scheme becomes equivalent to the schemes that do not distinguish
between the average and point values.  Note, that this procedure does not
affect the asymptotic order of convergence of the scheme since the difference between
the cell averaged and cell centred values is $\Order(h^2)$.

Occasionally, during the evolution negative rest-mass density or
internal energy might occur. If a negative rest-mass density
(internal energy) implies an imaginary sound speed, we set the sound speed
to unity (zero) and continue the evolution. This adjustment of the sound speed
only affects the diffusive flux used in the approximate HLL Riemann solver.
Most of such negative rest-mass density and internal energy occurrences are due to
inaccuracy of trial Runge-Kutta time steps and do not appear on the final
Runge-Kutta time steps. Such behaviour is acceptable since at the end
we get a high order accurate answer, and it does not matter that some of the
intermediate steps that lead to this answer were unphysical.
For the same reason we allow negative internal energy/baryon densities to occur at the final
Runge-Kutta time steps. The idea is the same: to temporarily allow the solution
to go out of the physical states space with the hope of it finally reaching a
well-defined accurate state. For instance, negative internal energy/baryon
densities may occur in front of a strong shock, however, we find that once the
shock passes through, all of the states that were temporarily unphysical are
well-defined.  Lastly, we implement a minor diffusive correction to the internal energy
under special circumstances in order the improve the fidelity of the internal energy near
shocks (as for a few cells near the shock in the caustic test in section~\ref{sec_test_1d_caustics}). If the
internal energy is negative but all surrounding values are larger,
then the internal energy is considered to be an isolated failure and its value is chosen
to be the smallest of the surrounding values in a way that leads to a symmetric result in multiple dimensions.

Finally, we apply some diffusive corrections to the primitive quantities used to obtain the
flux.  In rare cases that no state of primitive quantities can be found that
corresponds to the state of the conserved quantities (i.e. an
inversion from conserved to primitive quantities does not exist) we use
the average of the existing surrounding states of the primitive quantities.
This situation occurred only twice in
the two-dimensional relativistic Riemann problem (section~\ref{sec_test_2d_relriem})
in the highly relativistic part of the flow ($\gamma \approx 25$) and did not occur in any other
test problems we have performed.
If for all surrounding grid cells no primitive states
can be found, we use the average of states of primitive quantities in the surrounding
grid cells from the previous time step.

Note that the above features only alter the state of primitive quantities that
are used in order to compute interface fluxes, therefore, the scheme remains
conservative since the values of the conserved quantities are affected only
through the fluxes.

\subsection{Catastrophic cancellation in nonrelativistic problems}
\label{sec_cancellaton}

In order to be able to successfully study both highly relativistic and
nonrelativistic problems and avoid being severely limited by finite
numerical precision\footnote{Even though we
normally use the code at double precision, we have made it capable of
performing computations at long double precision.  For this, we use the long double
versions of exponential, trigonometric, \etc, functions from Cephes Math Library
Release 2.7 and use the -long\_double option for the Intel compiler.},
we have to use the a special technique for performing
the rest-mass subtraction in~\eqref{eq_integrals} and~\eqref{eq_fluxes}.
For instance, in the calculation of conserved quantities from the primitive ones,
given a $4$-velocity $u^\mu$, we need to find the kinetic energy, \ie~the difference between the total
bulk motion energy and the rest-mass energy in the coordinate basis frame,
\begin{equation} \label{eq_catastr_cancel1}
\rho u^t (-u_t - 1).
\end{equation}
A similar procedure is followed in~\citet{alo99} for the special relativistic equations of motion.
Further, the inverse process of computing the primitive quantities from
conserved quantities involves computing a similar expression
\begin{equation}
\rho \gamma (\gamma - 1).
\label{eq_catastr_cancel2}
\end{equation}
For nonrelativistic flows the value of $\gamma$ may get so close to unity that
within machine precision its numerical value is $1.0$.
This is why we cannot simply plug in such a value for $\gamma$ in the above
formula to compute the kinetic energy.
For double precision, this happens for $3$-velocity values of $v
\lesssim 10^{-8}$. We alleviate the above problem
by rewriting part of expression~\eqref{eq_catastr_cancel2}
for kinetic energy with the help of the velocity information:
\begin{equation} \label{eq_catastr_cancel_nonrel}
\gamma - 1 = \frac{1 - \sqrt{1-v^2}}{\sqrt{1-v^2}} = \frac{\gamma v^2}{1 + \sqrt{1-v^2}}
    = \frac{\gamma^2 v^2}{1 + \gamma},
\end{equation}
where the final expression does not contains catastrophic cancellations and
due to the particular choice of frame the Lorentz factor has
the conventional form $\gamma = (1-v^2)^{-1/2}$~\citep{nob06}.

In equation~\eqref{eq_catastr_cancel1}
we can similarly rewrite the problematic part as follows:
\begin{equation} \label{eq_catastr_cancel_rel}
- u_t-1 = - g_{it} u^i + 2\phi\frac{1-g_{tt}}{1+\hat\gamma}+\frac{\hat\gamma^2 \hat v^2}{1+\hat\gamma},
\end{equation}
where $\hat\gamma = - g_{tt} u^t$ and
$\hat v^2 = 2g_{it}v^i -2\phi + g_{ij}v^i v^j$ play the roles
of the $\gamma$-factor and the square of the velocity,
$\phi = -(1+g_{tt})/2$ is the gravitational potential and
is small compared to unity for problems involving nonrelativistic
gravity, so for such problems we separately store its value.
Similarly, we separately compute and store the value of $(g^{tt}-1)$,
which appears when converting the total energy with rest-mass subtracted
from the lab frame to the normal observer frame~\citep{nob06,mig07}.

The general expression \eqref{eq_catastr_cancel_rel} avoids catastrophic cancellations both for relativistic
problems and for problems involving nonrelativistic velocities and
gravitational fields.  We use this expression when computing the conserved
quantities from the primitive ones as well as for the inverse process.
Since expression~\eqref{eq_catastr_cancel_rel} is written in an arbitrary frame, it
involves two additional terms as compared to~\eqref{eq_catastr_cancel_nonrel}:
the first one appears for a metric
that has space-time mixing, the second one is due to a nonzero gravitational
potential, and the third one corresponds to~\eqref{eq_catastr_cancel_nonrel}.

\subsection{Catastrophic cancellation in ultrarelativistic problems}
\label{sec_cancellatonrel}

For the ultrarelativistic regime one can show that the inversion method we
use~\citep{mig07} that converts conserved to primitive quantities
leads to well-defined relative errors in the primitive quantities for a machine accurate
set of conserved quantities.   In the ultrarelativistic hydrodynamic case,
for a conserved lab-frame energy density ($E$), conserved mass density ($D$), and
conserved momentum density ($P^\alpha$), the Lorentz factor is given by
\begin{equation}
\gamma = \frac{E}{\sqrt{E^2-P^2}} ,
\end{equation}
which can be used to obtain the rest-mass density
\begin{equation}
\rho = \frac{D}{\gamma} ,
\end{equation}
the relative 4-velocity
\begin{equation}
\tilde{u}^\alpha = \gamma \tilde{v}^\alpha = \gamma \frac{P^\alpha}{E} ,
\end{equation}
and the quantity $\chi\equiv u_g + p_g$,
\begin{equation}
\chi = \frac{E}{\gamma^2} - \frac{D}{\gamma} ,
\end{equation}
which with an equation of state can be used to determine both $u_g$ and $p_g$.
For this inversion we have neglected the term in the energy equation proportional
to $p_g$ compared to $\gamma^2(\rho+\chi)$ and below we state
how this limits the remaining arguments.

Given $E$, $P$, and $D$ with an accuracy limited by a machine error of
$\epsilon_\text{machine}$, one can show that the relative errors in primitive
quantities are given by
\begin{align}
\frac{d\gamma}{\gamma} &\sim \left(\frac{\gamma}{\gamma_\text{max}}\right)^2 \equiv \epsilon_\text{machine} \gamma^2,\\
\frac{dv^\alpha}{v^\alpha} &\sim \epsilon_\text{machine} ,\\
\frac{d\chi}{\chi} &\sim \left(\frac{\rho+\chi}{\chi}\right)\left(\frac{\gamma}{\gamma_\text{max}}\right)^2 ,\\
\frac{d\rho}{\rho} &\sim  \frac{d\gamma}{\gamma} .
\end{align}
These error estimates are valid as long as
\begin{equation}
\frac{\gamma}{\gamma_\text{max}} \gg \left(\frac{\chi}{\rho\gamma^2_\text{max}}\right)^{1/4}.
\end{equation}
This implies a
smaller than order unity error for the primitive quantities when
\begin{align}
\gamma &\lesssim \gamma_\text{max} \equiv \epsilon_\text{machine}^{-1/2} ,\\
\chi &\gtrsim \chi_{\rm min} \equiv \frac{\rho(\gamma/\gamma_{\rm max})^2}{1-(\gamma/\gamma_{\rm max})^2} ,
\end{align}
and the error in $\rho$ only sets the maximum allowed $\gamma$.  For example, using double precision on a 32-bit machine
gives conserved quantities that have at best a relative error
of $\epsilon_\mathrm{machine}\approx 2.2\times 10^{-16}$ giving
\begin{align}
\gamma_{\rm max} &= 6.7\times 10^7 .
\end{align}
For example, the ultrarelativistic test discussed in~\citet{alo99} (see
section~\ref{sec_test_1d_rel_105}) with $\gamma=2.24\times 10^5$ and $p=7.63\times 10^{-6}$
for $\Gamma=4/3$ gives $\chi \gtrsim \chi_\text{min}$, so their
test is at the limit of resolving the pre-shock pressure for double precision.
The post-shock region will be resolved as long as the pre-shock Lorentz factor is resolved.

Finally, an additional source of error can be incurred when an iterative inversion method uses the expression
\begin{equation}
\gamma = (1-v^2)^{-1/2} ,
\end{equation}
as compared to
\begin{equation}
\gamma = \sqrt{1+\tilde{u}^2} ,
\end{equation}
since repeated use of the prior expression leads to a cumulated catastrophic
errors not present when using the latter expression. For this reason, in practice the
GRMHD iterative solver of~\citet{nob06} is limited to $\gamma\lesssim 10^5$
when using double precision.

\bibliographystyle{mn2e}

\label{lastpage}
\end{document}